\documentclass[a4paper,11pt]{article}
\pdfoutput=1 

\usepackage{jheppub} 

\usepackage[T1]{fontenc} 

\usepackage{caption} 
\usepackage{slashed} 
\usepackage{verbatim} 
\DeclareMathOperator{\sgn}{sgn} 

\title{\boldmath 
Aspects of compactification on a linear dilaton background}

\author[a,b]{I. Antoniadis,}
\author[c]{C. Markou,}
\author[a]{and F. Rondeau}

\affiliation[a]{Laboratoire de Physique Th\'eorique et Hautes Energies - LPTHE\\ Sorbonne Universit\'e, CNRS, 4 Place Jussieu, 75005 Paris, France}
\affiliation[b]{Institute for Theoretical Physics, KU Leuven, Celestijnenlaan 200D, B-3001 Leuven, Belgium}
\affiliation[c]{Max-Planck-Institut f\"ur Physik (Werner-Heisenberg-Institut)\\F\"ohringer Ring 6, 80805 Munich, Germany}

\emailAdd{antoniad@lpthe.jussieu.fr}
\emailAdd{cmarkou@mpp.mpg.de}
\emailAdd{francois.rondeau@lpthe.jussieu.fr}

\abstract{We consider the most general Kaluza-Klein (KK) compactification on $S^1/\mathbb{Z}_2$ of a five dimensional ($5D$) graviton-dilaton system, with a non-vanishing dilaton background varying linearly along the fifth dimension. We show that this background produces a Higgs mechanism for the KK vector coming from the $5D$ metric, which becomes massive by absorbing the string frame radion. The $\mathcal{N}=2$ minimal supersymmetric extension of this model, recently built as the holographic dual of Little String Theory, is then re-investigated. An analogous mechanism can be considered for the $4D$ vector coming from the (universal) $5D$ Kalb-Ramond two-form. Packaging the two massive vectors into a spin-$3/2$ massive multiplet, it is shown that the massless spectrum arranges into a $\mathcal{N}=1$, $D=4$ supersymmetric theory. This projection is compatible with an orbifold which preserves half of the original supersymmetries already preserved by the background. The description of the partial breaking $\mathcal{N}=2\rightarrow\mathcal{N}=1$ in this framework, with only vector multiplets and no hypermultiplets, remains an interesting open question which deserves further investigation.}

\begin{document}

\begin{flushright}

\hfill{MPP--2021--102}

\vspace{15mm}

\end{flushright}
 
\maketitle
\flushbottom

\newpage
\section{Introduction}
\label{sec:intro}

Little String Theory (LST) is a 6-dimensional non-gravitational theory, obtained for instance in type IIB or heterotic string theory by considering a stack of $k$ coincident (Neveu-Schwarz) NS5-branes, in the limit of vanishing string coupling constant $g_S$ \cite{Berkooz:1997cq, Review_LST}. With closed string amplitudes being proportional to $g_S$, in this limit the branes and bulk dynamics decouple. However, and in constrast with the D-brane case, the gauge coupling of the effective $U(k)$ gauge theory on the NS5-branes is independent of $g_S$ and therefore remains finite in the limit $g_S\rightarrow 0$. We are left with an interacting non-gravitational theory living on the NS5-branes.

Being a non-local and strongly coupled theory without any Lagrangian description, LST is easier studied through its 7-dimensional holographic dual, obtained in the near horizon limit of the NS5-branes. It is a weakly coupled string theory on the background $\mathbb{R}^{5,1}\times\mathbb{R}_y$, with the dilaton varying linearly in the coordinate $y$ of the real line $\mathbb{R}_y$ \cite{Duality_LST_LD}. The main features of LST phenomenology can be captured in a simpler model where two dimensions of the bulk are compactified on $T^2$, leading to a gravitational holographic dual of LST on $\mathbb{R}^{3,1}\times\mathbb{R}_y$ whose action is a simple graviton-dilaton model with a runaway scalar potential \cite{Antoniadis:2001sw, Pheno_LST}.

The vanishing limit of the string coupling constant leads to an interesting phenomenological application of LST in the context of the hierarchy problem~\cite{Pheno_LST, Baryakhtar:2012wj}. 
The string scale can be indeed separated from the Planck mass to much lower energies, such as in the (multi)TeV region using an ultra-weak string coupling. The
hierarchy problem then amounts to explain the smallness of the coupling~\cite{Radion_pheno}. This question has also been addressed more recently in the so-called clockwork mechanism~\cite{Choi:2015fiu}, which was shown to have as continuum limit the five-dimensional ($5D$) linear dilaton theory~\cite{Giudice:2016yja, Craig:2017cda}.

In order to obtain a finite string coupling and thus a four-dimensional Planck mass, the extra dimension $\mathbb{R}_y$ is compactified on a segment $S^1/\mathbb{Z}_2$. The dimensional reduction on a linear dilaton (LD) background reveals very distinguishable features. In particular the generic spectrum, such as that of the 
graviton, is 
a massless $4D$ zero mode with a flat wave-function along the extra dimension, followed by an infinite tower of Kaluza-Klein (KK) states starting from a mass gap fixed by the slope of the linear dilaton. In this paper, we first extend this analysis to the KK gauge sector of the metric which has not been studied so far. It turns out that the situation is different from the scalar and spin-2 excitations: due to the non-trivial background, we show that the zero mode of the KK vector acquires a mass by absorbing the scalar radion, while its wave-function is no longer flat along the extra dimension but rather localised around the origin and away exponentially suppressed. This result follows from a theoretical gauge symmetry analysis, which generalises the standard KK reduction in the more general case where the fields may depend on the extra coordinate. Within the gravitational sector, the reduction is performed in the ADM formalism~\cite{ADM}, crucially taking into account the Gibbons-Hawking boundary terms at the end-points of the interval.

Since the holographic dual of LST is actually a string theory, one can study its effective supergravity theory, a construction which has been first carried out in \cite{Kehagias:2017grx, IA_CM}. It relies on a $U(1)_R$ gauging of the $SU(2)$ R-symmetry of $\mathcal{N}=2$, $D=5$ supergravity coupled to one vector multiplet~\cite{GST, Ungauged_5D_sugra, Gauged_5D_sugra}. Besides minimal couplings between the gauge field and the fermions, the gauging generates fermion mass terms as well as a scalar potential, both highly constrained by supersymmetry. The family of the scalar potentials obtained in this way is parametrised by two independent parameters. The observation made in \cite{Kehagias:2017grx, IA_CM} is that the vanishing of one parameter precisely reproduces the runaway potential leading to the LD background solution. Here we revisit this construction in order to show that the five-dimensional supergravity theory with this property is actually unique. 

In addition to possible phenomenological implications mentioned above, the gravity dual of LST can also be used as a framework to study supersymmetry breaking, which will be the viewpoint adopted in the second part of our paper. The crucial point is that the LD background preserves ${1}/{2}$ of the original supersymmetries~\cite{IA_CM}. We shall show that the NS5-branes, already introduced at the bosonic level by the junction conditions, remain consistent in a supersymmetric context, namely they preserve the remaining supersymmetries, and this choice of branes is unique. Motivated by the massive vector field obtained in the KK reduction of the metric, we perform the full dimensional reduction of the bosonic sector of the supergravity action in order to find a similar mechanism in the Maxwell sector. Obtaining a second massive vector, one can arrange the massive (non KK) spectrum into a massive $4D$ ${\cal N}=1$ spin-${3}/{2}$ multiplet which contains half of the total degrees of freedom. With the two free parameters of the theory, the slope $\alpha$ of the LD background and the compactification radius $L$, one can then consider two different limits. 
\begin{itemize}
\item
The low energy limit $|\alpha|\rightarrow\infty$, $L\rightarrow 0$, where all masses are sent to infinity and only the massless spectrum remains, which we show to be described by an effective $D=4$, $\mathcal{N}=1$ supergravity. 
Moreover, such a truncation is consistent with a $\mathcal{N}=2\rightarrow\mathcal{N}=1$ orbifold projection, under which half of the degrees of freedom are assigned a $\mathbb{Z}_2$-odd parity. 
\item
The intermediate limit where $\alpha$ is kept finite and $L\rightarrow 0$, keeping all massive and massless zero modes and projecting out only the KK excitations. The possibility to describe a consistent $4D$ supergravity theory in this limit is not clear and
remains a non-trivial open question. This would potentially lead to a very interesting example of partial supersymmetry breaking $\mathcal{N}=2\rightarrow\mathcal{N}=1$ using only vector multiplets.
\end{itemize}

This article is organised as follows. In Section \ref{sect:LD_model}, we review the main characteristics of the five dimensional graviton-dilaton system, in the presence of a runaway scalar potential corresponding to a non-critical string, stressing some important points often left implicit in the litterature. The compactification of this model on a LD background is then performed in Section \ref{sect:KK_reduction}, shedding light in particular on the KK gauge sector. The minimal supersymmetrisation of this model, seen as the holographic dual of LST, is then introduced in Section \ref{sec:minimal_susy_extension}. We list the different supergravity theories coupled to one vector multiplet whose gauging reproduces the scalar potential of LST and show that they are all equivalent. We then study the supersymmetries preserved both by the background and by the NS5-branes sitting at the boundaries. The complete compactification down to $D=4$ is presented in Section \ref{sect:compactified_theory}, where we highlight a similar Higgs mechanism for a second vector, and show that a $\mathcal{N}=1$ supersymmetric theory can be obtained in the massless limit. Our conclusions are presented in Section \ref{sect:conclusion}. Finally, the paper contains four appendices summarising our conventions and notations (Appendix \ref{sect:append_conventions}), revewing some important aspects of General Relativity on a manifold with boundaries (Appendix \ref{sect:append_ADM}) as well as the formalism of $\mathcal{N}=2$, $D=5$ supergravity, together with its $U(1)_R$ gauging (Appendix \ref{sect:append_5D_sugra}). The last Appendix \ref{sect:append_heterotic_string} identifies the $5D$ supergravity studied here with the effective action of a (non-critical) heterotic string theory. 

Our new results are presented in Sections \ref{sect:KK_reduction}, 
\ref{sec:runaway_scal_pot}, \ref{sect:susy_branes} and 
\ref{sect:compactified_theory}, as well as in Appendix \ref{sect:append_ADM_2}.

\section{The Linear Dilaton model}
\label{sect:LD_model}
\subsection{The $5D$ theory on $\mathbb{R}^{1,3}\times S^1/\mathbb{Z}_2$}
\label{sect:5D_theory}
The work carried out in this paper is based on the five-dimensional dilaton-gravity theory whose action is given by
\begin{equation}\label{eq:action_string_frame_bulk}
S=\int d^5x\sqrt{-g}e^{-\sqrt{3}\phi}\left[\frac{1}{2}\mathcal{R}^{(5)}[g_{MN}]+\frac{3}{2}(\partial\phi)^2-\Lambda\right],
\end{equation}
where $g_{MN}$ is the five-dimensional metric in the string frame (not to be confused with the Einstein frame metric $G_{MN}$ which will be introduced below), $\phi$ the dilaton field, and $\Lambda$ a constant introducing a runaway dilaton potential, characteristic of non-critical string theory. Varying $S$ with respect to $\phi$ and $g^{MN}$ yields the equation of motion of the dilaton and the five-dimensional Einstein equations, respectively given by:
\begin{eqnarray}
\label{eq:dilaton_eom_bulk}
\sqrt{-g}&e^{-\sqrt{3}\phi}&\left\{\frac{\sqrt{3}}{2}\mathcal{R}^{(5)}-\frac{3\sqrt{3}}{2}(\partial\phi)^2+3\Box_5\phi-\sqrt{3}\Lambda\right\}=0,\\
\label{eq:g_{MN}_eom}
\sqrt{-g}&e^{-\sqrt{3}\phi}&\left\{R_{MN}-\frac{1}{2}g_{MN}\mathcal{R}^{(5)}+\sqrt{3}\nabla_M\partial_N\phi-\sqrt{3}g_{MN}g^{PQ}\nabla_P\partial_{Q}\phi\right.\nonumber\\
&&\qquad\left.+\frac{3}{2}g_{MN}(\partial\phi)^2+g_{MN}\Lambda\right\}=0.
\end{eqnarray}
One can easily check that these equations are solved by the five-dimensional Minkowski metric, in addition of a linearly varying dilaton along the fifth direction $y$, breaking the $5D$ Poincar\'e invariance into a $4D$ one,
\begin{eqnarray}
g_{MN}&=&\eta_{MN},\\
\label{eq:lin_dil_sol}
\phi&=&\alpha y,
\end{eqnarray}
provided that the bulk ``cosmological constant'' $\Lambda$ and the parameter $\alpha$ are related by
\begin{equation}\label{eq:relation_lambda_alpha}
\Lambda=-\frac{3}{2}\alpha^2.
\end{equation}

In order to have a finite four-dimensional Planck mass, the fifth direction $\mathbb{R}_y$ must be compactified, on a space chosen to be an interval $S^1/\mathbb{Z}_2$ of length $L$, in complete analogy with the Randall-Sundrum model~\cite{Randall:1999ee}. The $\mathbb{Z}_2$ symmetry which must have the background therefore imposes to replace the solution \eqref{eq:lin_dil_sol} by
\begin{equation}
\phi=\alpha|y|.
\end{equation}
Denoting here and all along this paper differentiation with respect to the fifth coordinate $y$ with a prime, we have $\phi^{'}=\alpha\sgn y$\footnote{The sign function is defined as yielding $\pm 1$ depending of the sign of its argument. An ambiguity remains at $0$, and it is worth noticing that as long as $\sgn 0\neq 0$, no inconsistency appears and one can arbitrarily choose $\sgn 0=\pm 1$.} and $\phi^{''}=2\alpha\left[\delta(y)-\delta(y-L)\right]$, so that boundary terms will arise from the terms $\Box_5\phi$ and $g^{PQ}\nabla_P\partial_{Q}\phi$ of the equations of motion. Consistency of the equations of motion then requires to add boundary terms to the action \eqref{eq:action_string_frame_bulk}, which is modified into:
\begin{eqnarray}\label{eq:action_string_frame_boundary}
S&=&\int d^5x\left\{\sqrt{-g}e^{-\sqrt{3}\phi}\left[\frac{1}{2}\mathcal{R}^{(5)}[g_{MN}]+\frac{3}{2}(\partial\phi)^2-\Lambda\right]\right.\nonumber\\
&&\qquad\left.-\sqrt{-g_1}e^{-\alpha_1\phi}V_1\delta(y)-\sqrt{-g_2}e^{-\alpha_2\phi}V_2\delta(y-L)\right\},
\end{eqnarray}
with $g_1$ and $g_2$ the determinant of the induced metrics at the two four-dimensional boundaries. The additional terms describe branes located at the fixed points $y=0$ and $y=L$ of the orbifold $S^1/\mathbb{Z}_2$, and contain four parameters: $\alpha_1$ and $\alpha_2$ characterizing the nature of the branes, and their tensions $V_1$ and $V_2$. As we are now going to show, these parameters can be fully determined by the classical equations of motion following from the action \eqref{eq:action_string_frame_boundary}. 

The dilaton equation of motion \eqref{eq:dilaton_eom_bulk} is modified into
\begin{eqnarray}\label{eq:dilaton_eom}
\sqrt{-g}&e^{-\sqrt{3}\phi}&\left\{\frac{\sqrt{3}}{2}\mathcal{R}^{(5)}-\frac{3\sqrt{3}}{2}(\partial\phi)^2+3\Box_5\phi-\sqrt{3}\Lambda\right\}\nonumber\\
&-&\alpha_1\sqrt{-g_1}e^{-\alpha_1\phi}V_1\delta(y)-\alpha_2\sqrt{-g_2}e^{-\alpha_2\phi}V_2\delta(y-L)=0,
\end{eqnarray}
while the five-dimensional Einstein equations split into equations for the 4-dimensional scalar $g_{55}$, the 4-dimensional metric $g_{\mu\nu}$ and the 4-dimensional vector $g_{\mu 5}$ as:
\begin{eqnarray}
\label{eq:g_{55}_eom}
&R_{55}&-\frac{1}{2}g_{55}\mathcal{R}^{(5)}+\sqrt{3}\nabla_5\partial_5\phi-\sqrt{3}g_{55}g^{PQ}\nabla_P\partial_{Q}\phi+\frac{3}{2}g_{55}(\partial\phi)^2+g_{55}\Lambda=0,\\
\label{eq:g_{munu}_eom}
&\sqrt{-g}&e^{-\sqrt{3}\phi}\left\{R_{\mu\nu}-\frac{1}{2}g_{\mu\nu}\mathcal{R}^{(5)}+\sqrt{3}\nabla_{\mu}\partial_{\nu}\phi-\sqrt{3}g_{\mu\nu}g^{PQ}\nabla_P\partial_{Q}\phi+\frac{3}{2}g_{\mu\nu}(\partial\phi)^2+g_{\mu\nu}\Lambda\right\}\nonumber\\
&+&g_{\mu\nu}\left\{\sqrt{-g_1}e^{-\alpha_1\phi}V_1\delta(y)+\sqrt{-g_2}e^{-\alpha_2\phi}V_2\delta(y-L)\right\}=0,\\
\label{eq:g_{mu5}_eom}
&R_{\mu 5}&-\frac{1}{2}g_{\mu 5}\mathcal{R}^{(5)}+\sqrt{3}\nabla_{\mu}\partial_5\phi-\sqrt{3}g_{\mu 5}g^{PQ}\nabla_P\partial_{Q}\phi+\frac{3}{2}g_{\mu 5}(\partial\phi)^2+g_{\mu 5}\Lambda=0,
\end{eqnarray}
where greek indices $\mu,\nu\dots$ denote only $4D$ spacetime. The dilaton and 4-dimensional graviton equations of motion \eqref{eq:dilaton_eom} and \eqref{eq:g_{munu}_eom} can be respectively rewritten as:
\begin{eqnarray}
\frac{\sqrt{3}}{2}&\mathcal{R}^{(5)}&-\frac{3\sqrt{3}}{2}(\partial\phi)^2+3\Box_5\phi-\sqrt{3}\Lambda-\alpha_1g_{55}^{-\frac{1}{2}}e^{(\sqrt{3}-\alpha_1)\phi}V_1\delta(y)\nonumber\\
&-&\alpha_2g_{55}^{-\frac{1}{2}}e^{(\sqrt{3}-\alpha_2)\phi}V_2\delta(y-L)=0,
\end{eqnarray}
and
\begin{eqnarray}
R_{\mu\nu}&-&\frac{1}{2}g_{\mu\nu}\mathcal{R}^{(5)}+\sqrt{3}\nabla_{\mu}\partial_{\nu}\phi-\sqrt{3}g_{\mu\nu}g^{PQ}\nabla_P\partial_{Q}\phi+\frac{3}{2}g_{\mu\nu}(\partial\phi)^2+g_{\mu\nu}\Lambda\nonumber\\
&+&g_{\mu\nu}g_{55}^{-\frac{1}{2}}\left\{e^{(\sqrt{3}-\alpha_1)\phi}V_1\delta(y)+e^{(\sqrt{3}-\alpha_2)\phi}V_2\delta(y-L)\right\}=0.
\end{eqnarray}
It is then straightforward to check that they are still solved by $g_{MN}=\eta_{MN}$ and $\phi=\alpha|y|$, provided that the bulk ``cosmological constant'' $\Lambda$ and the parameter $\alpha$ are related by\footnote{One sees here a first inconsistency the choice $\sgn 0=0$ would imply: a cosmological constant $\Lambda$ constant everywhere except at $y=0$ where it would vanish, while $\sgn 0=\pm 1$ remain consistent.}
\begin{equation}
\Lambda=-\frac{3}{2}\alpha^2\sgn^2y.
\end{equation}
In addition, the boundary terms at $y=0$ and $y=L$ lead to four equations relating $\alpha$, $\alpha_1$, $\alpha_2$, $V_1$ and $V_2$,
\begin{align}
\label{eq:bound_cond_1}
6\alpha&=\alpha_1V_1	&&& 6\alpha=-\alpha_2e^{(\sqrt{3}-\alpha_2)\alpha L}V_2,\\
\label{eq:bound_cond_1}
2\sqrt{3}\alpha&=V_1	&&& 2\sqrt{3}\alpha=-e^{(\sqrt{3}-\alpha_2)\alpha L}V_2,
\end{align}
which are obviously solved by
\begin{eqnarray}
\label{eq:brane_antibrane_sol}
V_1&=&2\sqrt{3}\alpha=-V_2,\\
\label{eq:NS5brane_sol}
\alpha_1&=&\alpha_2=\sqrt{3}.
\end{eqnarray}
Eq. \eqref{eq:brane_antibrane_sol} thus shows that consistency of the equations of motion requires a system of branes of opposite tensions to set at the fixed points $y=0$ and $y=L$ of the $S^1/\mathbb{Z}_2$ orbifold, similarly to the Randall-Sundrum model.\footnote{In the string context, a negative contribution to the tension can be provided by orientifolds.}  Eq. \eqref{eq:NS5brane_sol} tells us about the nature of these branes: from the action \eqref{eq:action_string_frame_boundary}, one sees that $\alpha_1=\alpha_2=\sqrt{3}$ correspond to a brane tension $\propto \frac{1}{g_s^2}$, with $g_s^2=e^{\sqrt{3}\phi}$ the string coupling constant. Therefore, the solution at the boundaries arising from the classical equations of motion consists of NS5-branes, as expected from the LST string theory approach.

Having found the nature of the branes sitting at the two boundaries, one can then move from the string frame to the Einstein frame metric by means of the Weyl transformation
\begin{equation}\label{eq:5D_Weyl_transfo}
G_{MN}=e^{-\frac{2}{\sqrt{3}}\phi}g_{MN},
\end{equation}
where $G_{MN}$ is the Einstein frame metric and $g_{MN}$ the string frame one. The Einstein frame bulk action $S_{\mathcal{M}}$ then reads 
\begin{equation}\label{eq:bulk_action}
S_{\mathcal{M}}=S_{EH}+S_{dil}+S_{\Lambda},
\end{equation}
with the Einstein-Hilbert action $S_{EH}$, the dilaton action $S_{dil}$ and the $5D$ cosmological constant action $S_{\Lambda}$ given by:
\begin{eqnarray}
\label{eq:einstein_hilbert_action}
S_{EH}&=&\frac{1}{2}\int d^5x\sqrt{-G}\mathcal R^{(5)}[G_{MN}],\\
S_{dil}&=&-\frac{1}{2}\int d^5x\sqrt{-G}G^{MN}\partial_M\phi\partial_N\phi,\\
\label{eq:scalar_potential}
S_{\Lambda}&=&-\int d^5x\sqrt{-G}e^{\frac{2}{\sqrt{3}}\phi}\Lambda.
\end{eqnarray}
On a bounded manifold, it is well-known that the Einstein-Hilbert action must be supplemented by the so-called Gibbons-Hawking (GH) boundary action $S_{GH}$, given in the Einstein frame by the integral over the boundary of the square root of the determinant of the induced metric on it times the trace of its extrinsic curvature tensor\footnote{In the string frame, the background metric is trivial, so that the GH action does not modify the analysis on the nature of the branes carried out above and can thus be neglected. However, in the Einstein frame, second derivatives of the background metric now contain delta functions, and we need the GH term in order to cancel them and clearly identify the spectrum of the KK vector $G_{\mu 5}$, as done below.}. Although the GH term is seldom considered in the literature dealing with this linear dilaton model, it is an important piece which will turn out to be crucial in the Kaluza-Klein reduction performed in Section \ref{sect:KK_reduction}. Thus, in addition to the bulk action \eqref{eq:bulk_action}, the five dimensional LD model presented here contains also a boundary action $S_{\partial\mathcal{M}}$ given by
\begin{equation}\label{eq:boundary_action}
S_{\partial\mathcal{M}}=\sum_{i=1}^{2}\left(S_{NS5_i}+S_{GH_i}\right),
\end{equation}
with, in the Einstein frame: 
\begin{eqnarray}
\label{eq:NS5_brane_action}
S_{NS5_i}&=&-V_i\int d^5x\sqrt{-g}e^{\frac{1}{\sqrt{3}}\phi}\delta(y-y_i),\\
\label{eq:GH_action}
S_{GH_i}&=&\int d^4x\sqrt{-g}K_i.
\end{eqnarray}
The indice $i$ labels the two four-dimensional boundaries of $\mathcal{M}$, located at $y=y_i$. $K_i$ is the trace of their extrinsic curvature tensor, $g$ the determinant of the $4D$ induced metric on them, and the constants $\Lambda$, $V_1$ and $V_2$ have been computed in \eqref{eq:relation_lambda_alpha} and \eqref{eq:brane_antibrane_sol}, and are given by $\Lambda=-\frac{3}{2}\alpha^2$, $V_1=-V_2=2\sqrt{3}\alpha$.

The total $5D$ action we will work with in Section \ref{sect:KK_reduction} is then
\begin{equation}
S=S_{\mathcal{M}}+S_{\partial\mathcal{M}},
\end{equation}
with $S_{\mathcal{M}}$ and $S_{\partial\mathcal{M}}$ respectively given by \eqref{eq:bulk_action} and \eqref{eq:boundary_action}.

\subsection{Spectrum of bosonic fields on a LD background}
The aim of this subsection is to remind some general results regarding the spectrum of bosonic fields on a linear dilaton background, and then motivate the first part of the work carried out in this paper, which will be described in Section \ref{sect:KK_reduction}. We start with the case of a bulk scalar and vector fields in five dimensions, an analysis which can be found for instance in \cite{Farakos, Clockwork_SM}, before moving to the spin-0 and spin-1 spectrum of the $5D$ dilaton-gravity theory described in the previous subsection.
\subsubsection*{Scalar field on a LD background}
\label{subsect:spin0_LD_background}
Let us first consider a given $5D$ massless scalar field $\chi$ on a LD background, distinct from the dilaton field $\phi$ of the previous section. The string frame Lagrangian of such a scalar $\chi$ would be obtained by merely  adding the standard kinetic term $-\frac{1}{2}(\partial\chi)^2$ in the bracket of the Lagrangian \eqref{eq:action_string_frame_bulk}, which yields at quadratic order
\begin{equation}\label{eq:scalar_field_lagrangian}
\mathcal{L}=-\frac{1}{2}e^{-Qy}\partial_M\chi\partial^M\chi,
\end{equation}
where $Q$ is a constant related to $\alpha$ in \eqref{eq:lin_dil_sol}, but kept as an arbitrary parameter in this subsection. The equation of motion for $\chi$ following from \eqref{eq:scalar_field_lagrangian} is easily found to be
\begin{equation}
\Box_4\chi+\chi^{''}-Q\chi^{'}=0.
\end{equation}
Considering the mode expansion
\begin{equation}
\chi(x,y)=\sum_{n=0}^{\infty}f_{(n)}(y)e^{ip_n\cdot x},
\end{equation}
we get for the internal wave functions $f_{(n)}(y)$ of the Kaluza-Klein modes:
\begin{equation}\label{eq:internal_wavefunctions_diff_eq}
f_{(n)}''(y)-Qf_{(n)}'(y)-p_n^2f_{(n)}(y)=0,~\forall n\geq 0,
\end{equation}
with $p_n^2=-m_n^2$. The most general solution reads
\begin{equation}
f_{(n)}(y)=Ae^{x_+y}+Be^{x_-y},
\end{equation}
where $A$ and $B$ are two constants, and $x_{\pm}$ are given by $x_{\pm}=\frac{Q\pm\sqrt{Q^2-4m_n^2}}{2}$. Imposing Neumann boundary conditions at $y=0$ and $y=L$, $\left.\partial_5 f_{(n)}(y) \right|_{y=0,L}=0$, it is then easy to see that the lowest mode compatible with the boundary conditions is massless, $m_0=0$, with a constant internal wave function, $f_0(y)=$ constant. We then have excited modes with masses $m_n^2=\left(\frac{n\pi}{L}\right)^2+\frac{Q^2}{4}, \forall n>0$, with wavefunctions given by
\begin{equation}\label{eq:solution_Nbc}
f^N_{(n)}(y)=e^{\frac{Q}{2}y}\left(\sin\frac{n\pi y}{L}-\frac{2n\pi}{QL}\cos\frac{n\pi y}{L}\right),~\forall n>0,
\end{equation}
up to an overall normalisation constant, irrelevant for the discussion here. In the case of Neumann boundary conditions, a $5D$ massless bulk scalar therefore gives rise, from the $4D$ point of view, to a single massless zero mode with constant wavefunction, followed by discrete KK excited states on top of a mass gap fixed by $Q$:
\begin{equation}\label{eq:scalar_field_spectrum}
\left\{
\begin{array}{r c l}
m_0 &=& 0, \\
m_n^2&=&\left(\frac{n\pi}{L}\right)^2+\frac{Q^2}{4},~\forall n>0,
\end{array}
\right.
\end{equation}
which is a distinctive feature of the linear dilaton background.

\subsubsection*{Vector field on a LD background}
\label{subsect:spin1_LD_background}
A very similar analysis can be carried out for a given $5D$ massless bulk vector field $A_M$ in the LD background. The equations of motion for $A_M$, following from the Lagrangian
\begin{equation}
\mathcal{L}=-\frac{1}{4}e^{-Qy}F_{MN}F^{MN}=-\frac{1}{2}e^{-Qy}\left(\partial_MA_N\partial^MA^N-\partial_MA_N\partial^NA^M\right),
\end{equation}
read
\begin{equation}
\partial_MF^{MN}-QF^{5N}=0,
\end{equation}
which split into:
\begin{eqnarray}
\label{A_5_eom}
&\Box_4A_5&-\partial_5\partial_{\mu}A^{\mu}=0,\\
\label{A_mu_eom}
&\Box_4A_{\nu}&+A_{\nu}^{''}-QA_{\nu}^{'}-\partial_{\nu}\left(\partial_{\mu}A^{\mu}+\partial_5A^5-QA^5\right)=0.
\end{eqnarray}
Similarly to the scalar case described above, we expand $A_{\mu}$ in modes
\begin{equation}
A_{\mu}(x,y)=\sum_{n=0}^{\infty}f_{(n)}(y)A_{\mu}^{(n)}(x),
\end{equation}
and impose in addition the gauge condition
\begin{equation}\label{eq:vector_gauge_cond}
\partial_5A^5=QA_5,
\end{equation}
which in the flat case $Q=0$ reduces to the standard gauge of toroidal KK compactification where $A_5$ is a function of $x$ only. For the zero mode with a flat internal wave function $f_0(y)=$ constant, one gets the equation of motion
\begin{equation}
\Box_4 A_{\mu}^{(0)}(x)-\partial_{\nu}\partial_{\mu}A^{\mu (0)}(x)=0,
\end{equation} 
which describes a massless vector with its remaining 4-dimensional gauge symmetry. Applying $\partial^{\nu}$ to \eqref{A_mu_eom}, we see that the KK modes $A_{\mu}^{(n)}$, $n>0$, satisfy $\partial_{\mu}A^{\mu (n)}=0$, so that the differential equation satisfied by the internal profiles $f_{(n)}(y)$ reads:
\begin{equation}
f_{(n)}''(y)-Qf_{(n)}'(y)-p_n^2f_{(n)}(y)=0,~\forall n\geq 0,
\end{equation}
with $p_n^2=-\Box_4 A_{\mu}^{(n)}$. This is the same equation as obtained previously in the scalar case, and we thus conclude that the spin-1 KK masses are given by:
\begin{equation}
m_n^2=\left(\frac{n\pi}{L}\right)^2+\frac{Q^2}{4},~\forall n>0.
\end{equation}
Regarding the scalar $A_5$, the gauge condition \eqref{eq:vector_gauge_cond} imposes the zero mode of $A_5$ to be of the form $A^{(0)}_5(x,y)=e^{Qy}\tilde A_5(x)$, describing 
a massless 4-dimensional scalar as follows from the equation of motion \eqref{A_5_eom}.

The dimensional reduction of a given 5-dimensional vector field on a LD background therefore leads to a 4-dimensional massless scalar, together with a 4-dimensional massless vector followed by massive KK vector excitations, as in the toroidal case. The LD background prints distinguishable features both on the scalar sector, through the exponential profile of its zero mode, as well as in the vector sector, through the mass gap above which the KK tower of massive states starts.

\subsection*{Dilaton-gravity sector on a LD background}
In the five-dimensional LD model introduced in Section \ref{sect:5D_theory}, the spectrum for the four-dimensional graviton has been computed in \cite{Pheno_LST}. Starting from the $5D$ Einstein-frame metric parametrisation
\begin{equation}\label{eq:metric_param_graviton_pheno}
G_{MN}=e^{-\frac{2}{\sqrt{3}}\alpha|y|}
\begin{pmatrix}
\eta_{\mu\nu}+h_{\mu\nu} & 0 \\
0 & 1 
\end{pmatrix},
\end{equation}
the spectrum of the gravitational excitations $h_{\mu\nu}$, for Neumann boundary conditions, has been found to be the same as the ones described in the above subsections: a massless zero mode with wavefunction independent of $y$, followed by discrete KK modes with masses $m_n^2=\left(\frac{n\pi}{L}\right)^2+\frac{3}{4}\alpha^2,~\forall n>0$.

The spectrum analysis in the scalar sector has been carried out in \cite{Radion_pheno}. Putting the $4D$ vector and tensor fluctuations to zero, the metric parametrisation considered here, at linear order in the scalars, is of the form
\begin{equation}\label{eq:metric_param_radion_pheno}
G_{MN}=e^{-\frac{2}{\sqrt{3}}\alpha|y|}
\begin{pmatrix}
(1+2\sigma)\eta_{\mu\nu} & 0 \\
0 & 1+2r 
\end{pmatrix},
\end{equation}
where $\sigma$ and $r$ are respectively the trace of the $4D$ metric excitations $h_{\mu\nu}$ and the radion $G_{55}$. A crucial point to notice is that the three scalar fluctuations $\delta\phi$, $\sigma$ and $r$ are not independent, but instead related by the following two constraints \cite{Constraints}:
\begin{eqnarray}
&r&=-2\sigma,\\
\sqrt{3}\alpha &r&+3\sgn y\sigma^{'}+\alpha\delta\phi=0.
\end{eqnarray}
The first one is the usual constraint on the trace of the metric tensor, corresponding to the Weyl transformation one has to perform on the four dimensional metric in order to bring its kinetic term into a canonical form. The second one is only relevant in the case of a LD background\footnote{In the flat case $\alpha=0$, it simply imposes $r^{'}=0$, which is the usual gauge choice for the radion in standard Kaluza-Klein reduction on a circle.}: in the general case $\alpha\neq 0$, it implies that only one linear combination of the scalars $r$ and $\delta\phi$ is dynamical \cite{Physical_scalar}, while the orthogonal combination can be eliminated by a gauge choice on part of the 5-dimensional diffeomorphisms, and is thus unphysical. The spectrum for the remaining physical scalar is given in \cite{Radion_pheno}. Although the analysis has been done in the more general case where the radion is stabilised, the unstabilised case we are considering here can be easily recovered: one finds exactly the same spectrum as for the $4D$ metric, namely a massless zero mode with wavefunction independent of $y$, followed by massive KK excitations with masses $m_n^2=\left(\frac{n\pi}{L}\right)^2+\frac{3}{4}\alpha^2,~\forall n>0$.

So far two points were left out in the literature: what is happening to the vanishing 0-mode scalar degree of freedom, which must be recovered as a physical degree of freedom in the limit $\alpha\rightarrow 0$, as well as the analysis for the Kaluza-Klein vector $G_{\mu 5}$? The aim of the next section is to clarify these two points. We will show that the zero mode of the unphysical scalar is actually absorbed by the zero mode of the KK vector, providing it with a mass via a gauge-fixing that is reminiscent of the St\"uckelberg term.
To this purpose, we first have to find the most general parametrisation for the metric tensor $G_{MN}$ including the KK vector $G_{\mu 5}$, which reduces to \eqref{eq:metric_param_graviton_pheno} or \eqref{eq:metric_param_radion_pheno} in the relevant limits, and whose components transform appropriately under four-dimensional diffeomorphisms. We also want this parametrisation to be valid not only at the linear level in the scalars, but to all orders in order to be able to find not only the scalar spectrum but also the full scalar potential of the dimensionally reduced $4D$ theory.

\section{Kaluza-Klein reduction on a linear dilaton background}
\label{sect:KK_reduction}

\subsection{$4D$ residual gauge symmetry}
\label{sect:residual_gauge_sym}

The five-dimensional Einstein frame metric $G_{MN}$ is written, in the most general case, as function of $x$ and $y$ according to
\begin{equation}
G_{MN}=
\begin{pmatrix}
G_{\mu\nu} & G_{\mu 5} \\
G_{\nu 5} & G_{55} 
\end{pmatrix}
(x,y).
\end{equation}
In order to parametrise the form of the metric as well as its $y$ dependence compatible with the dimensional reduction $D=5\rightarrow D=4$, we start from the $5D$ diffeomorphism transformations, with parameter $\xi^P=(\xi^{\mu}(x,y),\xi^5(x,y))$. Under $\xi^P$, the transformation of $G_{MN}$,
\begin{equation}\label{eq:5D_diff}
\delta G_{MN}=\xi^P\partial_PG_{MN}+G_{MP}\partial_N\xi^P+G_{NP}\partial_M\xi^P,
\end{equation}
splits for $G_{\mu\nu}$, $G_{\mu 5}$ and $G_{55}$ into:
\begin{eqnarray}
\label{eq:Gmunu_transfo}
\delta G_{\mu\nu}&=&\xi^{\rho}\partial_{\rho}G_{\mu\nu}+2G_{\rho(\mu}\partial_{\nu)}\xi^{\rho}+\xi^5\partial_5G_{\mu\nu}+2G_{5(\mu}\partial_{\nu)}\xi^{5},\\
\label{eq:Gmu5_transfo}
\delta G_{\mu 5}&=&\xi^{\nu}\partial_{\nu}G_{\mu 5}+G_{\mu\nu}\partial_5\xi^{\nu}+G_{5\nu}\partial_{\mu}\xi^{\nu}+\xi^5\partial_5G_{\mu 5}+G_{\mu 5}\partial_5\xi^5+G_{55}\partial_{\mu}\xi^5,\\
\label{eq:G55_transfo}
\delta G_{55}&=&\xi^{\mu}\partial_{\mu}G_{55}+2G_{\mu 5}\partial_5\xi^{\mu}+\xi^5\partial_5G_{55}+2G_{55}\partial_5\xi^5.
\end{eqnarray}
Let us first consider the $4D$ diffeomorphisms, parametrised by the 4-vector $\xi^{\mu}(x,y)$. From \eqref{eq:Gmunu_transfo}, one sees that $G_{\mu\nu}$ already transforms as a rank-2 tensor under $\xi^{\mu}$, namely
\begin{equation}
\delta_{\xi^{\rho}} G_{\mu\nu}=\xi^{\rho}\partial_{\rho}G_{\mu\nu}+2G_{\rho(\mu}\partial_{\nu)}\xi^{\rho}.
\end{equation}
The transformation of $G_{\mu 5}$ reads:
\begin{equation}
\delta_{\xi^{\rho}} G_{\mu 5}=\xi^{\nu}\partial_{\nu}G_{\mu 5}+G_{5\nu}\partial_{\mu}\xi^{\nu}+G_{\mu\nu}\partial_5\xi^{\nu}.
\end{equation}
The last term, being independent of $G_{\mu 5}$, is a shift in the transformation which can be used in order to gauge fix to zero the KK excitations of $G_{\mu 5}$, as long as $\partial_5\xi^{\nu}\neq 0$. Since all functions $\xi^{\nu}(x,y)$ which are not constant in $y$ can be used for this purpose, we end up with a residual $4D$ diffeomorphisms invariance parametrised by all functions $\xi^{\mu}$ constant in $y$. Under such $\xi^{\mu}(x)$, one sees that $G_{\mu 5}$ and $G_{55}$ transform indeed correctly as a $4D$ vector and a Lorentz scalar respectively.

We next turn to the $U(1)_{KK}$ transformations, parametrised by the $4D$ Lorentz scalar function $\xi^{5}(x,y)$. Let us first recall that the equations of motion are solved, in the Einstein frame, by the background metric
\begin{equation}
\bar G_{MN}=e^{-\frac{2}{\sqrt{3}}\alpha|y|}
\begin{pmatrix}
\eta_{\mu\nu} & 0 \\
0 & 1 
\end{pmatrix}.
\end{equation}
We thus define the radion $r$, which is the fluctuation of $G_{55}$ around the background solution through: 
\begin{equation}\label{eq:def_r}
G_{55}(x,y)=e^{-\frac{2}{\sqrt{3}}\alpha|y|}e^{2r(x,y)}.
\end{equation}
The $\xi^5$ part of the transformation \eqref{eq:G55_transfo} on $r$ then reads:
\begin{equation}\label{eq:r_transfo}
\delta_{\xi^5}r=\partial_5\xi^5-\frac{\alpha}{\sqrt{3}}\sgn y\xi^5+\xi^5\partial_5r.
\end{equation}
Defining
\begin{equation}\label{eq:def_Bmu}
K_{\mu}\equiv\frac{G_{\mu 5}}{G_{55}},
\end{equation}
and using the transformations \eqref{eq:Gmu5_transfo} for $G_{\mu 5}$ and \eqref{eq:G55_transfo} for $G_{55}$, one finds:
\begin{equation}
\delta_{\xi^5}K_{\mu}=\partial_{\mu}\xi^5-K_{\mu}\partial_5\xi^5+\xi^5\partial_5K_{\mu}.
\end{equation}
In order for $K_{\mu}$ to transform as a gauge field under $U(1)_{KK}$, and assuming that the $x$ and $y$ dependences of $\xi^5$ and $K_{\mu}$ factorize, one sees that they must satisfy
\begin{equation}\label{eq:relation_xi_B}
\xi^5=f(y)\tilde \xi^5(x),\qquad K_{\mu}=f(y)\tilde K_{\mu}(x),
\end{equation}
where $f(y)$ is an arbitrary function of $y$, and $\tilde\xi^5(x)$ and $\tilde K_{\mu}(x)$ two arbitrary functions of $x$.
Similarly, defining
\begin{equation}
g_{\mu\nu}\equiv \frac{G_{\mu\nu}}{G_{55}}-K_{\mu}K_{\nu},
\end{equation}
we have 
\begin{equation}\label{eq:g_munu_transfo}
\delta_{\xi^5}g_{\mu\nu}=\xi^5\partial_5g_{\mu\nu}-2g_{\mu\nu}\partial_5\xi^5.
\end{equation}

In a similar way that we have used a subset of the functions $\xi^{\mu}(x,y)$ to gauge fix the KK modes of the vector $G_{\mu 5}$, one can now use a subset of the functions $\xi^{5}(x,y)$ in order to gauge fix a series of KK modes of the scalars. From the $5D$ point of view, two scalar fields are present: the radion $r$ and the dilaton $\delta\phi$, whose transformations with respect to $\xi^5$ are respectively given by:
\begin{eqnarray}
\label{eq:r_transfo_2}
\delta_{\xi^5}r&=&\partial_5\xi^5-\frac{\alpha}{\sqrt{3}}\sgn y\xi^5+\xi^5\partial_5r,\\
\delta_{\xi^5}(\delta\phi)&=&\alpha\sgn y\xi^5+\xi^5\partial_5\delta\phi.
\end{eqnarray}
One can then choose to gauge fix the excitations of $r$, of $\delta\phi$, or of a combination of both. All possibilities are obviously physically equivalent, and our choice will be motivated by the requirement that the $4D$ metric must remain inert under the residual gauge freedom which has not been fixed. The canonically normalised $4D$ metric $\tilde g_{\mu\nu}$ is related to $g_{\mu\nu}$ by
\begin{equation}
g_{\mu\nu}=e^{-3r}\tilde g_{\mu\nu}.
\end{equation}
From the transformations \eqref{eq:g_munu_transfo} and \eqref{eq:r_transfo_2}, one finds the transformation of $\tilde g_{\mu\nu}$ under $\xi^5$,
\begin{equation}\label{eq:tilde_g_munu_transfo}
\delta_{\xi^5}\tilde g_{\mu\nu}=\partial_5\xi^5\tilde g_{\mu\nu}-\sqrt{3}\alpha\sgn y\xi^5\tilde g_{\mu\nu}+\xi^5\partial_5\tilde g_{\mu\nu}.
\end{equation}
This transformation motivates us to consider the linear combination $r-\frac{2}{\sqrt{3}}\delta\phi$, which transforms under $\xi^5$ according to:
\begin{equation}
\delta_{\xi^5}\left(r-\frac{2}{\sqrt{3}}\delta\phi\right)=\partial_5\xi^5-\sqrt{3}\alpha\sgn y \xi^5+\xi^5\partial_5\left(r-\frac{2}{\sqrt{3}}\delta\phi\right).
\end{equation}
Again $\forall\xi^5(x,y)\neq e^{\sqrt{3}\alpha |y|}\tilde\xi^5(x)$, the two first terms of the transformation are non zero constant shifts, which can thus be used to gauge fix the KK modes of the scalar $r-\frac{2}{\sqrt{3}}\delta\phi$. Therefore, we end up with a residual gauge freedom which cannot be fixed for now, associated with the parameter 
\begin{equation}
\xi^5\equiv e^{\sqrt{3}\alpha |y|}\tilde\xi^5(x).
\end{equation}
Under such $\xi^5$, one immediately verifies that the constant terms in the transformation \eqref{eq:tilde_g_munu_transfo} of $\tilde g_{\mu\nu}$ vanish. Then, the zero mode of $\tilde g_{\mu\nu}$, independent of $y$, does not transform under $\xi^5$ as required. Having found the form of $\xi^5$, we deduce from \eqref{eq:relation_xi_B} that
\begin{equation}\label{eq:KK_U(1)_vector}
K_{\mu}=e^{\sqrt{3}\alpha|y|}\tilde K_{\mu}(x).
\end{equation}
The final parametrisation of the $5D$ metric $G_{MN}$ therefore reads:
\begin{equation}\label{eq:final_metric}
G_{MN}=e^{-\frac{2}{\sqrt{3}}\alpha|y|}e^{2r}
\begin{pmatrix}
e^{-3r}\tilde g_{\mu\nu}(x)+e^{2\sqrt{3}\alpha|y|}\tilde K_{\mu}\tilde K_{\nu}(x) & e^{\sqrt{3}\alpha|y|}\tilde K_{\mu}(x) \\
e^{\sqrt{3}\alpha|y|}\tilde K_{\nu}(x) & 1 
\end{pmatrix},
\end{equation}
whose inverse is given by:
\begin{equation}\label{eq:final_inverse_metric}
G^{MN}=e^{\frac{2}{\sqrt{3}}\alpha|y|}e^{-2r}
\begin{pmatrix}
e^{3r}\tilde g^{\mu\nu}(x) & -e^{\sqrt{3}\alpha|y|}e^{3r}\tilde K^{\mu}(x) \\
-e^{\sqrt{3}\alpha|y|}e^{3r}\tilde K^{\nu}(x) & 1+e^{2\sqrt{3}\alpha|y|}e^{3r}\tilde K^2(x) 
\end{pmatrix},
\end{equation}
with $\tilde K^{\mu}=\tilde g^{\mu\nu}\tilde K_{\nu}$, $\tilde K^2=\tilde g^{\lambda\sigma}\tilde K_{\lambda}\tilde K_{\sigma}$. The metric $G_{MN}$ thus reduces in the appropriate limits to the two forms \eqref{eq:metric_param_graviton_pheno} or \eqref{eq:metric_param_radion_pheno} previously studied in the literature, namely by turning off the spin-1 excitations, as well as the scalar or the graviton fluctuations.

At that point, let us summarise the approach followed here. Using the gauge freedom associated to the $\xi^{\mu}(x,y)$, we first set to zero the KK modes of the vector $G_{\mu 5}$, ending with a residual $4D$ diffeomorphism invariance associated with the vectors $\xi^{\mu}(x)$. Under such $\xi^{\mu}(x)$, all the fields $G_{\mu\nu}$, $G_{\mu 5}$ and $G_{55}$ respectively transform as a $4D$ rank-2 tensor, Lorentz vector and Lorentz scalar. Next, we have used the gauge freedom associated to the $\xi^{5}(x,y)$ in order to get rid of the KK modes of the scalar $r-\frac{2}{\sqrt{3}}\delta\phi$. The remaining $U(1)_{KK}$ gauge transformation is associated with the scalar function $\xi^5\equiv e^{\sqrt{3}\alpha|y|}\tilde\xi^5(x)$, under which the different fields transform according to:
\begin{eqnarray}
\delta_{\xi^5}\tilde g_{\mu\nu}&=&\xi^5\partial_5\tilde g_{\mu\nu},\\
\delta_{\xi^5}\tilde K_{\mu}&=&\partial_{\mu}\tilde\xi^5(x),\\
\delta_{\xi^5}r&=&\frac{2\alpha}{\sqrt{3}}\sgn y\xi^5+\xi^5\partial_5r,\\
\delta_{\xi^5}(\delta\phi)&=&\alpha\sgn y\xi^5+\xi^5\partial_5\delta\phi.
\end{eqnarray}
Denoting by $\varphi_{-}$ and $\varphi_{+}$ the orthogonal combinations 
\begin{equation}\label{eq:ortho_comb}
\varphi_{-}\equiv r-\frac{2}{\sqrt{3}}\delta\phi,\qquad \varphi_{+}\equiv r+\frac{1}{\sqrt{3}}\delta\phi,
\end{equation}
their transformations under $\xi^5$ are given by:
\begin{eqnarray}
\delta_{\xi^5}\varphi_{-}&=&\xi^5\partial_5\varphi_{-},\\
\label{eq:Stuckelberg_transfo}
\delta_{\xi^5}\varphi_{+}&=&\sqrt{3}\alpha\sgn y\xi^5+\xi^5\partial_5\varphi_{+}.
\end{eqnarray}
The shift in the transformation of $\varphi_{+}$ resembles that of a Goldstone mode, and can be used in order to eliminate it through a suitable $U(1)_{KK}$ gauge transformation. The breaking of the remaining $U(1)_{KK}$ invariance will be reflected in the emergence of a mass term for the KK vector $K_{\mu}$ in a manner that is reminiscent of the St\"uckelberg term, which is why we will be referring to $\varphi_{+}$ as the St\"uckelberg field in the following. The aim of the next two subsections is to compute the dimensionally reduced action in order to clarify the origin of the mass term for the KK vector, and then find the scalar potential associated to the zero mode of the remaining physical scalar $\varphi_{-}$.

At this point, it is instructive to move from the Einstein to the string frame in order to get more physical intuition about the fields $\varphi_+$ and $\varphi_-$. The Einstein frame metric \eqref{eq:final_metric}, given in the basis $(\varphi_-,\varphi_+)$ by
\begin{equation}
G_{MN}=e^{-\frac{2}{\sqrt{3}}\alpha|y|}e^{\frac{2}{3}(\varphi_-+2\varphi_+)}
\begin{pmatrix}
e^{-(\varphi_-+2\varphi_+)}\tilde g_{\mu\nu}(x)+e^{2\sqrt{3}\alpha|y|}\tilde K_{\mu}\tilde K_{\nu}(x) & e^{\sqrt{3}\alpha|y|}\tilde K_{\mu}(x) \\
e^{\sqrt{3}\alpha|y|}\tilde K_{\nu}(x) & 1 
\end{pmatrix},
\end{equation}
can be brought, by means of the $5D$ Weyl transformation \eqref{eq:5D_Weyl_transfo}, to the string frame metric
\begin{equation}
G_{MN}^S=
\begin{pmatrix}
g_{\mu\nu}^S(x)+e^{2\varphi_+}e^{2\sqrt{3}\alpha|y|}\tilde K_{\mu}\tilde K_{\nu}(x) & e^{2\varphi_+}e^{\sqrt{3}\alpha|y|}\tilde K_{\mu}(x) \\
e^{2\varphi_+}e^{\sqrt{3}\alpha|y|}\tilde K_{\nu}(x) & e^{2\varphi_+}
\end{pmatrix},
\end{equation}
from which we deduce that $\varphi_+$ is actually the string frame radion. The $4D$ string frame metric $g_{\mu\nu}^S$ is related to the $4D$ Einstein frame metric $\tilde g_{\mu\nu}$ by the $4D$ Weyl transformation
\begin{equation}
g_{\mu\nu}^S=e^{-\varphi_-}\tilde g_{\mu\nu}.
\end{equation}
Regarding $\varphi_-$, we remind that in our normalisation in which the $5D$ action reads
\begin{equation}
S^{(5)}=\int d^5x \sqrt{-G^S}e^{-\sqrt 3\delta\phi}\frac{1}{2}\left[\mathcal R^{(5)}[G^S]+3(\partial\delta\phi)^2+\dots\right],
\end{equation}
the $4D$ dilaton $\delta\phi_4$ is defined such that the $4D$ action is
\begin{equation}
S^{(4)}=\int d^4x \sqrt{-g^S}e^{-\sqrt 3\delta\phi_4}\frac{1}{2}\left[\mathcal R^{(4)}[g^S]+3(\partial\delta\phi_4)^2+\dots\right].
\end{equation}
From the above equation then it follows that $\varphi_-$ is simply proportionnal to the $4D$ dilaton according to
\begin{equation}
\varphi_-=-\sqrt{3}\delta\phi_4.
\end{equation}

\subsection{St\"uckelberg ``mechanism''}
\label{sect:stuckelberg_mech}
Let us now compute the four-dimensional effective bosonic action from the five-dimensional one introduced in Section \ref{sect:5D_theory}, whose structure is reminded here for convenience:
\begin{equation}
S=S_{EH}+S_{dil}+S_{\Lambda}+\sum_{i=1}^{2}\left(S_{NS5_i}+S_{GH_i}\right).
\end{equation} 
The three first terms correspond to the bulk action given in \eqref{eq:einstein_hilbert_action}-\eqref{eq:scalar_potential}, while the last two are the boundary terms given in \eqref{eq:NS5_brane_action}-\eqref{eq:GH_action}. The computation of the dilaton kinetic term $S_{dil}$, cosmological constant $S_{\Lambda}$ and NS5-brane contributions $S_{NS5_i}$ is straightforward. Using the general metric decomposition found in \eqref{eq:final_metric}, its inverse \eqref{eq:final_inverse_metric} and $\phi=\alpha|y|+\delta\phi$, we get:
\begin{eqnarray}
\label{eq:dilaton_action}
S_{dil}&=&-\frac{1}{2}\int d^5xe^{-\sqrt{3}\alpha|y|}\sqrt{-\tilde g}\left\{e^{-3r}\alpha^2+\left(\alpha\sgn yK_{\mu}-\partial_{\mu}\delta\phi\right)^2\right.\nonumber\\
&&\qquad\left.+(e^{-3r}+\tilde K^2)\left((\delta\phi^{'})^2+2\alpha\sgn y\delta\phi^{'}\right)-2K^{\mu}\partial_{\mu}\delta\phi\delta\phi^{'}\right\} ,\\
S_{\Lambda}&=&\frac{3}{2}\alpha^2\int d^5xe^{-\sqrt{3}\alpha|y|}\sqrt{-\tilde g}e^{-r+\frac{2}{\sqrt{3}}\delta\phi},\\
S_{NS5_i}&=&\pm 2\sqrt{3}\alpha\int d^5xe^{-\sqrt{3}\alpha|y|}\sqrt{-\tilde g}e^{-2r+\frac{1}{\sqrt{3}}\delta\phi}\delta(y-y_i).
\end{eqnarray}

The computation of the gravitational action $S_G\equiv S_{EH}+S_{GH}$ is more involved. We perform the analysis in analogy to the ADM framework, splitting the five dimensional spacetime $\mathcal{M}$ into timelike slices of constant $y$. The general framework of a $d+1$ spacetime decomposition with boundaries is reviewed in Appendix \ref{sect:append_ADM_1}, and the particular computation for the metric \eqref{eq:final_metric} is detailed in Appendix \ref{sect:append_ADM_2}. The total gravitational action \eqref{eq:final_gravitational_action}, up to quadratic order in $K_{\mu}$, reads:
\begin{eqnarray}
S_G
\label{eq:gravitational_action}
&=&\int d^5xe^{-\sqrt{3}\alpha|y|}\sqrt{-\tilde g}\left\{e^{-3r}\left[2\alpha^2+2\sqrt{3}\alpha\sgn yr^{'}+\frac{3}{2}(r^{'})^2\right]\right.\nonumber\\
&&\quad\left.-\frac{3}{2}\tilde K^{\mu}\partial_{\mu}r^{'}-\frac{1}{4}e^{3r}\tilde g^{\mu\nu}\tilde g^{\rho\sigma}F_{\mu\rho}F_{\nu\sigma}-\left(\alpha\sgn yK_{\mu}-\frac{\sqrt{3}}{2}\partial_{\mu}r\right)^2\right\}\,,
\end{eqnarray}
where we have ignored the term $\tilde K^{\mu}\partial_{\mu}(r^{'}r)$ which is irrelevant for the discussion here, used \eqref{eq:KK_U(1)_vector} to write $(K^2)^{'}=2\sqrt{3}\alpha\sgn y K^2$, and finally arranged the radion kinetic term, the mass term for $K_{\mu}$ and their mixing into a perfect square. From \eqref{eq:dilaton_action} and \eqref{eq:gravitational_action}, one sees that the dilaton and gravitational actions contain the terms
\begin{eqnarray}
S_{dil}+S_G&\supset&\int d^5xe^{-\sqrt{3}\alpha|y|}\sqrt{-\tilde g}\left\{-\frac{3}{2}\left[\alpha\sgn y K_{\mu}-\frac{1}{\sqrt{3}}\partial_{\mu}\varphi_{+}\right]^2-\frac{1}{4}\left(\partial_{\mu}\varphi_{-}\right)^2\right\}
\end{eqnarray}
where we have introduced back the combinations $\varphi_{-}$ and $\varphi_{+}$ defined in \eqref{eq:ortho_comb}. Using the gauge transformation \eqref{eq:Stuckelberg_transfo} to fix the St\"uckelberg field $\varphi_{+}$ to zero, one immediately reads the mass term for the $U(1)_{KK}$ boson $K_{\mu}$:
\begin{equation}
m_{U(1)_{KK}}^2=\frac{3}{2}\alpha^2.
\end{equation}

The dimensional reduction from five to four dimensions of the metric tensor on a linear dilaton background therefore provides a very peculiar spectrum, parametrised by two parameters: the slope of the linear dilaton $\alpha$, and the radius of compactification $L$. In the KK excitation modes, the situation is similar to the flat case ($\alpha=0$): a scalar combination $\varphi_{-}$ of the radion and the dilaton as well as the $U(1)_{KK}$ vector $K_{\mu}$ have infinite towers of states which can be gauge fixed to zero, providing three additional polarisations to the excitation modes of the $4D$ graviton, which acquire masses $\propto 1/L$. The difference from the flat case comes from the mass gap above the zero mode, on top of which the tower of KK states starts. Proportionnal to $\alpha$, this gap is a characteristic feature of the LD background, and vanishes in the flat limit $\alpha\rightarrow 0$, thus recovering the usual case of toroidal compactification. In the zero mode sector, the $4D$ graviton as well as $\varphi_{-}$ have a massless zero mode. The zero mode of the scalar orthogonal combination $\varphi_{+}$ is absorbed by the zero mode of $K_{\mu}$, which acquires a mass $\propto \alpha$, in sharp contrast with the flat case where all the zero modes of the different components of the metric remain massless. The main result here is thus that the residual $U(1)_{KK}$ symmetry, which remains in standard KK compactification on a flat background $\alpha=0$, is here broken at a scale $m_{U(1)_{KK}}^2=\frac{3}{2}\alpha^2$ fixed by the slope $\alpha$ of the LD background.

One may wonder if the low-energy limit that corresponds to decouple all massive states, not only the KK modes, by sending formally the LD slope  $|\alpha|\to\infty$, corresponds to an orbifold reduction. Indeed, the $y$-parity projects out the KK vector boson $K_\mu$ but keeps the dilaton and radion which should also be projected, as it plays the role of the longitudinal polarisation of $K_\mu$ (in the string frame). A possible answer is to extend the orbifold by the T-duality which in the string frame inverts the radion and thus makes $\varphi_+$ odd. Consistency of the theory then requires to add also the NS antisymmetric tensor $B_{MN}$ since the vector $B_{5\mu}$ is exchanged with $G_{5\mu}$ under T-duality (to linear order). This will be done indeed in the supersymmetric case that we study in the following.

\subsection{Effective scalar potential}
\label{sect:eff_scal_pot}
Having found the mass term for the KK vector $K_{\mu}$ as well as its associated St\"uckelberg field, we can now focus on the scalar sector, putting $K_{\mu}=0$. Combining the results found above, the total scalar action, before gauge fixing $\varphi_{+}$ to zero, is given by:
\begin{eqnarray}\label{eq:scalar_action}
S_{\text{scalar}}&=&\int d^5xe^{-\sqrt{3}\alpha|y|}\sqrt{-\tilde g}\left\{-\frac{1}{2}(\partial_{\mu}\delta\phi)^2-\frac{3}{4}(\partial_{\mu}r)^2 \right.\nonumber\\
&&\qquad\left.+e^{-3r}\left[\frac{3}{2}\alpha^2\left(1+e^{2(r+\frac{1}{\sqrt{3}}\delta\phi)}\right)-2\sqrt{3}\alpha e^{r+\frac{1}{\sqrt{3}}\delta\phi}(\delta(y)-\delta(y-L))\right.\right.\nonumber\\
&&\qquad\qquad\left.\left.+\frac{3}{2}(r^{'})^2-\frac{1}{2}(\delta\phi^{'})^2+2\sqrt{3}\alpha\sgn yr^{'}-\alpha\sgn y\delta\phi^{'}\right]\right\}.
\end{eqnarray}
Varying it with respect to $\delta\phi$ and $r$, we respectively get the equations of motion:
\begin{eqnarray}
\label{eq:delta_phi_eom}
&\Box_4&\delta\phi+e^{-3r}\delta\phi^{''}-3\alpha\sgn ye^{-3r}\left(r+\frac{1}{\sqrt{3}}\delta\phi\right)^{'}-3e^{-3r}r^{'}\delta\phi^{'}\nonumber\\
&-&\sqrt{3}\alpha^2e^{-3r}\left(1-e^{2(r+\frac{1}{\sqrt{3}}\delta\phi)}\right)+2\alpha e^{-3r}\left(1-e^{r+\frac{1}{\sqrt{3}}\delta\phi}\right)(\delta(y)-\delta(y-L)=0,\\
\label{eq:r_eom}
&\Box_4&r-2e^{-3r}r^{''}+2\sqrt{3}\alpha\sgn ye^{-3r}\left(r+\frac{1}{\sqrt{3}}\delta\phi\right)^{'}+e^{-3r}\left((\delta\phi^{'})^2+3(r^{'})^2\right)\nonumber\\
&+&\alpha^2e^{-3r}\left(1-e^{2(r+\frac{1}{\sqrt{3}}\delta\phi)}\right)-\frac{8}{\sqrt{3}}\alpha e^{-3r}\left(1-e^{r+\frac{1}{\sqrt{3}}\delta\phi}\right)(\delta(y)-\delta(y-L)=0.
\end{eqnarray}
Integrating these two equations on a infinitesimal interval $[y_i-\epsilon, y_i+\epsilon]$, where $y_i$ denotes the location of the two branes, we find the jump conditions for the derivatives of the scalar fluctuations $\delta\phi$ and $r$:
\begin{equation}
\Delta\delta\phi^{'}=-2\alpha\left(1-e^{r+\frac{1}{\sqrt{3}}\delta\phi}\right),\qquad \Delta r^{'}=-\frac{4\alpha}{\sqrt{3}}\left(1-e^{r+\frac{1}{\sqrt{3}}\delta\phi}\right),
\end{equation}
where $\Delta X\equiv X(y_i+\epsilon)-X(y_i-\epsilon)$. In terms of the fields $\varphi_{-}$ and $\varphi_{+}$ introduced in \eqref{eq:ortho_comb}, the equations of motion \eqref{eq:delta_phi_eom} and \eqref{eq:r_eom} yields for $\varphi_{-}$:
\begin{eqnarray}
&\Box_4&\varphi_{-}-2e^{-3r}\varphi_{+}^{''}+4\sqrt{3}\alpha\sgn ye^{-3r}\varphi_{+}^{'}+3e^{-3r}\varphi_{+}^{'2}+3\alpha^2e^{-3r}\left(1-e^{2\varphi_{+}}\right)\nonumber\\
&-&4\sqrt{3}\alpha e^{-3r}\left(1-e^{\varphi_{+}}\right)(\delta(y)-\delta(y-L)=0,
\end{eqnarray}
with the jump condition
\begin{equation}
\Delta\varphi_{-}^{'}=0.
\end{equation}
Therefore, imposing for the zero modes the gauge condition 
\begin{equation}
\varphi_{+}=0,
\end{equation}
the equation of motion of the zero mode of $\varphi_{-}$ simplifies into the one of a massless $4D$ scalar field, 
\begin{equation}
\Box_4\varphi_{-}=0.
\end{equation}

It follows that the zero mode of the only physical scalar field $\varphi_{-}$ is not only massless, but has also vanishing scalar potential. This is expected from the fact that the interbrane distance $L$ has not been stabilised, and $\varphi_{-}$ plays the role of the modulus for this parameter. The vanishing of the full scalar potential can of course be checked at the level of the action \eqref{eq:scalar_action}, first noticing that $\frac{3}{2}(r^{'})^2-\frac{1}{2}(\delta\phi^{'})^2=0$ in the gauge $\varphi_{+}=r+\frac{1}{\sqrt{3}}\delta\phi=0$. Then, regarding the last two terms in $S_{\text{scalar}}$, they read $2\sqrt{3}\alpha\sgn yr^{'}-\alpha\sgn y\delta\phi^{'}=\sqrt{3}\alpha\sgn y\varphi_{+}^{'}+\sqrt{3}\alpha\sgn y\varphi_{-}^{'}$. The first term vanishes due to the gauge condition, while the second is zero on the zero mode of $\varphi_{-}$ which is independent of $y$. Hence, for the zero mode of the physical scalar $\varphi_{-}$, the last line of \eqref{eq:scalar_action} vanishes and we end up with
\begin{eqnarray}
S_{\text{scalar}}\!\!=\!\!\!\int\! d^5xe^{-\sqrt{3}\alpha|y|}\sqrt{-\tilde g}\left\{\!-\frac{1}{4}(\partial_{\mu}\varphi_{-})^2\!+\!e^{-3r}\!\left[3\alpha^2-2\sqrt{3}\alpha(\delta(y)-\delta(y-L))\right]\right\}.
\end{eqnarray}
Denoting $\mathcal{V}$ the volume factor 
\begin{equation}
\mathcal{V}\equiv\int_{-L}^{L}dy e^{-\sqrt{3}\alpha|y|}=2\int_{0}^{L}dy e^{-\sqrt{3}\alpha y}=2\frac{1-e^{-\sqrt{3}\alpha L}}{\sqrt{3}\alpha},
\end{equation}
the integration of the second term yields $\int_{-L}^{L} dy e^{-\sqrt{3}\alpha|y|}\sqrt{-\tilde g}e^{-3r}3\alpha^2=3\alpha^2\sqrt{-\tilde g}e^{-3r}\mathcal{V}=2\sqrt{3}\alpha\sqrt{-\tilde g}e^{-3r}\left(1-e^{-\sqrt{3}\alpha L}\right)$, which is cancelled by the delta function terms. Hence, the full scalar potential of the zero mode of $\varphi_{-}$ vanishes, and the effective $4D$ scalar action is simply given by the kinetic term of $\varphi_{-}$, in agreement with the analysis of the equations of motion carried out above:
\begin{equation}
S_{\text{scalar}}=-\frac{\mathcal{V}}{4}\int d^4x\sqrt{-\tilde g}~\tilde g^{\mu\nu}\partial_{\mu}\varphi_{-}\partial_{\nu}\varphi_{-}.
\end{equation}

\section{Minimal supersymmetric extension}
\label{sec:minimal_susy_extension}
The minimal supersymmetric extension of the bosonic linear dilaton model introduced above has been constructed in \cite{Kehagias:2017grx, IA_CM}. It is based on a gauging of $\mathcal{N}=2$ supergravity coupled to one vector multiplet along the $U(1)_R$ subgroup of the $SU(2)_R$ symmetry group, a construction holographically dual to Little String Theory. 

The formalism of ungauged $\mathcal{N}=2$, $D=5$ supergravity coupled to an arbitrary number $n_V$ of vector multiplets is reminded in Appendix \ref{sect:append_ungauged_sugra}, and its abelian gauging in Appendix \ref{sect:append_gauged_sugra}. Following \cite{IA_CM}, we will restrict ourselves to the case $n_V=1$, and thus only one physical real scalar $\varphi^1$ is present, associated to the dilaton degree of freedom. The multiplet content of the theory is then
\begin{equation}
\left(e_M^m~,~\psi_M^i~,~A_M^0\right),\qquad\left(A_M^1~,~\lambda^i~,~\varphi^1=e^{\frac{\phi}{\sqrt 3}}\right),
\end{equation}
with $\psi_M^i$ the two gravitini, $A_M^0$ the graviphoton, $A_M^1$ the $U(1)$ vector field, $\lambda^i$ the two dilatini and $\phi$ the canonically normalized dilaton. The coordinates of the $n_V+1=2$-dimensional embedding scalar manifold are called $\xi^1\equiv\varphi^1$ and $\xi^0$, the latter being an unphysical scalar field which will be fixed in terms of the physical one $\varphi^1$ after imposing the constraint 
\begin{equation}
\mathcal{F}\equiv\beta^3C_{IJK}\xi^I\xi^J\xi^K=1,
\end{equation} 
where $C_{IJK}$ are completely symmetric real constants which define the theory, and with $\beta\equiv\sqrt{\frac{2}{3}}$. The functions $\xi^I(\phi)$ are related to the $h^I(\phi)$ introduced in Appendix \ref{sect:append_ungauged_sugra} by $\xi^I(\phi)=\beta^{-1}h^I(\phi)$. Our conventions and notations are summarised in Appendix \ref{sect:append_conventions}.

\subsection{Runaway scalar potential from $5D$ gauged supergravity}
\label{sec:runaway_scal_pot}

In this subsection we want to come back on the work carried out in \cite{Kehagias:2017grx, IA_CM} in order to classify all possible $D=5$, $\mathcal{N}=2$ supergravity theories, coupled to $n_V=1$ vector multiplet, whose $U(1)_R$ gauging exactly produces the runaway scalar potential of the non-critical string. The approach followed here is to start from the wanted scalar potential and trace back the construction carried out in \cite{GST} towards the different allowed $5D$ prepotentials, which completely define the theory. In the $n_V=1$ case, the two functions $P_0$ and $P_{\tilde a}$ introduced by the gauging and defined in Appendix \ref{sect:append_gauged_sugra} are expressed in terms of two arbitrary constants $A$ and $B$ by:
\begin{equation}\label{eq:P0_Px_functions}
P_0=Ae^{-\frac{2}{\sqrt 3}\phi}+Be^{\frac{\phi}{\sqrt 3}},\qquad P_{\tilde a}=Ae^{-\frac{2}{\sqrt 3}\phi}-\frac{B}{2}e^{\frac{\phi}{\sqrt 3}},
\end{equation}
so that the full scalar potential $P$ reads:
\begin{equation}
P=-P_0^2+P_{\tilde a}P^{\tilde a}=-3B\left(Ae^{-\frac{\phi}{\sqrt 3}}+\frac{B}{4}e^{\frac{2}{\sqrt 3}\phi}\right)\,,
\end{equation}
where ${\tilde a}$ denotes the flat indices of the scalar manifold, which take just one value if $n_V=1$;
$\phi$ is the canonically normalised dilaton whose kinetic term is $-\frac{1}{2}(\partial\phi)^2$. The scalar metric $g_{xy}=g_{xx}$ in the $n_V=1$ case is therefore simply given in this basis by $g_{xx}=1$, and similarly for the scalar manifold einbein $f^{\tilde a}_x$ defined in \eqref{eq:nv_bein_def}. In the following we will thus not distinguish flat indices $\tilde a$ and curved indices $x$ of the scalar manifold, and simply write $P_x$. The $P_0$ and $P_x$ can be seen as the coordinates of the constants $v_I$ defined in \eqref{eq:direction_gauging} in the basis $(h_I,h_I^{'})$ according to
\begin{equation}\label{eq:diff_eq_h_I}
v_I=\frac{1}{2}P_0h_I(\phi)+\frac{\sqrt 3}{2}P_xh_I^{'}(\phi),~\forall I=0,1,
\end{equation}
where the prime denotes differentiation with respect to $\phi$. Knowing from \eqref{eq:P0_Px_functions} the $P_0$ and $P_x$, \eqref{eq:diff_eq_h_I} is a first-order differential equation for the $h_I$, whose general solution is given by
\begin{equation}\label{eq:general_solution_h_I}
h_I(\phi)=\left(\frac{4v_I}{3B}+2AC_I\right)e^{-\frac{\phi}{\sqrt 3}}-BC_Ie^{\frac{2}{\sqrt 3}\phi},~\forall I=0,1,
\end{equation}
with $C_I$ the constants of integration. The completely symmetric constants $C_{IJK}$, which enter the $5D$ prepotential $\mathcal{F}$ through $\mathcal{F}=C_{IJK}h^Ih^Jh^K$, are then related to the $h_I$ and their derivatives according to:
\begin{equation}\label{eq:C_IJK_expression}
C_{IJK}=h_Ih_Jh_K-\frac{3}{2\beta^2}h_{(I}^{'}h_J^{'}h_{K)}+\frac{1}{\sqrt 2\beta^3}h_I^{'}h_J^{'}h_K^{'}.
\end{equation}
Plugging the result \eqref{eq:general_solution_h_I} and its derivative into the relation \eqref{eq:C_IJK_expression}, we then find the four independent constants $C_{IJK}$ expressed in terms of the six constants $A$, $B$, $v_I$ and $C_I$ by:
\begin{equation}
C_{IJJ}=-9B\left(AC_J+\frac{2v_J}{3B}\right)\left[2\left(AC_I+\frac{2v_I}{3B}\right)C_J+\left(AC_J+\frac{2v_J}{3B}\right)C_I\right],~\forall I,J=\{0,1\}.
\end{equation}

After these general considerations, we now focus on the particular case $A=0$, $B\neq 0$, which reproduces the runaway potential of the non-critical string\footnote{The $A$ term cannot be considered as a string loop correction either.}
\begin{equation}\label{eq:runaway_scalar_pot}
P=-P_0^2+P_xP^x=-\frac{3}{4}B^2e^{\frac{2}{\sqrt 3}\phi},
\end{equation}
with $P_0=Be^{\frac{\phi}{\sqrt 3}}$, $P_x=-\frac{B}{2}e^{\frac{\phi}{\sqrt 3}}$, and gives for the constants $C_{IJK}$:
\begin{equation}\label{eq:C_IJJ_expression}
C_{IJJ}=-\frac{4v_J}{B}\left(2v_IC_J+v_JC_I\right),~\forall I,J=\{0,1\}.
\end{equation}
Since we are interested in $n_V=1$ gauged supergravity theories leading to the runaway scalar potential \eqref{eq:runaway_scalar_pot}, we need to investigate which choices of the different constants are compatible with the system of the four equations \eqref{eq:C_IJJ_expression}.

Without lost of generality, one can impose as a starting point $C_{011}$ to be non-vanishing, which from \eqref{eq:C_IJJ_expression} implies $v_1\neq 0$. Two cases can then be considered, depending if $C_1$ vanishes or not.
\begin{itemize}
\item If $C_1=0$, it implies $C_{111}=0$, and necessarily $C_0\neq 0$ since $C_{011}\neq 0$. But there remains a freedom on the choice of $v_0$ which can vanish or not, leading to $C_{100}=C_{000}=0$ or $C_{100}\neq 0, C_{000}\neq 0$.
\item In the second case $C_1\neq 0$, we have $C_{111}\neq 0$. There are then three subcases, depending on the choice of $v_0$ and $C_0$. If $v_0=0$, since $C_{011}\neq 0$, we must have $C_0\neq 0$. But if $v_0$ is non vanishing, the two possibilities for $C_0$ to vanish or not are allowed, respectively leading to $(C_{000}=0, C_{100}\neq 0)$ and $(C_{000}\neq 0,C_{100}\neq 0)$.
\end{itemize}
These results are summarised in Table \ref{tab:possible_cases}. We end up with five different possibilities for the choice of the constants $C_{IJK}$, and therefore five different theories leading to the scalar potential \eqref{eq:runaway_scalar_pot}. The associated values for the constants $v_I$ and $C_I$ are given in the second vertical part of the table, from which one can find all relevant quantities, among which the functions $h_I$ and the kinetic matrix $G_{IJ}$.
\begin{center}
\begin{tabular}{|c|cccc||cccc|}
   \hline
   $~$ & $C_{011}$ & \qquad $C_{111}$ & \qquad $C_{100}$ & \qquad $C_{000}$ & $v_1$ & \qquad $C_1$ & \qquad $v_0$ & \qquad $C_0$ \qquad \\
   \hline
   $(a)$ & $\neq 0$ & \qquad $0$ & \qquad $0$ & \qquad $0$ & $\neq 0$ & \qquad $0$ & \qquad $0$ & \qquad $\neq 0$ \\
   \hline
   $(b)$ & $\neq 0$ & \qquad $0$ & \qquad $\neq 0$ & \qquad $\neq 0$ & $\neq 0$ & \qquad $0$ & \qquad $\neq 0$ & \qquad $\neq 0$ \\
   \hline
   $(c)$ & $\neq 0$ & \qquad $\neq 0$ & \qquad $0$ & \qquad $0$ & $\neq 0$ & \qquad $\neq 0$ & \qquad $0$ & \qquad $\neq 0$ \\
   \hline
   $(d)$ & $\neq 0$ & \qquad $\neq 0$ & \qquad $\neq 0$ & \qquad $0$ & $\neq 0$ & \qquad $\neq 0$ & \qquad $\neq 0$ & \qquad $0$ \\
   \hline
   $(e)$ & $\neq 0$ & \qquad $\neq 0$ & \qquad $\neq 0$ & \qquad $\neq 0$ & $\neq 0$ & \qquad $\neq 0$ & \qquad $\neq 0$ & \qquad $\neq 0$ \\
   \hline
\end{tabular}
\captionof{table}{The five different supergravity theories (defined by the $C_{IJK}$ on the left part) coupled to one vector multiplet whose $U(1)_R$ gauging reproduces the potential of LST, and the associated constants (on the right part).}
\label{tab:possible_cases}
\end{center}

However, while being a priori distinct, these five cases have actually the same physical content. The scalar potential as well as the scalar metric having been fixed, it is sufficient to check that the actions of the vectors are equivalent for these different cases. According to \eqref{eq:ungauged_lagrangian}, the gauge field part of the action contains the kinetic term $-\frac{1}{4}G_{IJ}F^IF^J$ and the topological term $C_{IJK}A^IF^JF^K$, where spacetime indices, respectively contracted with the $5D$ metric and $5D$ Levi-Civita tensor, have been left implicit in both cases. In the simplest case $(a)$, the vector kinetic matrix $G_{IJ}^{(a)}$ and topological term $\mathcal{L}_{CS}^{(a)}$ reads:
\begin{eqnarray}
\label{eq:G_IJ_a}
G_{IJ}^{(a)}&=&
\begin{pmatrix}
3B^2C_0^2e^{\frac{4}{\sqrt{3}}\phi} & 0 \\
0 & \frac{8v_1^2}{3B^2}e^{-\frac{2}{\sqrt{3}}\phi}
\end{pmatrix},\\
\label{eq:topological_terms_a}
\mathcal{L}_{CS}^{(a)}&=&C_{011}A^0F^1F^1+2C_{011}A^1F^1F^0.
\end{eqnarray}
Starting with the case $(e)$, the kinetic metric and the topological terms are respectively given by:
\begin{eqnarray}
\label{eq:G_IJ_e}
G_{IJ}^{(e)}&=&
\begin{pmatrix}
\frac{8v_0^2}{3B^2}e^{-\frac{2}{\sqrt{3}}\phi}+3B^2C_0^2e^{\frac{4}{\sqrt{3}}\phi} & \frac{8v_0v_1}{3B^2}e^{-\frac{2}{\sqrt{3}}\phi}+3B^2C_0C_1e^{\frac{4}{\sqrt{3}}\phi} \\
\frac{8v_0v_1}{3B^2}e^{-\frac{2}{\sqrt{3}}\phi}+3B^2C_0C_1e^{\frac{4}{\sqrt{3}}\phi} & \frac{8v_1^2}{3B^2}e^{-\frac{2}{\sqrt{3}}\phi}+3B^2C_1^2e^{\frac{4}{\sqrt{3}}\phi}
\end{pmatrix},\\
\label{eq:topological_terms_e}
\mathcal{L}_{CS}^{(e)}&=&C_{011}A^0F^1F^1+2C_{011}A^1F^1F^0+C_{100}A^1F^0F^0+2C_{100}A^0F^1F^0\nonumber\\
&&\qquad +C_{111}A^1F^1F^1+C_{000}A^0F^0F^0.
\end{eqnarray}
Since $C_0\neq 0$ and $v_1\neq 0$, we are free to consider the field redefinitions
\begin{equation}
A_M^0\rightarrow A_M^0-\frac{C_1}{C_0}A_M^1,\qquad A_M^1\rightarrow A_M^1-\frac{v_0}{v_1}A_M^0,
\end{equation}
which bring $G_{IJ}^{(e)}$ into the diagonal form
\begin{equation}
G_{IJ}^{(e)}\rightarrow G_{IJ}^{(e)}=
\begin{pmatrix}
3B^2\left(C_0-C_1\frac{v_0}{v_1}\right)^2e^{\frac{4}{\sqrt{3}}\phi} & 0 \\
0 & \frac{8}{3B^2}\left(v_1-v_0\frac{C_1}{C_0}\right)^2e^{-\frac{2}{\sqrt{3}}\phi}
\end{pmatrix},
\end{equation}
while the Chern-Simons terms are sent to:
\begin{equation}
\mathcal{L}_{CS}^{(e)}\rightarrow\mathcal{L}_{CS}^{(e)}=\left(C_{011}+\frac{20v_0v_1C_1}{B}-\frac{12v_0^2C_1^2}{BC_0}+\frac{4v_0^3C_1^3}{Bv_1C_0^2}\right)\left(A^0F^1F^1+2A^1F^0F^1\right)\,.
\end{equation}
These are indeed equivalent to \eqref{eq:G_IJ_a}-\eqref{eq:topological_terms_a} of the case $(a)$. The theories $(b)$ and $(c)$ are immediately obtained from $(e)$ by simply turning off $C_1$ or $v_0$, respectively. Finally, for the case $(d)$ where $C_0=0$, one has to consider instead the field redefinition
\begin{equation}
A_M^0\rightarrow A_M^0-\frac{v_1}{v_0}A_M^1,
\end{equation}
which brings $G_{IJ}^{(d)}$ and $\mathcal{L}_{CS}^{(d)}$ into
\begin{eqnarray}
G_{IJ}^{(d)}&\rightarrow& G_{IJ}^{(d)}=
\begin{pmatrix}
\frac{8v_0^2}{3B^2}e^{-\frac{2}{\sqrt{3}}\phi} & 0 \\
0 & 3B^2C_1^2e^{\frac{4}{\sqrt{3}}\phi}
\end{pmatrix},\\
\mathcal{L}_{CS}^{(d)}&\rightarrow& \mathcal{L}_{CS}^{(d)}=C_{100}A^1F^0F^0+2C_{100}A^0F^0F^1.
\end{eqnarray}
This simply corresponds to an exchange between the $0$ and $1$ vectors compared to the case $(a)$, as can be immediately checked from Eqs. \eqref{eq:G_IJ_a}-\eqref{eq:topological_terms_a}. 

Hence, after appropriate field redefinitions, the five theories of Table~\ref{tab:possible_cases} turn out to be physically equivalent, and we can then restrict without lost of generality to the simplest one $(a)$, where $C_{111}=C_{100}=C_{000}=0$. Its prepotential
\begin{equation}
\mathcal{F}=\xi^0(\varphi^1)^2
\end{equation}
has been considered in the five-dimensional heterotic string theory compactified on $K_3\times S^1$ in \cite{heterotic_string_dual_5D}, where $1/{\xi^0}$ has been identified with the heterotic string coupling. It corresponds to the choice of the only non-vanishing constant $C_{IJK}$ 
\begin{equation}\label{eq:C_IJK_coeff}
C_{011}=\frac{1}{3\beta^3}.
\end{equation}
In particular, one can easily find from \eqref{eq:C_IJK_coeff} the gauge kinetic matrix
\begin{equation}
G_{IJ}=
\begin{pmatrix}
\frac{1}{2}e^{\frac{4}{\sqrt{3}}\phi} & 0 \\
0 & e^{-\frac{2}{\sqrt{3}}\phi} 
\end{pmatrix},
\end{equation}
as well as the various functions $h^I$, $h_I$, $h^I_{\tilde a}$ and $h_I^{\tilde a}$ defined in Appendix \ref{sect:append_ungauged_sugra}:
\begin{align}
h^0 &= \sqrt{\frac{2}{3}} e^{-\frac{2}{\sqrt{3}}\phi}, & 
h^1 &= \sqrt{\frac{2}{3}} e^{\frac{1}{\sqrt{3}}\phi}, \\
h_0 &= \frac{1}{\sqrt 6} e^{\frac{2}{\sqrt{3}}\phi}, &
h_1 &= \sqrt{\frac{2}{3}} e^{-\frac{1}{\sqrt{3}}\phi}, \\
\label{eq:function_h^I_1}
h^0_{\tilde 1} &= \frac{2}{\sqrt 3} e^{-\frac{2}{\sqrt{3}}\phi}, & 
h^1_{\tilde 1} &= -\frac{1}{\sqrt{3}}e^{\frac{\phi}{\sqrt{3}}}, \\
h_0^{\tilde 1} &= \frac{1}{\sqrt{3}} e^{\frac{2}{\sqrt{3}}\phi}, & 
h_1^{\tilde 1} &= -\frac{1}{\sqrt{3}}e^{-\frac{\phi}{\sqrt{3}}}.
\end{align}

The minimal supersymmetric extension of the dilaton-gravity theory we are working with is therefore described in the Einstein frame by the Lagrangian $\mathcal{L}=\mathcal{L}^{(0)}+\mathcal{L}^{'}$, where $\mathcal{L}^{(0)}$ is the Lagrangian of the ungauged theory given in \eqref{eq:ungauged_lagrangian} and $\mathcal{L}^{'}$ the part added by the gauging \eqref{eq:additional_lagrangian}. From the results for $G_{IJ}$ and $P$ found above, it is easy to see that its bosonic part is given by:
\begin{eqnarray}\label{eq:total_bosonic_action}
e^{-1}\mathcal{L}^{\text{bos}}&=&\frac{1}{2}\mathcal R^{(5)}[G_{MN}]-\frac{1}{2}\partial_M\phi\partial^M\phi-e^{\frac{2}{\sqrt{3}}\phi}\Lambda\nonumber\\
&&\qquad-\frac{1}{8}e^{\frac{4}{\sqrt{3}}\phi}F_{MN}^0F^{MN0}-\frac{1}{4}e^{-\frac{2}{\sqrt{3}}\phi}F_{MN}^1F^{MN1}\\
&&\qquad\qquad+\frac{e^{-1}}{6\sqrt{6}}C_{011}\epsilon^{MNPQR}\left(A_M^0F_{NP}^1F_{QR}^1+2A_M^1F_{NP}^1F_{QR}^0\right).\nonumber
\end{eqnarray}
In addition of the terms on the first line, already present in the analysis of the Sections \ref{sect:LD_model} and \ref{sect:KK_reduction}, the Lagrangian also contains the kinetic terms for the two gauge bosons as well as a five-dimensional Chern-Simons term, with $F_{MN}^0$ and $F_{MN}^1$ the abelian field strengths of the graviphoton and the $U(1)$ vector respectively. From the $5D$ point of view, the only effect of the gauging in the bosonic sector appears in the scalar potential. 

This theory is invariant under $\mathcal{N}=2$ supersymmetries in four dimensions, namely has $8$ real conserved supercharges. In the following subsection \ref{sect:BPS_solution}, we remind that the background solution $\phi=\alpha|y|$ preserves ${1}/{2}$ of the supersymmetries, and compute the direction of the unbroken one. A second source of supersymmetry breaking can then arise from the compactification of the theory on $S^1/\mathbb{Z}_2$, and from the introduction of branes on the boundaries. We will then check in subsection \ref{sect:susy_branes} that the supersymmetry preserved by the LD background is also preserved by the branes located at the singular points of the orbifold. 

\subsection{The LD background as a ${1}/{2}$-BPS solution}
\label{sect:BPS_solution}
In the vacuum of the theory where all fermions and vectors vanish, the relevant parts of the supersymmetry transformations for the gravitini and dilatini read:
\begin{eqnarray}
\label{eq:gravitini_susy_transfo}
\delta(\epsilon_1,\epsilon_2)\psi_{Mi}&=&\partial_M\epsilon_i+\frac{1}{4}\omega_M^{mn}(e)\gamma_{mn}\epsilon_i-\frac{i\alpha}{2\sqrt{3}}e^{\frac{\phi}{\sqrt{3}}}e_M^m\gamma_m\epsilon_{ij}\delta^{jk}\epsilon_k,\\
\label{eq:dilatini_susy_transfo}
\delta(\epsilon_1,\epsilon_2)\lambda_i&=&-\frac{i}{2}e^{M}_m\gamma^m\partial_M\phi\epsilon_i+\frac{\alpha}{2}e^{\frac{\phi}{\sqrt{3}}}\epsilon_{ij}\delta^{jk}\epsilon_k,
\end{eqnarray}
where the last terms proportional to $\alpha$ arise from the gauging. In order to compute their explicit forms in the background, we recall the background metric and frame field in the Einstein frame:
\begin{equation}
G_{MN}=e^{-\frac{2}{\sqrt{3}}\alpha|y|}\eta_{MN},\qquad e^m_M=e^{-\frac{1}{\sqrt{3}}\alpha|y|}\delta^m_M.
\end{equation}
The relation between the spin-connection $\omega_M^{mn}(e)$ and the frame field $e^m_M$,
\begin{equation}
\omega_M^{mn}(e)=2e^{N[m}\partial_{[M}e_{N]}^{n]}-e^{N[m}e^{n]Q}e_{Mq}\partial_Ne_Q^q,
\end{equation}
leads to the following components for the spin-connection:
\begin{subequations}\label{eq:spin_connection_components}
\begin{align}
\omega_{\mu}^{ab}(e) &=\omega_5^{ab}(e)=\omega_5^{a5}(e)=0, \\
\omega_{\mu}^{a5}(e) &= -\frac{\alpha}{\sqrt{3}}|y|^{'}\delta_{\mu}^a=-\frac{\alpha}{\sqrt{3}}\sgn(y)\delta_{\mu}^a.
\end{align}
\end{subequations}
The transformations \eqref{eq:gravitini_susy_transfo} and \eqref{eq:dilatini_susy_transfo} in the background $\phi=\alpha|y|$ therefore read:
\begin{eqnarray}
\label{eq:gravitini_susy_transfo_2}
\delta(\epsilon_1,\epsilon_2)\psi_{Mi}&=&\partial_M\epsilon_i-\frac{\alpha}{2\sqrt{3}}\sgn y\delta^a_M\gamma_a\gamma_5\epsilon_i-\frac{i\alpha}{2\sqrt{3}}\delta_M^m\gamma_m\epsilon_{ij}\delta^{jk}\epsilon_k,\\
\label{eq:dilatini_susy_transfo_2}
\delta(\epsilon_1,\epsilon_2)\lambda_i&=&\frac{\alpha}{2}e^{\frac{\phi}{\sqrt{3}}}\left[-i\sgn y\gamma_5\epsilon_i+\epsilon_{ij}\delta^{jk}\epsilon_k\right].
\end{eqnarray}

Setting the transformations \eqref{eq:gravitini_susy_transfo_2} and \eqref{eq:dilatini_susy_transfo_2} to zero, we obtain the Killing equations which need to be solved in order to study the existence or not of preserved supersymmetries in the vacuum. We start with the Killing equations for the fifth components of the two gravitini, which form the system of coupled partial differential equations
\begin{subequations}\label{eq:psi5_Killing_eq}
\begin{align}
\delta(\epsilon_1,\epsilon_2)\psi_{51}&=\partial_5\epsilon_1-\frac{i\alpha}{2\sqrt{3}}\gamma_5\epsilon_2=0,\\
\label{eq:psi52_Killing_eq}
\delta(\epsilon_1,\epsilon_2)\psi_{52}&=\partial_5\epsilon_2+\frac{i\alpha}{2\sqrt{3}}\gamma_5\epsilon_1=0,
\end{align}
\end{subequations}
whose solutions are given by:
\begin{equation}
\epsilon_1(x,y)=e^{-\frac{\alpha}{2\sqrt{3}}|y|}\epsilon(x),~~~\epsilon_2(x,y)=i\sgn(y)e^{-\frac{\alpha}{2\sqrt{3}}|y|}\gamma_5\epsilon(x).
\end{equation}
Plugging these solutions into the Killing equations for the 4-dimensional gravitini,
\begin{equation}
\delta(\epsilon_1,\epsilon_2)\psi_{\mu i}=\partial_{\mu}\epsilon_i-\frac{\alpha}{2\sqrt{3}}\sgn(y)\delta_{\mu}^a\gamma_a\gamma_5\epsilon_i-\frac{i\alpha}{2\sqrt{3}}\delta_{\mu}^a\gamma_a\epsilon_{ij}\delta^{jk}\epsilon_k=0
\end{equation}
we find that they are satisfied provided that $\epsilon(x)$ is a constant spinor, $\epsilon(x)=\epsilon$. The solutions to the Killing equations are thus given by
\begin{subequations}\label{eq:Killing_spinors}
\begin{align}
\label{eq:Killing_spinor_1}
\epsilon_1(y)&=e^{-\frac{\alpha}{2\sqrt{3}}|y|}\epsilon,\\
\label{eq:Killing_spinor_2}
\epsilon_2(y)&=i\sgn(y)e^{-\frac{\alpha}{2\sqrt{3}}|y|}\gamma_5\epsilon,
\end{align}
\end{subequations}
and the direction of unbroken supersymmetry is obviously:
\begin{equation}\label{eq:direction_unbroken_susy}
\epsilon_2(y)=i\sgn(y)\gamma_5\epsilon_1(y).
\end{equation}
It is then straightforward to check that the dilatini Killing equations are also satisfied, namely $\delta(\epsilon_1,\epsilon_2=i\sgn y\gamma_5\epsilon_1)\lambda_1=\delta(\epsilon_1,\epsilon_2=i\sgn y\gamma_5\epsilon_1)\lambda_2=0$.

We thus conclude that the linear dilaton background $\phi=\alpha|y|$ breaks $\mathcal{N}=2$ supersymmetry to $\mathcal{N}=1$, with the 4 remaining real supercharges associated with the 4 components of the Killing spinor $\epsilon$.

In the following we will define the supersymmetric transformation in the direction of the preserved supersymmetry by:
\begin{equation}
\delta_L\equiv\delta(\epsilon_1,\epsilon_2=i\sgn(y)\gamma_5\epsilon_1),
\end{equation}
while the transformation in the direction of the broken supersymmetry would be given by $\delta_{NL}\equiv\delta(\epsilon_1,\epsilon_2=-i\sgn(y)\gamma_5\epsilon_1)$.

\subsection{Preserved supersymmetry and NS5-branes}
\label{sect:susy_branes}
As described in Section \ref{sect:5D_theory}, introducing a $\mathbb{Z}_2$ symmetry on the background solution produces discontinuous terms in the equations of motion, whose cancellation requires adding brane contributions to the original Lagrangian. In the Einstein frame, these brane Lagrangians are given by:
\begin{equation}\label{eq:branes_lagrangians}
\mathcal{L}_1=-2\sqrt{3}\alpha e^{(4)}e^{\frac{\phi}{\sqrt{3}}}\delta(y),\qquad
\mathcal{L}_2=2\sqrt{3}\alpha e^{(4)}e^{\frac{\phi}{\sqrt{3}}}\delta(y-L),
\end{equation}
with $e^{(4)}$ the Einstein frame four-dimensional vierbein induced on the branes. The aim of this subsection is to show how the boundary terms coming from the supersymmetric variation of the bulk Lagrangian are cancelled by the supersymmetric variations of the brane Lagrangians \eqref{eq:branes_lagrangians}, up to linear order in the dilatini $\lambda_i$ and gravitini $\psi_{M i}$. The part of the bulk Lagrangian whose supersymmetric variation brings terms linear in $\lambda$ reads:
\begin{equation}
\mathcal{L}=\mathcal{L}_{kin}^{\phi}+\mathcal{L}_{kin}^{\lambda}+\mathcal{L}_{\phi\lambda\psi}+\mathcal{L}_{\Lambda}^{(\alpha)}+\mathcal{L}_{\lambda\lambda}^{(\alpha)}+\mathcal{L}_{\lambda\psi}^{(\alpha)},
\end{equation} 
with, in the Einstein frame:
\begin{eqnarray}
e^{-1}\mathcal{L}_{kin}^{\phi}&=&-\frac{1}{2}\partial_M\phi\partial^M\phi,\\
e^{-1}\mathcal{L}_{kin}^{\lambda}&=&-\frac{1}{2}\bar{\lambda}^i\gamma^MD_M(\omega)\lambda_i=-\frac{1}{2}\bar{\lambda}^i\gamma^M(\partial_M+\frac{1}{4}\omega_M^{mn}\gamma_{mn})\lambda_i,\\
e^{-1}\mathcal{L}_{\phi\lambda\psi}&=&-\frac{i}{2}\partial_N\phi\bar{\lambda}^i\gamma^M\gamma^N\psi_{Mi},\\
e^{-1}\mathcal{L}_{\Lambda}^{(\alpha)}&=&\frac{3}{2}\alpha^2e^{\frac{2}{\sqrt{3}}\phi},\\
e^{-1}\mathcal{L}_{\lambda\lambda}^{(\alpha)}&=&-\frac{i\alpha}{4\sqrt{3}}e^{\frac{\phi}{\sqrt{3}}}\bar{\lambda}^i\lambda^j\delta_{ij},\\
e^{-1}\mathcal{L}_{\lambda\psi}^{(\alpha)}&=&\frac{\alpha}{2}e^{\frac{\phi}{\sqrt{3}}}\bar{\lambda}^i\gamma^M\psi_M^j\delta_{ij},
\end{eqnarray}
and where the superscript $^{(\alpha)}$ means that the corresponding terms arise from the gauging. The relevant parts of the supersymmetric transformations of the dilaton, dilatini and gravitini are respectively given by:
\begin{eqnarray}
\label{eq:dilaton_susy_transfo}
\delta\phi&=&\frac{i}{2}\bar{\epsilon}^i\lambda_i,\\
\delta\lambda^i&=&-\frac{i}{2}\slashed{\partial}\phi\epsilon^i-\frac{\alpha}{2}e^{\frac{\phi}{\sqrt{3}}}\delta^{ij}\epsilon_j,\\
\delta\psi_M^i&=&D_M(\omega)\epsilon^i+\frac{i\alpha}{2\sqrt{3}}\delta_M^m\gamma_m\delta^{ij}\epsilon_j.
\end{eqnarray}
In the bulk, the variation of the dilaton and dilatini kinetic terms, as well as $\mathcal{L}_{\phi\lambda\psi}$, yields:
\begin{eqnarray}
\label{eq:susy_var_dil_kin}
e^{-1}\delta\mathcal{L}_{kin}^{\phi}&=&-\frac{i}{2}\partial_M\phi\bar{\epsilon}^i\partial^M\lambda_i=\frac{i}{2}\partial^M\partial_M\phi\bar{\epsilon}^i\lambda_i,\\
e^{-1}\delta\mathcal{L}_{kin}^{\lambda}
\label{eq:susy_var_gaug_kin}
&=&\frac{i}{2}\partial^M\partial_M\phi\bar{\lambda}^i\epsilon_i+\frac{i}{2}\partial_N\phi\bar{\lambda}^i\gamma^M\gamma^ND_M(\omega)\epsilon_i\nonumber\\
&&\qquad-\frac{\alpha}{2\sqrt{3}}e^{\frac{\phi}{\sqrt{3}}}\bar{\lambda}^i\slashed{\partial}\phi\epsilon_k\epsilon_{ij}\delta^{jk}-\frac{\alpha}{2}e^{\frac{\phi}{\sqrt{3}}}\bar{\lambda}^i\slashed{D}(\omega)\epsilon_k\epsilon_{ij}\delta^{jk},\\
\label{eq:susy_var_phi_lambda_psi}
e^{-1}\delta\mathcal{L}_{\phi\lambda\psi}&=&-\frac{i}{2}\partial_N\phi\bar{\lambda}^i\gamma^M\gamma^ND_M(\omega)\epsilon_i+\frac{3\alpha}{4\sqrt{3}}e^{\frac{\phi}{\sqrt{3}}}\bar{\lambda}^i\slashed{\partial}\phi\epsilon_k\epsilon_{ij}\delta^{jk}.
\end{eqnarray}
One can already check that without the gauging, i.e. taking $\alpha=0$, and using $\bar{\epsilon}^i\lambda_i=-\bar{\lambda}^i\epsilon_i$, the three above variations cancel, $\delta(\mathcal{L}_{kin}^{\phi}+\mathcal{L}_{kin}^{\lambda}+\mathcal{L}_{\phi\lambda\psi})=0$ as expected. The gauging produces extra terms in $\delta\mathcal{L}_{kin}^{\lambda}$ and $\delta\mathcal{L}_{\phi\lambda\psi}$, proportional to $\alpha$, which are cancelled by the variation of $\mathcal{L}_{\Lambda}^{(\alpha)}+\mathcal{L}_{\lambda\lambda}^{(\alpha)}+\mathcal{L}_{\lambda\psi}^{(\alpha)}$, as it can be checked by computing:
\begin{eqnarray}
\label{eq:susy_var_Lambda}
e^{-1}\delta\mathcal{L}_{\Lambda}^{(\alpha)}&=&\frac{i\sqrt{3}}{2}\alpha^2e^{\frac{2}{\sqrt{3}}\phi}\bar{\epsilon}^i\lambda_i,\\
\label{eq:susy_var_lambda_lambda}
e^{-1}\delta\mathcal{L}_{\lambda\lambda}^{(\alpha)}&=&-\frac{\alpha}{4\sqrt{3}}e^{\frac{\phi}{\sqrt{3}}}\bar{\lambda}^i\slashed{\partial}\phi\epsilon^j\delta_{ij}+\frac{i\alpha^2}{4\sqrt{3}}e^{\frac{2}{\sqrt{3}}\phi}\bar{\lambda}^i\epsilon_k\delta^{k}_i\\
e^{-1}\delta\mathcal{L}_{\lambda\psi}^{(\alpha)}&=&\frac{\alpha}{2}e^{\frac{2}{\sqrt{3}}\phi}\bar{\lambda}^i\gamma^M\delta\psi_M^j\delta_{ij}\\
\label{eq:susy_var_lambda_psi}
&=&\frac{\alpha}{2}e^{\frac{\phi}{\sqrt{3}}}\bar{\lambda}^i\slashed{D}(\omega)\epsilon^j\delta_{ij}+\frac{5i\alpha^2}{4\sqrt{3}}e^{\frac{2}{\sqrt{3}}\phi}\bar{\lambda}^i\epsilon_k\delta^{k}_i.
\end{eqnarray}
From \eqref{eq:susy_var_dil_kin}, \eqref{eq:susy_var_gaug_kin}, \eqref{eq:susy_var_phi_lambda_psi}, \eqref{eq:susy_var_Lambda}, \eqref{eq:susy_var_lambda_lambda} and \eqref{eq:susy_var_lambda_psi}, we see that $\delta(\mathcal{L}_{kin}^{\phi}+\mathcal{L}_{kin}^{\lambda}+\mathcal{L}_{\phi\lambda\psi}+\mathcal{L}_{\Lambda}^{(\alpha)}+\mathcal{L}_{\lambda\lambda}^{(\alpha)}+\mathcal{L}_{\lambda\psi}^{(\alpha)})=0$ at linear order in $\lambda$.

We now consider the boundary terms on $S^1/\mathbb{Z}_2$. They arise from integrations by part done in the bulk variations, as well as from the brane Lagrangians \eqref{eq:branes_lagrangians}. In the bulk analysis carried out above, two integrations by parts have been done, in the variations of the dilatini and dilaton kinetic terms:

(i) In $\delta\mathcal{L}_{kin}^{\lambda}$, the integration by parts brings a total derivative of the form $\partial_5(ee^{\frac{\phi}{\sqrt{3}}}\bar{\lambda}^i\gamma^5\delta\lambda_i)$. At the linear level in the fluctuations $\bar{\lambda}_i$, we evaluate $\delta\lambda_i$ at the background level, which vanishes in the direction of the preserved supersymmetry
\begin{equation}
\delta_L\lambda_i\equiv\delta(\epsilon_1,\epsilon_2=i\sgn y \gamma_5\epsilon_1)\lambda_i=0.
\end{equation}
We therefore conclude that the dilatini kinetic term does not bring additional boundary contributions in the variation of the Lagrangian.

(ii) In $\delta\mathcal{L}_{kin}^{\phi}$, the integration by parts yields
\begin{eqnarray}
\delta\mathcal{L}_{kin}^{\phi}
\label{eq:total_deriv_phi}
&=&-\frac{i}{2}\partial_5\left(e\partial^5\phi\bar{\epsilon}^i\lambda_i\right)+\frac{i}{2}e\partial^M\partial_M\phi\bar{\epsilon}^i\lambda_i.
\end{eqnarray}
The first term is a total derivative integrated on the interval $[-L,L]$. The integrand $\partial_5\left(e\partial^5\phi\bar{\epsilon}^i\lambda_i\right)$ being an even function of $y$, its integral on $[-L,L]$ is $2$ times its integral on $[0,L]$, giving, at linear order in $\lambda_i$, $-i\alpha\left[e^{-\sqrt{3}\alpha|y|}\bar{\epsilon}^i\lambda_i\right]_{y=0}^{y=L}$. The second term of \eqref{eq:total_deriv_phi} cancels with the first one of \eqref{eq:susy_var_gaug_kin}, like in the bulk analysis, and we thus conclude that in the direction of the preserved supersymmetry $\epsilon_2=i\sgn y \gamma_5\epsilon_1$, $\delta(\mathcal{L}_{kin}^{\phi}+\mathcal{L}_{kin}^{\lambda})$ brings the boundary contribution
\begin{equation}\label{eq:bulk_bound_terms_lin_lambda}
\delta(\mathcal{L}_{kin}^{\phi}+\mathcal{L}_{kin}^{\lambda})=-i\alpha\left[e^{-\sqrt{3}\alpha|y|}\bar{\epsilon}^i\lambda_i\right]_{y=0}^{y=L}.
\end{equation}
It is then straightforward to check that this contribution is indeed cancelled by the supersymmetric variation of the brane Lagrangians \eqref{eq:branes_lagrangians}. Using \eqref{eq:dilaton_susy_transfo}, the variation of the dilaton $\phi$ yields:
\begin{equation}
\delta_{\phi}\mathcal{L}_1=-i\alpha e^{(4)}e^{\frac{\phi}{\sqrt{3}}}\bar{\epsilon}^i\lambda_i\delta(y),\qquad
\delta_{\phi}\mathcal{L}_2=i\alpha e^{(4)}e^{\frac{\phi}{\sqrt{3}}}\bar{\epsilon}^i\lambda_i\delta(y-L).
\end{equation}
Since we are interested in the variations linear in the fluctuations $\lambda_i$, we replace $\phi$ and $e^{(4)}$ by their background values $\phi=\alpha|y|$ and $e^{(4)}=e^{-\frac{4}{\sqrt{3}}\alpha|y|}$, leading to
\begin{equation}
\delta_{\phi}\mathcal{L}_1=-i\alpha e^{-\sqrt{3}\alpha|y|}\bar{\epsilon}^i\lambda_i\delta(y),\qquad
\delta_{\phi}\mathcal{L}_2=i\alpha e^{-\sqrt{3}\alpha|y|}\bar{\epsilon}^i\lambda_i\delta(y-L),
\end{equation}
so that:
\begin{equation}\label{eq:branes_bound_terms_lin_lambda}
\delta_{\phi}(\mathcal{L}_1+\mathcal{L}_2)=i\alpha\left[e^{-\sqrt{3}\alpha|y|}\bar{\epsilon}^i\lambda_i\right]_{y=0}^{y=L},
\end{equation}
which exactly cancels the boundary terms \eqref{eq:bulk_bound_terms_lin_lambda} coming from the bulk variations.

Finally, we consider the supersymmetric variations linear in the gravitini $\psi_{Mi}$. The part of the bulk Lagrangian whose supersymmetric variation brings terms linear in $\psi_{M i}$ is simply the Einstein-Hilbert action,
\begin{equation}
\mathcal{L}_{EH}=\frac{1}{2}e\mathcal{R}^{(5)}(\omega)=\frac{1}{2}ee^{M}_m e^{N}_n{R_{MN}}^{mn}(\omega),
\end{equation}
whose supersymmetric variation contains two terms:
\begin{equation}\label{eq:susy_var_EH}
\delta\mathcal{L}_{EH}=-\frac{1}{2} e\left(R_{MN}-\frac{1}{2}G_{MN}\mathcal{R}^{(5)}\right)\bar{\epsilon}^i\gamma^{M}\psi^{N}_i+\frac{1}{2}ee^{M}_m e^{N}_n\delta{R_{MN}}^{mn}(\omega).
\end{equation}
From $\delta{R_{MN}}^{mn}(\omega)=D_M\delta{\omega_N}^{mn}-D_N\delta{\omega_M}^{mn}$, we see that the second term is a total derivative whose integral would vanish in the absence of boundaries. Taking into account the boundaries in the fifth direction, it remains the total derivative $\partial_5(e e^5_n e^N_p\delta{\omega_{N}}^{np})$. Contracting the general variation of the spin connection $\omega_{Mnp}$
\begin{equation}
e^n_Ne^p_P\delta\omega_{Mnp}=(D_{[M}\delta e_{N]}^n)e_{Pn}-(D_{[N}\delta e_{P]}^n)e_{Mn}+(D_{[P}\delta e_{M]}^n)e_{Nn}
\end{equation}
with $G^{PM}$, we see that $e^5_n e^N_p\delta{\omega_{N}}^{np}=0$ and thus the second term of \eqref{eq:susy_var_EH} vanishes. In order to evaluate the first one at linear order in the perturbation $\psi^{N}_i$, we plug the background values in the Einstein tensor, which yields
\begin{equation}
R_{MN}-\frac{1}{2}G_{MN}\mathcal{R}^{(5)}=-2\sqrt{3}\alpha\eta_{MN}[\delta(y)-\delta(y-L)]+2\sqrt{3}\alpha\delta^5_M\delta^5_N[\delta(y)-\delta(y-L)]+\dots,
\end{equation}
where the dots denote bulk terms. In the contraction with $\bar{\epsilon}^i\gamma^{M}\psi^{N}_i$, the terms with $M=N=5$ cancel, and after considering the background value $e=e^{-\frac{5}{\sqrt{3}}\alpha|y|}$ and writing $\gamma^{\mu}=e^{\mu}_a\gamma^a=e^{\frac{\alpha}{\sqrt{3}}|y|}\delta^{\mu}_a\gamma^{a}$, $\psi^{\nu}_i=G^{\nu\rho}\psi_{\rho i}=e^{\frac{2}{\sqrt{3}}\alpha|y|}\eta^{\nu\rho}\psi_{\rho i}$, we end with
\begin{equation}\label{eq:bulk_bound_terms_lin_psi}
\delta\mathcal{L}_{EH}=\sqrt{3}\alpha e^{-\frac{2}{\sqrt{3}}\alpha|y|}\bar{\epsilon}^i\delta^{\mu}_a\gamma^{a}\psi_{\mu i}[\delta(y)-\delta(y-L)].
\end{equation}
Again we want to check that this contribution is cancelled by the supersymmetric variation of the brane Lagrangians \eqref{eq:branes_lagrangians}. From the variation of the determinant of the four-dimensional vierbein
\begin{equation}
\delta e^{(4)}=\frac{1}{2} e^{(4)}\bar{\epsilon}^i \gamma^{\mu}\psi_{\mu i},
\end{equation}
we deduce that
\begin{equation}
\delta_{e}\mathcal{L}_1=-\sqrt{3}\alpha e^{(4)}e^{\frac{\phi}{\sqrt{3}}}\bar{\epsilon}^i\gamma^{\mu}\psi_{\mu i}\delta(y),\qquad
\delta_{e}\mathcal{L}_2=\sqrt{3}\alpha e^{(4)}e^{\frac{\phi}{\sqrt{3}}}\bar{\epsilon}^i\gamma^{\mu}\psi_{\mu i}\delta(y-L).
\end{equation}
Since we are interested in the variations linear in the fluctuations $\psi_{\mu i}$, we replace the other fields by their background values $\phi=\alpha|y|$, $e^{(4)}=e^{-\frac{4}{\sqrt{3}}\alpha|y|}$, and write again $\gamma^{\mu}=e^{\mu}_a\gamma^a=e^{\frac{\alpha}{\sqrt{3}}|y|}\delta^{\mu}_a\gamma^{a}$, which leads to:
\begin{equation}\label{eq:branes_bound_terms_lin_psi}
\delta_{e}\mathcal{L}_1=-\sqrt{3}\alpha e^{-\frac{2}{\sqrt{3}}\alpha|y|}\bar{\epsilon}^i\delta^{\mu}_a\gamma^{a}\psi_{\mu i}\delta(y),\qquad
\delta_{e}\mathcal{L}_2=\sqrt{3}\alpha e^{-\frac{2}{\sqrt{3}}\alpha|y|}\bar{\epsilon}^i\delta^{\mu}_a\gamma^{a}\psi_{\mu i}\delta(y-L).
\end{equation}
Again, these brane variations exactly cancel the boundary terms \eqref{eq:bulk_bound_terms_lin_psi} coming from the bulk variation of $\mathcal{L}_{EH}$. We thus conclude that the original $\mathcal{N}=1$ supersymmetry preserved by the linear dilaton background on $\mathbb{R}^{1,4}$ remains preserved after the compactification of the fifth direction on $S^1/\mathbb{Z}_2$, provided the branes added at the two boundaries of the interval are NS5-branes.

\section{Compactified $D=4$ effective theory}
\label{sect:compactified_theory}
The total bosonic Lagrangian of the $\mathcal{N}=2$, $D=5$ supergravity theory introduced in Section \ref{sec:runaway_scal_pot} has been written in \eqref{eq:total_bosonic_action}. The compactification of the graviton-dilaton system performed in Section \ref{sect:KK_reduction} has revealed two important features: on the LD background, the KK vector $K_{\mu}$ coming from the $5D$ metric becomes massive by absorbing a scalar combination $\varphi_+$, identified with the string frame radion, while only the orthogonal combination $\varphi_-$, identified as the $4D$ dilaton, remains massless, with a vanishing effective scalar potential. Since we know that the LD background breaks $1/2$ of the original supersymmetries, we should be able to write the effective $\mathcal{N}=1$, $D=4$ supergravity after identifying all massive states and decoupling them from the massless spectrum. To this purpose, in the following two subsections we will first dimensionally reduce the remaining part of the bosonic action, taking into account that the zero modes of the fields may depend on the compactified coordinate $y$, in contrast with standard KK reduction. Then, in the scalar sector we will first need to identify the massless $4D$ scalars and complexify them in a consistent way to form a chiral matter multiplet. In the vector sector, since we already know that the KK vector $K_{\mu}$ becomes massive, we will need to identify a second massive vector, so that both of them could form the bosonic content of a massive spin-${3}/{2}$ multiplet, decoupled from the spectrum in the low energy limit.
 
\subsection{General considerations on dimensional reduction}
The five-dimensional action of the two vectors $A_M^0$ and $A_M^1$ contains, in addition to the kinetic terms $S_{kin}$, a Chern-Simons term $S_{CS}$,
\begin{equation}
S=S_{kin}+S_{CS},
\end{equation}
with
\begin{eqnarray}
\label{eq:5D_kinetic_term}
S_{kin}&=&\int d^5x\sqrt{-G}\left\{-\frac{1}{8}e^{\frac{4}{\sqrt{3}}\phi}\hat F_{MN}^0\hat F^{MN0}-\frac{1}{4}e^{-\frac{2}{\sqrt{3}}\phi}\hat F_{MN}^1\hat F^{MN1}\right\},\\
S_{CS}&=&\int d^5x\frac{1}{6\sqrt{6}}C_{011}\hat \epsilon^{MNPQR}\left(\hat A_M^0\hat F_{NP}^1\hat F_{QR}^1+2\hat A_M^1\hat F_{NP}^1\hat F_{QR}^0\right),
\end{eqnarray} 
where we have denoted all the five dimensional quantities with a hat in order to distinguish them from their four dimensional counterparts. The standard dimensional reduction of a given tensor field is usually performed in the vielbein formalism rather than the metric one. Using local Lorentz transformations, one can write the $5D$ Einstein frame f\"{u}nfbein $\hat e_M^m$ and its inverse $\hat e_m^M$ as:
\begin{equation}\label{eq:funfbein_field_param}
\hat e_M^m=e^{-\frac{\alpha}{\sqrt 3}|y|}
\begin{pmatrix}
e^{-\frac{1}{2}r}\tilde e_{\mu}^a(x) & e^{r}K_{\mu}(x,y) \\
0 & e^{r}
\end{pmatrix},
\qquad
\hat e_m^M=e^{\frac{\alpha}{\sqrt 3}|y|}
\begin{pmatrix}
e^{-\frac{1}{2}r}\tilde e^{\mu}_a(x) & -e^{-\frac{1}{2}r}K_{a}(x,y) \\
0 & e^{-r}
\end{pmatrix},
\end{equation}
whose squares obviously reproduce the metric \eqref{eq:final_metric} and its inverse \eqref{eq:final_inverse_metric}. The dimensional reduction of the vector kinetic terms is then carried out in the following way: we first identify the five and four dimensional vectors $\hat A_a$ and $A_a$ on flat indices, and then we use the f\"{u}nfbein parametrisation \eqref{eq:funfbein_field_param} to relate five and four-dimensional vectors with curved indices:\begin{equation}
A_a\equiv\hat A_a=\hat e^M_a\hat A_M=\hat e^{\mu}_a\hat A_{\mu}+\hat e^{5}_a\hat A_5=e^{\frac{\alpha}{\sqrt 3}|y|}e^{\frac{1}{2}r}\tilde e_a^{\mu}(\hat A_{\mu}-A_5K_{\mu}),
\end{equation}
where we have defined $\hat A_5=A_5$. This construction automatically implies invariance of the $4D$ vector $A_{\mu}=\hat A_{\mu}-A_5K_{\mu}$ under the $U(1)_{KK}$\footnote{$\delta_{\xi^5}A_{\mu}=0$ holds only in the standard case when $A_{\mu}$ is independent of the compactified coordinate. If $A_{\mu}$ has a $y$-dependence, it transforms under the $U(1)_{KK}$ according to $\delta_{\xi^5}A_{\mu}=\xi^5\partial_5A_{\mu}$.}. A similar analysis for the field strength yields:
\begin{eqnarray}
F_{ab}\equiv \hat F_{ab}&=&e^{\frac{2}{\sqrt 3}\alpha|y|}e^{r}\tilde e_a^{\mu}\tilde e^{\nu}_b\left\{2\partial_{[\mu}A_{\nu]}+2A_5\partial_{[\mu}K_{\nu]}-2K_{[\mu}\partial_5(A_{\nu]}+A_5K_{\nu]})\right\},\\
\hat F_{a5}&=&e^{\frac{2}{\sqrt 3}\alpha|y|}e^{-\frac{1}{2}r}\tilde e^{\mu}_a\left\{\partial_{\mu}A_5-\partial_5(A_{\mu}+A_5K_{\mu})\right\}.
\end{eqnarray}
Putting everything together, using $\sqrt{-G}=e^{-\frac{5\alpha}{\sqrt 3}|y|}e^{-r}\sqrt{-\tilde g}$ and moving from the scalar basis $(r,\delta\phi)$ to $(\varphi_-,\varphi_+)$, we get the vector kinetic action in terms of the four-dimensional quantities:
\begin{eqnarray}\label{eq:exact_kinetic_action}
S_{kin}=&-&\frac{1}{8}\int d^5xe^{\sqrt 3\alpha|y|}\sqrt{-\tilde g}\left\{e^{2\varphi_+-\varphi_-}\left[F_{\mu\nu}(A^0)+A_5^0F_{\mu\nu}(K)-2K_{[\mu}\partial_5(A_{\nu]}^0+A_5^0K_{\nu]})\right]^2\right.\nonumber\\
&&\qquad \left. +2e^{-2\varphi_-}\left[\partial_{\mu}A_5^0-\partial_5(A_{\mu}^0+A_5^0K_{\mu})\right]^2\right\}\nonumber\\
&-&\frac{1}{4}\int d^5xe^{-\sqrt 3\alpha|y|}\sqrt{-\tilde g}\left\{e^{\varphi_-}\left[F_{\mu\nu}(A^1)+A_5^1F_{\mu\nu}(K)-2K_{[\mu}\partial_5(A_{\nu]}^1+A_5^1K_{\nu]})\right]^2\right.\nonumber\\
&&\qquad \left. +2e^{-2\varphi_+}\left[\partial_{\mu}A_5^1-\partial_5(A_{\mu}^1+A_5^1K_{\mu})\right]^2\right\},
\end{eqnarray}
where in the right-hand-side (RHS), contractions are made with the $4D$ metric $\tilde g_{\mu\nu}(x)$. The dimensional reduction of the Chern-Simons term works in a similar way. We first write its expression in terms of flat indices and then identify the four and five-dimensional Levi-Civita tensors to be equal on flat indices, $\hat \epsilon^{abcd5}=\epsilon^{abcd}$. This leads to:
\begin{eqnarray}
C_{IJK}\hat\epsilon^{MNPQR}\hat A_M^I\hat F_{NP}^J\hat F_{QR}^K&=&C_{IJK}\sqrt{-G}\hat\epsilon^{mnpqr}\hat A_m^I\hat F_{np}^J\hat F_{qr}^K\nonumber\\
&=&C_{IJK}\sqrt{-G}\epsilon^{abcd}\left(\hat F_{ab}^I\hat F_{cd}^J\hat A_5^K-4\hat F_{ab}^I\hat F_{c5}^J\hat A_d^K\right).
\end{eqnarray}
Using the expressions for $\hat F_{ab}$, $\hat F_{a5}$ and $\hat A_a$ obtained above, converting back flat into curved indices and integrating by parts, we obtain:
\begin{eqnarray}\label{eq:exact_CS_action}
&C_{IJK}&\hat\epsilon^{MNPQR}\hat A^I_M \hat F_{NP}^J \hat F_{QR}^K=C_{IJK}\epsilon^{\mu\nu\rho\sigma}\left\{3A_5^IF_{\mu\nu}^J(A)F_{\rho\sigma}^K(A)+3A_5^IA_5^JF_{\mu\nu}^K(A)F_{\rho\sigma}(K)\right.\nonumber\\
&+&\left.A_5^IA_5^JA_5^KF_{\mu\nu}(K)F_{\rho\sigma}(K)-4(F_{\mu\nu}^I(A)+A_5^IF_{\mu\nu}(K))K_{[\rho}\partial_5\hat A_{\sigma]}^JA_5^K\right.\nonumber\\
&+&\left.4(F_{\mu\nu}^I(A)+A_5^IF_{\mu\nu}(K))\partial_5\hat A_{\rho}^JA_{\sigma}^K+8K_{[\mu}\partial_5\hat A_{\nu]}^I(\partial_{\rho}A_5^J-\partial_5\hat A_{\rho}^J)A_{\sigma}^K \right\}.
\end{eqnarray}
The difference compared to standard KK compactification, where the zero modes of the fields are assumed to be independent of the compactified coordinates, lies in the terms proportional to $\partial_5(\dots)$: they must be kept in the framework of the LD background, since the latter may introduce an explicit $y$-dependence even on the zero modes of the fields. These considerations are general in the sense that they do not depend on the background, and can be used in other frameworks where the fields may have given dependences on the compactified coordinates.

\subsection{$\mathcal{N}=1$, $D=4$ effective theory}

Having performed the dimensional reduction of the spin-1 action, one can now find the $4D$ spectrum of the zero modes of the different fields, and in particular check that the massless spectrum arranges into a $\mathcal{N}=1$ supersymmetric effective theory. Since we are ultimately interested in the massless limit, we can set the massive Kaluza-Klein vector $K_{\mu}=0$ in the dimensionally reduced action obtained above. Up to quadratic order in $A_{\mu}^I$, $A_{5}^I$, the kinetic action \eqref{eq:exact_kinetic_action} then reads:
\begin{eqnarray}\label{eq:quadratic_spin_1_action}
S_{kin}&=&\int d^5x\sqrt{-\tilde g}\left\{-\frac{1}{8}e^{\sqrt{3}\alpha|y|}e^{-\varphi_-+2\varphi_+}F_{\mu\nu}^0F^{0\mu\nu}-\frac{1}{4}e^{-\sqrt{3}\alpha|y|}e^{\varphi_-}F_{\mu\nu}^1F^{1\mu\nu}\right.\\
&&\qquad \left.-\frac{1}{4}e^{\sqrt{3}\alpha|y|}e^{-2\varphi_-}(\partial_{\mu}A_5^0-\partial_5A_{\mu}^0)^2-\frac{1}{2}e^{-\sqrt{3}\alpha|y|}e^{-2\varphi_+}(\partial_{\mu}A_5^1-\partial_5A_{\mu}^1)^2\right\},\nonumber
\end{eqnarray}
while the Chern-Simons action \eqref{eq:exact_CS_action}, for $C_{011}\neq 0$ only, yields:
\begin{equation}\label{eq:4d_CS_term}
S_{CS}=\int d^5x\frac{1}{6\sqrt{6}}C_{011}\epsilon^{\mu\nu\rho\sigma}\left(3A_5^0F_{\mu\nu}^1F_{\rho\sigma}^1+6A_5^1F_{\mu\nu}^1F_{\rho\sigma}^0\right).
\end{equation}
In order to ease the comparison between the Einstein and Maxwell sectors, both of which should contribute to the $\mathcal{N}=1$ effective supergravity, we also recall the (quadratic) dimensionally reduced action of the Kaluza-Klein vector $K_{\mu}$ and the scalars $\varphi_-$ and $\varphi_+$ coming from the $5D$ dilaton and gravitational actions, obtained previously in Section \ref{sect:stuckelberg_mech}:
\begin{eqnarray}
S_{dil+G}&=&\int d^5xe^{-\sqrt{3}\alpha|y|}\sqrt{-\tilde g}\left\{-\frac{1}{4}e^{\varphi_-+2\varphi_+}F^2(K)\right.\nonumber\\
&&\qquad\left.-\frac{3}{2}\left[\alpha\sgn yK_{\mu}-\frac{1}{\sqrt 3}\partial_{\mu}\varphi_+\right]^2-\frac{1}{4}(\partial_{\mu}\varphi_-)^2\right\}\,.
\end{eqnarray}
The point is that we would like to determine which are the massless $\mathcal{N}=1$ multiplets and the corresponding truncation of the dimensionally reduced action.

In the scalar sector, the exponentials in front of the kinetic terms of $A_5^0$ and $A_5^1$ in \eqref{eq:quadratic_spin_1_action} tell us that $\varphi_-$ and $\varphi_+$ must be complexified with $A_5^0$ and $A_5^1$ respectively. From the action for $\varphi_-$ found in section \ref{sect:eff_scal_pot}, $S[\varphi_-]=-\frac{1}{4}\int d^5xe^{-\sqrt{3}\alpha|y|}\sqrt{-\tilde g}(\partial_{\mu}\varphi_-)^2$, one sees that the exponential of $y$ factorizes out of the kinetic terms of $A_5^0$ and $\varphi_-$ if and only if $A_5^0$ is of the form 
\begin{equation}
A_5^0=e^{-\sqrt{3}\alpha|y|}\tilde A_5^0(x),
\end{equation}
which gives for the scalar action:
\begin{equation}
S_{\text{scalar}}=-\frac{1}{4}\int d^5xe^{-\sqrt{3}\alpha|y|}\sqrt{-\tilde g}\left\{(\partial_{\mu}\varphi_-)^2+e^{-2\varphi_-}(\partial_{\mu}\tilde A_5^0(x))^2\right\}.
\end{equation}
Defining the complex scalar field 
\begin{equation}
S=e^{\varphi_-}+i\tilde A_5^0(x),
\end{equation}
the above scalar action can then be rewriten as
\begin{eqnarray}
S_{\text{scalar}}=-\int d^5xe^{-\sqrt{3}\alpha|y|}\sqrt{-\tilde g}\frac{\partial_{\mu}S\partial^{\mu}\bar S}{(S+\bar S)^2},
\end{eqnarray}
from which we deduce the Kahler potential to be $K(S,\bar S)=-\ln(S+\bar S)$.

In the vector sector, the $\mathcal{N}=1$ spectrum can be found by considering the $4D$ Chern-Simons term \eqref{eq:4d_CS_term}. Denoting $f(S)$ the holomorphic gauge kinetic function, the resulting $\mathcal{N}=1$ theory must have a topological term of the form $\text{Im} f F\wedge F$. Such term can only come from the first one of \eqref{eq:4d_CS_term} $A_5^0F^1F^1$, which is consistent with the fact that $A_5^1$, being complexified with $\varphi_+$, must ultimately be projected out of the spectrum. We therefore deduce that the $A_{\mu}^1$ vector remains massless and sit in a $\mathcal{N}=1$ vector multiplet, while $A_{\mu}^0$ must acquire a mass, by absorbing $A_5^1$. From the kinetic terms \eqref{eq:quadratic_spin_1_action}, one sees that such a mechanism can be obtained by assuming a mixing of the zero modes of the vectors of the form
\begin{equation}
A_{\mu}^1=\tilde A_{\mu}^1(x)+e^{\sqrt 3\alpha|y|} A_{\mu}^0(x).
\end{equation}
The factor in the exponential, a priori arbitrary, is found by requiring the two massive vectors $K_{\mu}$ and $A_{\mu}^0$ to have the same mass, since they should form together with the massive gravitino a massive $\mathcal{N}=1$ spin-$3/2$ multiplet, namely
\begin{equation}
m^2_{U(1)_{KK}}=m^2_{U(1)^0}=\frac{3}{2}\alpha^2.
\end{equation}

In the low energy limit, obtained by truncating the massive spectrum, the remaining bosonic action reads (we denote $F^1\equiv F(\tilde A^1(x))$:
\begin{equation}
S_{\text{bos.}}=\int d^5xe^{-\sqrt{3}\alpha|y|}\sqrt{-\tilde g}\left\{\partial_S\partial_{\bar S}K\partial_{\mu}S\partial^{\mu}\bar S-\frac{1}{4}\text{Re}fF_{\mu\nu}^1F^{1\mu\nu}+\frac{1}{8}\frac{\text{Im}f}{\sqrt{-\tilde g}}\epsilon^{\mu\nu\rho\sigma}F_{\mu\nu}^1F_{\rho\sigma}^1\right\},
\end{equation}
which, after integration on $y$, gives the bosonic Lagrangian of a $D=4$, $\mathcal{N}=1$ supersymmetric theory, defined by the Kahler potential $K$, superpotential $W$ and gauge kinetic function $f$ given by:
\begin{equation}
K(S,\bar S)=-\ln(S+\bar S),\qquad W(S)=0,\qquad f(S)=S.
\end{equation}

It is easy to check that these results are consistent with a $\mathcal{N}=1$ supersymmetric spectrum obtained from a standard orbifold compactification. The four-dimensional theory contains two $\mathcal{N}=2$ vector multiplets $(A_{\mu}^0,e^{\varphi_-}+i\tilde A_5^0)$ and $(A_{\mu}^1,e^{\varphi_+}+i A_5^1)$. Two truncations $\mathcal{N}=2\rightarrow\mathcal{N}=1$ can then be considered, putting $A_{\mu}^1$ and $e^{\varphi_-}+i\tilde A_5^0$ to zero, or $A_{\mu}^0$ and $e^{\varphi_+}+iA_5^1$ to zero. The first case is obviously excluded, since no topological term would remain in \eqref{eq:4d_CS_term}, which is consistent with the gauge $\varphi_+=0$ set in Section \ref{sect:stuckelberg_mech}, where we also identified $\varphi_-$ with the massless scalar field. We are thus forced to truncate the $A_{\mu}^0$ and $A_5^1$, by assigning them a $\mathbb{Z}_2$-odd parity, while $A_{\mu}^1$ and $A_5^0$ are kept even. In Appendix \ref{sect:append_heterotic_string} we show that the $5D$ vector $A_M^0$ is dual to the $5D$ Kalb-Ramond two form $B_{MN}$. Truncating $A_{\mu}^0$ therefore amounts to assigning an orbifold $\mathbb{Z}_2$-odd parity to $B_{\mu 5}$, and thus an even parity to $B_{\mu\nu}$, the $4D$ dual of $A_5^0$. Regarding the string frame radion $\varphi_+$, it is obviously even under the $\mathbb{Z}_2$ of the orbifold, and odd under the discrete $\mathbb{Z}_2$ subgroup of the T-duality group, which inverts the radius of compactification and interchanges at the same time $B_{\mu 5}$ with the KK vector $K_{\mu}$. The full truncation must therefore combine both $\mathbb{Z}_2$ transformations; it projects out half of the degrees of freedom of the original theory, namely $6$ bosonic and $6$ fermionic on-shell degrees of freedom. These results are summarised in Table \ref{tab:bosonic_fields_parity}.  
\begin{table}
\begin{center}
\begin{tabular}{|cc||cc||cc|}
   \hline
   \multicolumn{2}{|c||}{$\mathcal{N}=2$ supergravity multiplet} & \multicolumn{2}{|c||}{$\mathcal{N}=2$ vector multiplet} & \multicolumn{2}{|c|}{$\mathcal{N}=2$ vector multiplet} \\
   \hline
   \qquad $g_{\mu\nu}$ & \qquad $g_{\mu 5}$ & \qquad $A_{\mu}^0$ & \qquad $e^{\varphi_-}+i\tilde A_5^0$ & \qquad $A_{\mu}^1$ & \qquad $e^{\varphi_+}+iA_{5}^1$ \qquad \\
   \hline
   \qquad $+$ & \qquad $-$ & \qquad $-$ & \qquad $+$ & \qquad $+$ & \qquad $-$ \\
   \hline
\end{tabular}
\captionof{table}{Bosonic content of the $D=4$, $\mathcal{N}=2$ multiplets and their $\mathbb{Z}_2$-parity.}
\label{tab:bosonic_fields_parity}
\end{center}
\end{table}

The $\mathcal{N}=1$, $D=4$ multiplet content is then easily obtained after dimensional reduction of the supersymmetry (susy) transformations introduced in Section \ref{sect:append_ungauged_sugra}, using in particular the $5D$ frame field parametrisation \eqref{eq:funfbein_field_param}. In the following, we will also work with the linear combinations of the fermions $\lambda_+$ and $\lambda_-$ defined by
\begin{equation}
\lambda_+\equiv\lambda_1+i\sgn y\gamma_5\lambda_2,\qquad \lambda_-\equiv\lambda_1-i\sgn y\gamma_5\lambda_2.
\end{equation} 
From the susy transformation of the dilaton \eqref{eq:original_dilaton_susy_transfo}, we deduce the transformation of the physical scalar $\varphi_-$ in the direction of the preserved supersymmetry:
\begin{equation}
\delta_Le^{\varphi_-}=-\frac{i}{\sqrt{3}}e^{\varphi_-}\bar{\epsilon}^1\lambda_+.
\end{equation}
Since we know the $y$-dependence of the spinor $\bar{\epsilon}^1$,
\begin{equation}
\bar{\epsilon}^1=e^{-\frac{\alpha}{2\sqrt{3}}|y|}\bar{\epsilon},
\end{equation}
with $\epsilon$ the constant Killing spinor associated with the $\mathcal{N}=1$ preserved supersymmetry, we deduce that the zero mode of the spinor $\lambda_+$ must satisfy
\begin{equation}
\lambda_+=e^{\frac{\alpha}{2\sqrt{3}}|y|}\tilde \lambda_+(x),
\end{equation}
where $\tilde \lambda_+(x)$ is a spinor independent of $y$. Similarly, the transformations of the $5D$ vectors \eqref{eq:original_vector_susy_transfo} yields for the $4D$ scalars $A_5^0$:
\begin{equation}
\delta_L A_5^0=-\frac{1}{\sqrt{3}}e^{-\sqrt{3}\alpha|y|}e^{\varphi_-}\bar{\epsilon}^1\gamma_5\lambda_+.
\end{equation}
The combination $\bar{\epsilon}^1\gamma_5\lambda_+$ being independent of $y$, we find again that $A_5^0$ must be of the form $A_5^0=e^{-\sqrt{3}\alpha|y|}\tilde A_5^0(x)$, as already obtained above. 
The transformation of the complex scalar $S=e^{\varphi_-}+i\tilde A_5^0$ is thus given by:
\begin{equation}
\delta_L S=-\frac{i}{\sqrt 3}e^{\varphi_-}\bar{\epsilon}^1(1+\gamma_5)\lambda_+.
\end{equation}

In the vector sector, the susy transformations of $A_{\mu}^1$ in the direction of the preserved supersymmetry is given by:
\begin{equation}
\delta_L A_{\mu}^1=\frac{1}{2\sqrt{3}}e^{-\frac{1}{2}\varphi_-}\bar{\epsilon}^1\tilde e_{\mu}^a(x)\gamma_a\lambda_-.
\end{equation}
Similarly as for $\lambda_+$, using $\bar{\epsilon}^1=e^{-\frac{\alpha}{2\sqrt{3}}|y|}\bar{\epsilon}$ and the fact that $A_{\mu}^1$ is a function of $x$ only tell us that $\lambda_-$ must have a $y$-dependence of the form
\begin{equation}
\lambda_-=e^{\frac{\alpha}{2\sqrt{3}}|y|}\tilde \lambda_-(x).
\end{equation}
The transformations of the fermions $\lambda_+$ and $\lambda_-$ respectively read:
\begin{eqnarray}
\delta_L\lambda_+&=&\frac{i}{\sqrt{3}}e^{\frac{\alpha}{\sqrt{3}}|y|}e^{\frac{1}{6}(\varphi_-+2\varphi_+)}\tilde e^{\mu}_a(x)\gamma^a\partial_{\mu}(\varphi_-+2\varphi_+)\epsilon_1\\
&+&\frac{1}{\sqrt{3}}\tilde e^{\mu}_a(x)\gamma^a\gamma^5\left[e^{\frac{4\alpha}{\sqrt{3}}|y|}e^{-\frac{5}{6}\varphi_-+\frac{1}{3}\varphi_+}(\partial_{\mu}A_5^0-\partial_5A_{\mu}^0)-e^{\frac{\alpha}{\sqrt{3}}|y|}e^{\frac{1}{6}\varphi_--\frac{2}{3}\varphi_+}(\partial_{\mu}A_5^1-\partial_5A_{\mu}^1)\right]\epsilon_1,\nonumber\\
\delta_L\lambda_-&=&\frac{1}{2\sqrt{3}}\tilde e^{\mu}_a(x)\tilde e^{\nu}_b(x)\gamma^a\gamma^b\left[e^{\frac{4\alpha}{\sqrt{3}}|y|}e^{-\frac{1}{3}\varphi_-+\frac{4}{3}\varphi_+}F_{\mu\nu}^0-e^{\frac{\alpha}{\sqrt{3}}|y|}e^{\frac{2}{3}\varphi_-+\frac{1}{3}\varphi_+}F_{\mu\nu}^1\right]\epsilon_1.
\end{eqnarray}
After gauge fixing the scalars $\varphi_+$ and $A_5^1$ to zero, truncating $A_{\mu}^0$ and using $A_5^0=e^{-\sqrt 3\alpha|y|}\tilde A_5^0(x)$ and that $A_{\mu}^1=\tilde A_{\mu}^1(x)$ when $A_{\mu}^0=0$, they lead for the four-dimensional spinors $\tilde\lambda_+(x)$ and $\tilde\lambda_-(x)$:
\begin{eqnarray}
\delta_L\tilde \lambda_+(x)&=&\frac{i}{\sqrt{3}}e^{\frac{1}{6}\varphi_-}\tilde e^{\mu}_a(x)\gamma^a\partial_{\mu}\varphi_-\epsilon(x)+\frac{1}{\sqrt{3}}\tilde e^{\mu}_a(x)\gamma^a\gamma^5e^{-\frac{5}{6}\varphi_-}\partial_{\mu}\tilde A_5^0(x)\epsilon(x),\\
\delta_L\tilde\lambda_-(x)&=&-\frac{1}{2\sqrt{3}}e^{\frac{2}{3}\varphi_-}\tilde e^{\mu}_a(x)\tilde e^{\nu}_b(x)\gamma^a\gamma^bF_{\mu\nu}^1\epsilon(x).
\end{eqnarray}
The gravitational multiplet is then easily obtained from
\begin{equation}
\delta_L\tilde e_{\mu}^a(x)=\frac{1}{2}e^{\frac{\alpha}{\sqrt 3}|y|}e^{\frac{1}{6}\varphi_-+\frac{1}{3}\varphi_+}\bar\epsilon^1\gamma^a\psi_{\mu-},
\end{equation}
where we have defined $\psi_{\mu-}\equiv\psi_{\mu 1}-i\sgn y\gamma_5\psi_{\mu 2}$ and used $e_{\mu}^a=e^{-\frac{\alpha}{\sqrt 3}|y|}e^{-\frac{1}{2}r}\tilde e_{\mu}^a(x)$. The same argument as for $\lambda_{\pm}$ imposes the $4D$ gravitino $\psi_{\mu-}$ to be of the form
\begin{equation}
\psi_{\mu-}=e^{-\frac{\alpha}{2\sqrt 3}|y|}\tilde\psi_{\mu-}(x),
\end{equation}
so that the $4D$ transformation, after gauge fixing $\varphi_+=0$, reads:
\begin{equation}
\delta_L\tilde e_{\mu}^a(x)=\frac{1}{2}e^{\frac{1}{6}\varphi_-}\bar\epsilon(x)\gamma^a\tilde\psi_{\mu-}(x).
\end{equation}

Finally, one has to rescale the different fermions in order to get both canonical kinetic terms and standard susy transformations. Since we want to keep standard gravitino transformation of the form $\delta_L\psi_{\mu}=\partial_{\mu}\epsilon+\dots$, the parameter $\epsilon$ and the gravitino $\psi_{\mu}$ have to be rescaled by the same powers of $e^{\varphi_-}$, and we thus define the normalised fermions $\boldsymbol{\tilde \epsilon}(x)$ and $\boldsymbol{\tilde \psi}_{\mu-}(x)$ by:
\begin{equation}
\boldsymbol{\tilde \epsilon}(x)=e^{\frac{1}{12}\varphi_-}\tilde\epsilon(x),\qquad \boldsymbol{\tilde \psi}_{\mu-}(x)=e^{\frac{1}{12}\varphi_-}\tilde\psi_{\mu-}(x).
\end{equation}
The correct normalisations for the chiral fermion $\tilde\lambda_+$ and the gaugino $\tilde\lambda_-$ can be obtained from their kinetic terms, which after dimensional reduction are found to be:
\begin{equation}
\mathcal{L}_{kin}(\tilde\lambda_{\pm})=-\frac{1}{4}e^{-\sqrt{3}\alpha|y|}\sqrt{-\tilde g}e^{-\frac{1}{6}\varphi_--\frac{1}{3}\varphi_+}\bar{\tilde \lambda}_{\pm}(x)\tilde e_{\mu}^a(x)\gamma^a\partial_{\mu}\tilde\lambda_{\pm}(x).
\end{equation}
Defining the normalised fermions $\boldsymbol{\tilde \lambda_+}(x)$ and $\boldsymbol{\tilde \lambda_-}(x)$ by,
\begin{equation}
\boldsymbol{\tilde \lambda_+}(x)=e^{\frac{11}{12}\varphi_-}\tilde\lambda_+(x),\qquad \boldsymbol{\tilde \lambda_-}(x)=e^{-\frac{7}{12}\varphi_-}\tilde\lambda_-(x),
\end{equation}
their kinetic terms have now the correct powers of $e^{\varphi_-}$ matching with the ones of their bosonic partners, namely (in the gauge $\varphi_+=0$):
\begin{eqnarray}
\mathcal{L}_{kin}(\boldsymbol{\tilde\lambda_+})&=&-\frac{1}{4}e^{-\sqrt{3}\alpha|y|}\sqrt{-\tilde g}e^{-2\varphi_-}\boldsymbol{\bar{\tilde \lambda}_+}(x)\tilde e_{\mu}^a(x)\gamma^a\partial_{\mu}\boldsymbol{\tilde\lambda_+}(x),\\
\mathcal{L}_{kin}(\boldsymbol{\tilde\lambda_-})&=&-\frac{1}{4}e^{-\sqrt{3}\alpha|y|}\sqrt{-\tilde g}e^{\varphi_-}\boldsymbol{\bar{\tilde \lambda}_-}(x)\tilde e_{\mu}^a(x)\gamma^a\partial_{\mu}\boldsymbol{\tilde\lambda_-}(x).
\end{eqnarray}
With these normalisations, the different factors of $e^{\varphi_-}$ disappear in the susy transformations, and we end up with the standard $\mathcal{N}=1$ supersymmetric transformations involving only four-dimensional $x$-dependent fields:
\begin{align}
\delta_L S &= -\frac{i}{\sqrt 3}\boldsymbol{\bar{\epsilon}}(1+\gamma_5)\boldsymbol{\tilde \lambda_+}, & 
\delta_L A_{\mu}^1 &= \frac{1}{2\sqrt{3}}\boldsymbol{\bar{\epsilon}}\gamma_{\mu}\boldsymbol{\tilde \lambda_-}, \\
\delta_L [(1+\gamma_5)\boldsymbol{\tilde\lambda_+}] &=\frac{i}{\sqrt 3}(1+\gamma_5) \slashed{\partial}S\boldsymbol{\epsilon}, &
\delta_L \boldsymbol{\tilde\lambda_-} &= -\frac{1}{2\sqrt{3}}\gamma^{\mu}\gamma^{\nu}F_{\mu\nu}^1\boldsymbol{\epsilon},
\end{align}
where all $x$-dependences have been now left implicit for compactness. These results are summarised in Table \ref{tab:n=1_field_content}.

\begin{center}
\begin{tabular}{|cc||cc||cc|}
   \hline
   \multicolumn{2}{|c||}{$\mathcal{N}=1$ supergravity multiplet} & \multicolumn{2}{|c||}{$\mathcal{N}=1$ vector multiplet} & \multicolumn{2}{|c|}{$\mathcal{N}=1$ chiral multiplet} \\
   \hline
   \qquad $e_{\mu}^a$ & \qquad $\boldsymbol{\tilde \psi}_{\mu -}$ & \qquad $A_{\mu}^1$ & \qquad $\boldsymbol{\tilde \lambda_-}$ &   $S=e^{\varphi_-}+i\tilde A_5^0$ & $(1+\gamma_5)\boldsymbol{\tilde \lambda_+}$ \qquad \\
   \hline
   \qquad $-\frac{1}{3}$ & \qquad $-\frac{1}{6}$ & \qquad $0$ & \qquad $\frac{1}{6}$ & $(0~;~-1)$ & $\frac{1}{6}$ \\
   \hline
\end{tabular}
\captionof{table}{$\mathcal{N}=1$, $D=4$ field content}
\label{tab:n=1_field_content}
\end{center}
In the second row we have listed the constants $n$ which appear in the exponent of the internal wave functions of 
the zero modes of the five-dimensional fields $\Phi(x,y)$ 
through $\Phi^{(0)}(x,y)=e^{n\sqrt 3\alpha|y|}\tilde \Phi(x)$.

\section{Conclusion}
\label{sect:conclusion}
The work carried out in this paper analysed different aspects of the linear dilaton background arising from a runaway scalar potential in five dimensions, in relation to two different perspectives: compactification and supersymmetry breaking. 

On the one hand, we performed the KK compactification down to four dimensions of the dilaton-gravity action of a non-critical string, emphasising the new features emerging from the non-trivial background compared to the standard (toroidal) flat case. Besides the known mass gap between the $4D$ zero mode of the $5D$ fields and the lowest state of their KK tower, it has been shown how the LD background may induce an exponential profile of the wavefunction of the zero modes of some fields. In the case of spin-1, this behaviour brings
a vector mass term which breaks the residual gauge symmetry, inherited from the original higher dimensional one, at a scale proportional to the slope of the LD background. This mechanism has been explicitly described in the case of the KK vector $G_{\mu 5}$ arising from the $5D$ metric $G_{MN}$, through a gauge symmetry analysis which generalises the well-known results of the toroidal case.

On the other hand in a supersymmetric theory, the LD background behaves as a ${1}/{2}$-BPS solution and can thus be used to study supersymmetry breaking. This aspect has been investigated in the framework of the $\mathcal{N}=2$, $D=5$ gauged supergravity coupled to one vector multiplet, built as a holographic dual of Little String Theory.

Both aspects are then combined together by carrying out the compactification down to four dimensions of the full bosonic sector of the above $5D$ supergravity theory. We have shown how the lowest massive spectrum associated to the mass gap, induced by an exponential profile of the vector wavefunctions, can be decoupled from the massless sector in the low energy limit, so that the remaining degrees of freedom arrange into $\mathcal{N}=1$ multiplets under the supersymmetry preserved by the background. The corresponding supersymmetric truncation is consistent with a standard orbifold projection.

The possibility to describe a consistent $\mathcal{N}=2$, $D=4$ supergravity theory by including the above lowest massive modes is an interesting open problem. A starting point would be to examine if the LD background induces a gauging in the 4-dimensional $\mathcal{N}=2$ theory that can lead to a (spontaneous) partial supersymmetry breaking $\mathcal{N}=2\rightarrow\mathcal{N}=1$. This could bring new insights regarding the partial supersymmetry breaking in supergravity, which is highly restrictive and requires so far the use of hypermultiplets~\cite{Ferrara:1983gn, Ferrara:1995gu, Ferrara:1995xi} (see also~\cite{FVP}).

\appendix
\section{Conventions and notations}
\label{sect:append_conventions}
We adopt the metric convention $(-,+,+,+,+)$, and write the five-dimensional Minkowski space coordinates as $x^M=(x^{\mu},y)$. The indices are defined according to the following pattern:

\begin{align*}\label{eq:indices}
M,N...=0,...,3,5	&&& \text{$5D$ spacetime curved indices} \\
m,n...=0,...,3,5	&&& \text{$5D$ spacetime flat indices} \\
\mu,\nu...=0,...,3	&&& \text{$4D$ spacetime curved indices} \\
a,b...=0,...,3		&&& \text{$4D$ spacetime flat indices} \\
I,J...=0,...,n_V	&&& \text{vector field labels} \\
x,y...=1,...,n_V	&&& \text{scalar manifold curved indices} \\
\tilde{a},\tilde{b}...=1,...,n_V	&&& \text{scalar manifold flat indices} \\
i,j...=1,2		&&& \text{fundamental representation of $SU(2)_R$}
\end{align*}
Curved and flat indices in $5D$ ($4D$) are related through the f\"{u}nfbein $e_M^m$ (vierbein $e_{\mu}^a$) according to 
\begin{equation}
X_M=e_M^mX_m~~~(X_{\mu}=e_{\mu}^aX_a).
\end{equation}
Similarly, curved and flat indices of the scalar manifold are related through the $n_V$-bein $f^{\tilde a}_x$ according to
\begin{equation}
\varphi^x=(f^{\tilde a}_x)^{-1}\varphi^{\tilde a}=f_{\tilde a}^x\varphi^{\tilde a}.
\end{equation}
The five-dimensional Dirac matrices $\gamma^m$ satisfy the Clifford algebra
\begin{equation}
\left\{\gamma^m,\gamma^n\right\}=2\eta^{mn}.
\end{equation}
In particular, 
\begin{equation}
\left(\gamma^5\right)^2=\left(\gamma_5\right)^2=1.
\end{equation}
Antisymmetrized products of $\gamma$ matrices are defined with weight one, $\gamma^{m_1...m_r}=\gamma^{[m_1}...\gamma^{m_r]}$. In particular,
\begin{equation}
\gamma^{mn}\equiv\frac{1}{2}(\gamma^m\gamma^n-\gamma^n\gamma^m).
\end{equation} 

Spinors in $D=5$ dimensions can be equivalently described either in terms of unconstrained Dirac spinors, which have 4 complex components, or either in terms of pairs of spinors $\chi^i$, $i=1,...,\mathcal{N}=2k$ satisfying a reality condition. The pairs are called symplectic since the position of the indices is raised and lowered according to 
\begin{equation}
\chi^i=\Omega^{ij}\chi_j,\qquad \chi_i=\chi^j\Omega_{ji},
\end{equation}
where $\Omega_{ij}$ is a $2k\times 2k$ matrix satisfying
\begin{equation}
\Omega^{ik}\Omega_{kj}=-\delta^i_j.
\end{equation}
The Dirac and Majorana conjugates of a spinor $\chi^i$ are respectively defined by
\begin{equation}
\bar{\chi}^i\equiv(\chi_i)^{\dagger}\gamma^0,\qquad (\chi^i)^C\equiv(\chi^i)^{T}C,
\end{equation}
with $C$ the charge conjugation matrix, satisfying in five dimensions $\gamma_{M}^T=C\gamma_MC^{-1}$. The symplectic-Majorana condition then imposes the Dirac conjugate of $\chi^i$ to be equal to its Majorana conjugate, namely:
\begin{equation}\label{eq:symplectic_Majorana_cond}
\chi_i^{\dagger}\gamma^0=(\chi^i)^TC.
\end{equation}
Since \eqref{eq:symplectic_Majorana_cond} relates the components of $\chi^i$ to those of its complex conjugate, this is a reality condition which projects out half of the degrees of freedom of the $k$ pairs of spinors, ending with $2\times 4k$ real components. Both descriptions in terms of one Dirac spinor or one pair of symplectic-Majorana spinors are thus equivalent, both of them describing $8$ real off-shell degrees of freedom. In practice however, only the symplectic formulation is used, since it makes explicit the action of the R-symmetry group $USp(\mathcal{N})$ in $D=5$ dimensions.

For the case $\mathcal{N}=2$ we are considering in this article, $\Omega_{ij}=\epsilon_{ij}$. Since $\epsilon_{ij}$ is an $SU(2)$-invariant tensor, the indices $i,j=1,2$ are referred to the $SU(2)_R$ indices, therefore raised and lowered according to the NorthWest-SouthEast convention
\begin{equation}\label{eq:NW_SE_convention}
\chi^i=\epsilon^{ij}\chi_j,\qquad \chi_i=\chi^j\epsilon_{ji},
\end{equation}
and where we choose $\epsilon_{12}=\epsilon^{12}=1=-\epsilon_{21}=-\epsilon^{21}$. In particular, bilinears of fermions satisfy
\begin{equation}
\bar{\lambda}^i\chi_i=\bar{\chi}_i\lambda^i=-\bar{\chi}^i\lambda_i,
\end{equation}
where the first equality is a standard Majorana flip in $D=5$ dimension and the second one follows from \eqref{eq:NW_SE_convention}. Let us finally highlight that in five dimensions, the bilinear quantities $\bar\lambda^i\chi_i$ and $\bar\lambda^i\gamma_{\mu}\chi_i$ built from symplectic-Majorana spinors are pure imaginary and real respectively.

\section{Gravitational action on a bounded manifold}
\label{sect:append_ADM}
\subsection{$d+1$ decomposition and Gibbons-Hawking boundary term: a review}
\label{sect:append_ADM_1}
Let us consider a $d+1$-dimensional space-time manifold $\mathcal{M}$, equipped with a metric $\hat G_{MN}$ and foliated with a set of co-dimension $1$ hypersurfaces $\Sigma_y$. Such hypersurfaces can be defined by an arbitrary scalar field $y(x^M)$ such that $y$ is constant on each of these hypersurfaces. The unit normal vector $n_M$ to $\Sigma_y$ is then proportional to $\partial_My$, and normalized such that $n^Mn_M=\hat G^{MN}n_Mn_N=\epsilon$, with $\epsilon=+1$ if $\Sigma_y$ is timelike, and $\epsilon=-1$ if $\Sigma_y$ is spacelike. 

In order to relate the coordinate systems on each hypersurfaces, we consider a congruence of curves which intersect each $\Sigma_y$ once and only once. The coordinates $z^{\mu}$ on each hypersurfaces are now chosen such that points on the same curves have the same coordinates $z^{\mu}$ on the different slices $\Sigma_y$. Therefore, considering two infinitesimally closed hypersurfaces $\Sigma_y$ and $\Sigma_{y+dy}$, the vector $y^M$ tangent to the curve points from a point with coordinates $z^{\mu}$ on $\Sigma_y$ to a point with the same coordinates on $\Sigma_{y+dy}$. This construction therefore allows us to move from the original coordinate system $x^M$ to a new one $(y,z^{\mu})$. The vectors $y^M$ tangent to the curves and $E^{M}_{\mu}$ tangent to $\Sigma_y$ are respectively given by:
\begin{equation}
y^M=\left.\frac{\partial x^M}{\partial y}\right|_{z^{\mu}},\qquad E^{M}_{\mu}=\left.\frac{\partial x^M}{\partial z^{\mu}}\right|_{y}.
\end{equation}
The $E^{M}_{\mu}$ can be seen as a map from $\otimes^q{T_P}^{\star}(\mathcal{M})$, an arbitrary tensor product of the cotangent spaces of $\mathcal{M}$, to $\otimes^q{T_P}^{\star}(\Sigma_y)$, an arbitrary tensor product of the cotangent spaces of $\Sigma_y$, projecting any $X_{M_1...M_q}\in \otimes^q{T_P}^{\star}(\mathcal{M})$ down to a $X_{\mu_1...\mu_q}\in \otimes^q{T_P}^{\star}(\Sigma_y)$ through
\begin{equation}
E^{M}_{\mu}: X_{M_1...M_q}\longmapsto X_{\mu_1...\mu_q}=X_{M_1...M_q}E^{M_1}_{\mu_1}...E^{M_q}_{\mu_q}.
\end{equation}
With this projection map, one can define the induced metric $\hat g_{\mu\nu}$ and the extrinsic curvature $\hat K_{\mu\nu}$ respectively by
\begin{eqnarray}
\hat g_{\mu\nu}&=&\hat G_{MN}E^M_{\mu}E^N_{\nu},\\
\hat K_{\mu\nu}&=&E^{M}_{\mu}E^{N}_{\nu}\hat\nabla_Mn_N,
\end{eqnarray}
with $\hat\nabla_M$ the covariant derivative compatible with the metric $\hat G_{MN}$. While $\hat g_{\mu\nu}$ characterises the local intrinsic geometry of $\Sigma_y$, $\hat K_{\mu\nu}$ describes how the hypersurface is embedded in the ambient space $\mathcal{M}$, and the data of both is sufficient to completely characterise the geometry of $\Sigma_y$ embedded into $\mathcal{M}$.

The vector $y^M$ tangent to the curves can be decomposed in the basis provided by normal $n^M$ and tangent vectors $E^M_{\mu}$ according to
\begin{equation}
y^M=\hat Nn^M+\beta^{\mu}E^M_{\mu}.
\end{equation}
The components $\hat N$ and $\beta^{\mu}$ are respectively called the lapse function and the shift vector. Their physical interpretation follows from the construction described above: the lapse describes the orthogonal distance between the two slices, while the shift describes how the coordinate systems of the two infinitesimally closed hypersurfaces are transversally shifted one with respect to the other. Using
\begin{equation}
dx^M=\frac{\partial x^M}{\partial y}dy+\frac{\partial x^M}{\partial z^{\mu}}dz^{\mu}=(\hat Nn^M+\beta^{\mu}E^M_{\mu})dy+E^M_{\mu}dz^{\mu},
\end{equation}
it is easy to find the following decomposition of the line element in terms of the lapse $\hat N$, the shift $\beta^{\mu}$ and the induced metric $\hat g_{\mu\nu}$:
\begin{equation}
ds^2=\epsilon \hat N^2dy^2+\hat g_{\mu\nu}(\beta^{\mu}dy+dz^{\mu})(\beta^{\nu}dy+dz^{\nu}).
\end{equation}
This is the famous ADM decomposition, first introduced in \cite{ADM}, which corresponds to the ADM metric tensor
\begin{equation}\label{eq:ADM_metric}
\hat G_{MN}=
\begin{pmatrix}
\hat g_{\mu\nu} & \hat g_{\mu\rho}\beta^{\rho} \\
\hat g_{\nu\rho}\beta^{\rho} & \epsilon \hat N^2+\hat g_{\rho\sigma}\beta^{\rho}\beta^{\sigma} 
\end{pmatrix},
\end{equation}
whose inverse reads
\begin{equation}\label{eq:ADM_inverse_metric}
\hat G^{MN}=
\begin{pmatrix}
\hat g^{\mu\nu}+\epsilon \hat N^{-2}\beta^{\mu}\beta^{\nu} & -\epsilon \hat N^{-2}\beta^{\mu} \\
-\epsilon \hat N^{-2}\beta^{\nu} & \epsilon \hat N^{-2} 
\end{pmatrix}.
\end{equation}

We now suppose that $\mathcal{M}$ is bounded in the $y$ direction by two hypersurfaces $\Sigma_{y_1}$ and $\Sigma_{y_2}$ located at $y_1$ and $y_2$. The total gravitational action $S_G$ is now the sum of the Einstein-Hilbert bulk term $S_{EH}$ and the Gibbons-Hawking boundary term $S_{GH}$:
\begin{equation}
S_G=S_{EH}+S_{GH},
\end{equation}
with
\begin{eqnarray}
\label{eq:EH_action_M}
S_{EH}&=&\frac{1}{2}\int_{\mathcal{M}} d^{d+1}x\sqrt{-\hat G}R^{(d+1)}[\hat G_{MN}],\\
\label{eq:GH_action_dM}
S_{GH}&=&-\epsilon\int_{\Sigma_{y_1}} d^dx\sqrt{|\hat g|}\hat K+\epsilon\int_{\Sigma_{y_2}} d^dx\sqrt{|\hat g|}\hat K.
\end{eqnarray}
Here $\hat K$ is the trace of the extrinsic curvature, $\hat K\equiv \hat g^{\mu\nu}\hat K_{\mu\nu}$, and the relative sign between the two terms in the GH action comes from the fact that both normals to $\Sigma_{y_1}$ and $\Sigma_{y_2}$ point along the increasing $y$, and are thus directed respectively inward and outward of $\mathcal{M}$.\\    
Starting from the Gauss equation
\begin{equation}
R_{MNPQ}E^M_{\mu}E^N_{\nu}E^P_{\rho}E^Q_{\sigma}=R_{\mu\nu\rho\sigma}+\epsilon(\hat K_{\mu\sigma}\hat K_{\nu\rho}-\hat K_{\mu\rho}\hat K_{\nu\sigma}),
\end{equation}
which relates the tangential components of the curvature tensor of $\mathcal{M}$ on the LHS with the intrinsic and extrinsic curvature tensors of $\Sigma$ on the RHS, one can find the relation between the $(d+1)$-dimensional Ricci scalar $R^{(d+1)}$ evaluated on $\Sigma_y$ and the intrinsic curvature scalar $R^{(d)}$ of $\Sigma_y$:
\begin{equation}
R^{(d+1)}[\hat G_{MN}]=R^{(d)}[\hat g_{\mu\nu}]+\epsilon(\hat K^2-\hat K^{\mu\nu}\hat K_{\mu\nu})+2\epsilon\hat \nabla_M\left(n^N\hat \nabla_Nn^M-n^M\hat \nabla_Nn^N\right).
\end{equation}
The first two terms $R^{(d)}[\hat g_{\mu\nu}]$ and $(\hat K^2-\hat K^{\mu\nu}\hat K_{\mu\nu})$ depend respectively on the intrinsic and extrinsic geometry of $\Sigma_y$. The third term contains second normal derivatives, which will cancel with the GH boundary term as we are now going to show. In the foliated spacetime $\mathcal{M}=\Sigma_y\times S^1/\mathbb{Z}_2$, the EH action \eqref{eq:EH_action_M} reads:
\begin{multline}
\frac{1}{2}\int_{\mathcal{M}}d^{d+1}x\sqrt{-\hat G}R^{(d+1)}[\hat G_{MN}]=\frac{1}{2}\int_{y_1}^{y_2}dy\int_{\Sigma_y}d^dx\hat N\sqrt{|\hat g|}\left[R^{(d)}[\hat g_{\mu\nu}]+\epsilon(\hat K^2-\hat K^{\mu\nu}\hat K_{\mu\nu})\right]\\
+\epsilon\int_{\mathcal{M}}d^{d+1}x\sqrt{-\hat G}\hat \nabla_M\left[n^N\hat \nabla_Nn^M-n^M\hat \nabla_Nn^N\right].
\end{multline}
Using Gauss's theorem, the second term can be written as:
\begin{eqnarray}
I&\equiv&\epsilon\int_{\mathcal{M}}d^{d+1}x\sqrt{-\hat G}\hat \nabla_M\left[n^N\hat \nabla_Nn^M-n^M\hat \nabla_Nn^N\right],\nonumber\\
&=&\epsilon\int_{\partial\mathcal{M}}\left[n^N\hat \nabla_Nn^M-n^M\hat \nabla_Nn^N\right]d\Sigma_M.
\end{eqnarray}
$I$ is the sum of two contributions at each boundaries $\Sigma_{y_1}$ and $\Sigma_{y_2}$ of $\mathcal{M}$. On $\Sigma_{y_1}$ ($\Sigma_{y_2}$), the surface element is given by $d\Sigma_M=-\epsilon n_M\sqrt{|\hat g|}d^dx$ ($d\Sigma_M=+\epsilon n_M\sqrt{|\hat g|}d^dx$), where the minus (plus) sign takes into account the inward (outward) direction of the normal of $\Sigma_{y_1}$ ($\Sigma_{y_2}$). Using $n_Mn^M=\epsilon$ and $n_M\hat \nabla_Nn^M=0$, we find:
\begin{eqnarray}
I=\epsilon\int_{\Sigma_{y_1}}d^dx\sqrt{|\hat g|}\hat K-\epsilon\int_{\Sigma_{y_2}}d^dx\sqrt{|\hat g|}\hat K,
\end{eqnarray}
which is exactly cancelled by the GH boundary term \eqref{eq:GH_action_dM}. Hence, the total gravitational action $S_G=S_{EH}+S_{GH}$ reads:
\begin{equation}\label{eq:d+1_grav_action}
S_G=\frac{1}{2}\int_{y_1}^{y_2}dy\int_{\Sigma_y}d^dx\hat N\sqrt{|\hat g|}\left[R^{(d)}[\hat g_{\mu\nu}]+\epsilon(\hat K^2-\hat K^{\mu\nu}\hat K_{\mu\nu})\right].
\end{equation}
The $d$-dimensional Ricci scalar is given by:
\begin{eqnarray}\label{eq:d_dim_ricci_scalar}
R^{(d)}[\hat g_{\mu\nu}]&=&\frac{1}{2}\partial_{\mu}\left(\sqrt{|\hat g|}\hat g^{\mu\nu}\hat N\right)\partial_{\nu}\ln \hat g+\partial_{\mu}\hat N\partial_{\nu}\left(\sqrt{|\hat g|}\hat g^{\mu\nu}\right)-\frac{1}{2}\hat N\sqrt{|\hat g|} \hat \Gamma_{\mu\nu}^{\rho}\partial_{\rho}\hat g^{\mu\nu}\nonumber\\
&&\qquad-\partial_{\mu}\left(\hat N\hat g^{\mu\nu}\partial_{\nu}\sqrt{|\hat g|}+\hat N\partial_{\nu}(\sqrt{|\hat g|}\hat g^{\mu\nu})\right),
\end{eqnarray}
with $\hat \Gamma_{\mu\nu}^{\rho}$ the $d$-dimensional Christoffel symbols computed from the metric $\hat g_{\mu\nu}$. The last term in the above expression is a total derivative in the unbounded $x^{\mu}$ directions, and can thus be discarded. In order to express the extrinsic curvature tensor $\hat K_{\mu\nu}=E_{\mu}^ME_{\nu}^N\hat \nabla_Mn_N$ in terms of the induced metric $\hat g_{\mu\nu}$ and the ADM variables $\hat N$ and $\beta_{\mu}$, we start by writing the derivative of $\hat g_{\mu\nu}$ with respect to $y$ as:
\begin{equation}
\hat g_{\mu\nu}^{'}\equiv\mathcal{L}_y\hat g_{\mu\nu}=\mathcal{L}_y(\hat G_{MN}E^{M}_{\mu}E^{N}_{\nu})=(\mathcal{L}_y\hat G_{MN})E^{M}_{\mu}E^{N}_{\nu},
\end{equation}
where, in the last equality, we have used
\begin{equation}
\mathcal{L}_yE^M_{\mu}=[y,E_{\mu}]^M=[\partial_y,\partial_{x^{\mu}}]^M=0.
\end{equation}
The Lie derivative of the metric $\hat G_{MN}$ is given by:
\begin{eqnarray}
\mathcal{L}_y\hat G_{MN}&=&\hat \nabla_My_N+\hat \nabla_Ny_M\nonumber\\
&=&\hat \nabla_M(\hat Nn_N+\beta_N)+\hat \nabla_N(\hat Nn_M+\beta_M)\nonumber\\
&=&n_N\partial_M\hat N+n_M\partial_N\hat N+\hat N(\hat \nabla_Mn_N+\hat \nabla_Nn_M)+\hat \nabla_M\beta_N+\hat \nabla_N\beta_M.
\end{eqnarray}
Contracting with $E^M_{\mu}E^N_{\nu}$, the two first terms vanish since $E^M_{\mu}$ and $n_M$ are orthognal. The third term yields the extrinsic curvature, while the last ones give the intrinsic covariant derivative of the shift vector. We deduce the extrinsic curvature tensor $\hat K_{\mu\nu}$ expressed in terms of the induced metric $\hat g_{\mu\nu}$, the lapse $\hat N$ and shift $\beta^{\rho}$:
\begin{equation}\label{eq:extrinsic_curvature}
\hat K_{\mu\nu}=\frac{1}{2\hat N}\left(\hat g_{\mu\nu}^{'}-^{(d)}\hat \nabla_{\mu}(\hat g_{\nu\rho}\beta^{\rho})-^{(d)}\hat \nabla_{\nu}(\hat g_{\mu\rho}\beta^{\rho})\right),
\end{equation}
an expression which is directly used in the computation of the gravitational action in Section \ref{sect:stuckelberg_mech}, whose technical details are developed in the following subsection. 

\subsection{Computation in the framework of the LD background}
\label{sect:append_ADM_2}

We now want to compute the gravitational action in the framework of the Section \ref{sect:stuckelberg_mech}, on a $5D$ manifold with two boundaries along the fifth direction, with the metric \eqref{eq:final_metric} and its inverse \eqref{eq:final_inverse_metric} which we reproduce here\footnote{In this subsection we denote the KK vector $B_{\mu}$ instead of $K_{\mu}$ to avoid confusion with the extrinsic curvature, also called $K$.}:
\begin{eqnarray}\label{eq:final_metric_app}
G_{MN}&=&e^{-\frac{2}{\sqrt{3}}\alpha|y|}e^{2r(x,y)}
\begin{pmatrix}
e^{-3r(x,y)}\tilde g_{\mu\nu}(x)+B_{\mu}B_{\nu}(x,y) & B_{\mu}(x,y) \\
B_{\nu}(x,y) & 1 
\end{pmatrix},\\
\label{eq:final_inverse_metric_app}
G^{MN}&=&e^{\frac{2}{\sqrt{3}}\alpha|y|}e^{-2r}
\begin{pmatrix}
e^{3r}\tilde g^{\mu\nu}(x,y) & -g^{\mu\rho}B_{\rho}(x,y) \\
-g^{\nu\rho}B_{\rho}(x,y) & 1+g^{\lambda\sigma}B_{\lambda}B_{\sigma}(x,y) 
\end{pmatrix}.
\end{eqnarray}
Rescaling the ADM $4D$ induced metric $\hat g_{\mu\nu}$ according to $\hat g_{\mu\nu}\rightarrow \hat N^2\hat g_{\mu\nu}$, the ADM metric $\hat G_{MN}$ $\eqref{eq:ADM_metric}$ and its inverse $\hat G^{MN}$ \eqref{eq:ADM_inverse_metric}, in the case of timelike hypersurfaces ($\epsilon=1$) we are interested in, read:
\begin{equation}\label{eq:rescaled_ADM_metric}
\hat G_{MN}=\hat N^2
\begin{pmatrix}
\hat g_{\mu\nu} & \hat g_{\mu\rho}\beta^{\rho} \\
\hat g_{\nu\rho}\beta^{\rho} & 1+\hat g_{\rho\sigma}\beta^{\rho}\beta^{\sigma} 
\end{pmatrix},
\qquad
\hat G^{MN}=\hat N^{-2}
\begin{pmatrix}
\hat g^{\mu\nu}+\beta^{\mu}\beta^{\nu} & -\beta^{\mu} \\
-\beta^{\nu} & 1 
\end{pmatrix}.
\end{equation}
They turn out to be very similar to the parametrisations \eqref{eq:final_metric_app} and \eqref{eq:final_inverse_metric_app}, and are in fact equivalent, noticing the identification:
\begin{align}\label{eq:identification}
	\hat N^2(1+\hat g_{\rho\sigma}\beta^{\rho}\beta^{\sigma}) &= N^{-2}, &
	\hat N^{-2} &= N^2(1+g^{\rho\sigma}B_{\rho}B_{\sigma}),\nonumber\\
	\hat N^2\hat g_{\mu\rho}\beta^{\rho} &= N^{-2}B_{\mu}, &
	\hat N^{-2}\beta^{\mu} &= N^2g^{\mu\rho}B_{\rho},\\
	\hat N^2\hat g_{\mu\nu} &= N^{-2}(g_{\mu\nu}+B_{\mu}B_{\nu}), &
	\hat N^{-2}(\hat g^{\mu\nu}+\beta^{\mu}\beta^{\nu}) &= N^2g^{\mu\nu},\nonumber
\end{align}
with the overall factor $N$ given by
\begin{equation}
N=e^{\frac{\alpha}{\sqrt{3}}|y|-r}.
\end{equation}
One easily finds for the inverse metric of $\hat N^2\hat g_{\mu\nu}$, $\hat N^{-2}\hat g^{\mu\nu}=N^2(g^{\mu\nu}-xB^{\mu}B^{\nu})$, $x\equiv\frac{1}{1+B^2}$. 

The main result \eqref{eq:d+1_grav_action} of the previous subsection giving the total gravitational action $S_G=S_{EH}+S_{GH}$, in the case of timelike ($\epsilon=1$) hypersurfaces of dimension $d=4$, and for the rescaled $4D$ induced metric $\hat N^2\hat g_{\mu\nu}$ reads:
\begin{equation}
S_G=\frac{1}{2}\int_{y_1}^{y_2}dy\int_{\Sigma_y}d^4x\hat N\sqrt{|\hat N^8\hat g|}\left[R^{(4)}[\hat N^2\hat g_{\mu\nu}]+(\hat K^2-\hat K^{\mu\nu}\hat K_{\mu\nu})\right],
\end{equation}
where $\hat g\equiv \det \hat g_{\mu\nu}$, $\hat K_{\mu\nu}$ is the extrinsic curvature associated to the metric $\hat N^2\hat g_{\mu\nu}$, $\hat K^{\mu\nu}=\hat N^{-4}\hat g^{\mu\rho}\hat g^{\nu\sigma}\hat K_{\rho\sigma}$, and $\hat K=\hat N^{-2}\hat g^{\mu\nu}\hat K_{\mu\nu}$. According to the result \eqref{eq:extrinsic_curvature} derived earlier, the extrinsic curvature tensor is given by
\begin{equation}
\hat K_{\mu\nu}=\frac{1}{2\hat N}\left((\hat N^{2}\hat g_{\mu\nu})^{'}-^{(4)}\hat \nabla_{\mu}(\hat N^{2}\hat g_{\nu\rho}\beta^{\rho})-^{(4)}\hat \nabla_{\nu}(\hat N^{2}\hat g_{\mu\rho}\beta^{\rho})\right),
\end{equation}
where $^{(4)}\hat \nabla_{\mu}$ denotes the covariant derivative compatible with the metric $\hat N^2\hat g_{\mu\nu}$. Using the identification \eqref{eq:identification}, one can express the extrinsic curvature tensor in terms of the variables $N$, $g_{\mu\nu}$ and $B_{\rho}$. The computation yields:
\begin{eqnarray}
K_{\mu\nu}&=&\frac{1}{2}N\sqrt{1+B^2}\left\{\left[N^{-2}(g_{\mu\nu}+B_{\mu}B_{\nu})\right]^{'}-N^{-2}(\partial_{\mu}B_{\nu}+\partial_{\nu}B_{\mu})+2xN^{-2}\Gamma_{\mu\nu}^{\rho}B_{\rho}\right.\nonumber\\
&&\qquad\left.+2xN^{-2}B^{\rho}\left[B_{(\nu}F_{\mu)\sigma}+B_{\sigma}\partial_{(\mu}B_{\nu)}\right]+2xN^{-3}(g_{\mu\nu}+B_{\mu}B_{\nu})B^{\rho}\partial_{\rho}N\right\},
\end{eqnarray}
where $\Gamma_{\mu\nu}^{\rho}$ are the Christoffel symbols computed from the metric $g_{\mu\nu}$, and $F_{\mu\nu}$ is the field strength of $B_{\mu}$, $F_{\mu\nu}=\partial_{\mu}B_{\nu}-\partial_{\nu}B_{\mu}$. Since we are interested in the spectrum of the theory and especially in the mass term for the KK vector $B_{\mu}$, we work up to quadratic order in $B$. Ignoring interaction terms, we get for the extrinsic curvature 
\begin{eqnarray}
K_{\mu\nu}&=&\frac{1}{2}N\sqrt{1+B^2}\left\{\left[N^{-2}(g_{\mu\nu}+B_{\mu}B_{\nu})\right]^{'}-N^{-2}(\partial_{\mu}B_{\nu}+\partial_{\nu}B_{\mu})\right.\nonumber\\
&&\qquad\left.+2N^{-2}\Gamma_{\mu\nu}^{\rho}B_{\rho}+2N^{-3}g_{\mu\nu}B^{\rho}\partial_{\rho}N\right\}+\text{h.o.t.},
\end{eqnarray}
where here and in the following, $\text{h.o.t.}$ will denote higher order terms in $B$, but also terms of the form $B^2\times r$. From $K=N^2(g^{\mu\nu}-xB^{\mu}B^{\nu})K_{\mu\nu}=N^2(g^{\mu\nu}-B^{\mu}B^{\nu})K_{\mu\nu}+\text{h.o.t.}$, we get
\begin{eqnarray}
K^{2}&=&N^2(1+B^2)\left\{\frac{16}{3}\alpha^2+\frac{16\alpha}{\sqrt{3}}\sgn yr^{'}+4(r^{'})^2\right.\nonumber\\
&&\quad+\left(\frac{8\alpha}{\sqrt{3}}\sgn y+4r^{'}\right)\left(g^{\mu\nu}\partial_{\mu}B_{\nu}-B_{\rho}\Gamma^{\rho}_{\mu\nu}g^{\mu\nu}+4B^{\rho}\partial_{\rho}r\right)\nonumber\\
&&\qquad\left.+g^{\mu\nu}g^{\rho\sigma}\partial_{\mu}B_{\nu}\partial_{\rho}B_{\sigma}-\frac{4\alpha}{\sqrt{3}}\sgn y(B^2)^{'}\right\}+\text{h.o.t.}
\end{eqnarray}
Similarly, using $K^{\mu\nu}=N^4(g^{\mu\rho}-xB^{\mu}B^{\rho})(g^{\nu\sigma}-xB^{\nu}B^{\sigma})K_{\rho\sigma}=N^4(g^{\mu\rho}g^{\nu\sigma}-g^{\mu\rho}B^{\nu}B^{\sigma}-g^{\nu\sigma}B^{\mu}B^{\rho})K_{\rho\sigma}+\text{h.o.t.}$, one finds
\begin{eqnarray}
K^{\mu\nu}K_{\mu\nu}&=&\frac{1}{4}N^2(1+B^2)\left\{\frac{16}{3}\alpha^2+\frac{16\alpha}{\sqrt{3}}\sgn yr^{'}+4(r^{'})^2\right.\nonumber\\
&&\quad+\left(\frac{8\alpha}{\sqrt{3}}\sgn y+4r^{'}\right)\left(g^{\mu\nu}\partial_{\mu}B_{\nu}-B_{\rho}\Gamma^{\rho}_{\mu\nu}g^{\mu\nu}+4B^{\rho}\partial_{\rho}r\right)\nonumber\\
&&\qquad\left.+4g^{\mu\rho}g^{\nu\sigma}\partial_{\mu}B_{\nu}\partial_{\rho}B_{\sigma}-\frac{4\alpha}{\sqrt{3}}\sgn y(B^2)^{'}\right\}+\text{h.o.t.},
\end{eqnarray}
and therefore
\begin{eqnarray}\label{eq:intermediate_result_1}
K^2-K^{\mu\nu}K_{\mu\nu}&=&\frac{3}{4}N^2(1+B^2)\left\{\frac{16}{3}\alpha^2+\frac{16\alpha}{\sqrt{3}}\sgn yr^{'}+4(r^{'})^2\right.\nonumber\\
&&\quad+\left(\frac{8\alpha}{\sqrt{3}}\sgn y+4r^{'}\right)\left(g^{\mu\nu}\partial_{\mu}B_{\nu}-B_{\rho}\Gamma^{\rho}_{\mu\nu}g^{\mu\nu}+4B^{\rho}\partial_{\rho}r\right)\\
&&\qquad\left.+\frac{4}{3}(g^{\mu\nu}g^{\rho\sigma}-g^{\mu\rho}g^{\nu\sigma})\partial_{\mu}B_{\nu}\partial_{\rho}B_{\sigma}-\frac{4\alpha}{\sqrt{3}}\sgn y(B^2)^{'}\right\}+\text{h.o.t.}\nonumber
\end{eqnarray}
Regarding the determinant of the metric $\det(\hat N^2\hat g_{\mu\nu})=\det(N^{-2}(g_{\mu\nu}+B_{\mu}B_{\nu}))$, we use the relation $\det(g_{\mu\nu}+B_{\mu}B_{\nu})=(\det g_{\mu\rho})\det(\delta^{\rho}_{\nu}+B^{\rho}B_{\nu})=(\det g_{\mu\rho})(1+B^{\sigma}B_{\sigma})+\text{h.o.t.}$ Using $g_{\mu\rho}=e^{-3r}\tilde g_{\mu\rho}$, we deduce:
\begin{equation}
\sqrt{-\det(\hat N^2\hat g_{\mu\nu})}=N^{-4}e^{-6r}\sqrt{-\tilde g}\left(1+\frac{B^2}{2}\right)+\text{h.o.t.}
\end{equation}
From the identification \eqref{eq:identification}, we have $\hat N=N^{-1}(1+B^2)^{-1/2}=N^{-1}(1-\frac{B^2}{2})+\text{h.o.t.}$, so that 
\begin{equation}\label{eq:intermediate_result_2}
\hat N\sqrt{-\det(\hat N^2\hat g_{\mu\nu})}=N^{-5}e^{-6r}\sqrt{-\tilde g}+\text{h.o.t.}
\end{equation}
Combining the results \eqref{eq:intermediate_result_1} and \eqref{eq:intermediate_result_2}, we get:
\begin{eqnarray}
\frac{1}{2}\hat N\sqrt{|\hat N^8\hat g|}\left(K^2-K_{\mu\nu}K^{\mu\nu}\right)&=&e^{-\sqrt{3}\alpha|y|}e^{-3r}\sqrt{-\tilde g}\left\{2\alpha^2+2\sqrt{3}\alpha\sgn yr^{'}+\frac{3}{2}(r^{'})^2\right.\nonumber\\
&&\quad+\left(\sqrt{3}\alpha\sgn y+\frac{3}{2}r^{'}\right)\left(g^{\mu\nu}\partial_{\mu}B_{\nu}-B_{\rho}\Gamma^{\rho}_{\mu\nu}g^{\mu\nu}+4B^{\rho}\partial_{\rho}r\right)\nonumber\\
&&\qquad\left.+\frac{1}{2}(g^{\mu\nu}g^{\rho\sigma}-g^{\mu\rho}g^{\nu\sigma})\partial_{\mu}B_{\nu}\partial_{\rho}B_{\sigma}\right.\nonumber\\
&&\qquad\left.-\frac{\sqrt{3}\alpha}{2}\sgn y(B^2)^{'}+2\alpha^2B^2\right\}+\text{h.o.t.}
\end{eqnarray}
The third line is the kinetic term for $B_{\mu}$, $-\frac{1}{4}g^{\mu\nu}g^{\rho\sigma}F_{\mu\rho}F_{\nu\sigma}$, up to higher order terms of the form $B^2\times r$, as can be seen after integrations by part on the unbounded directions $x^{\mu}$, whose total derivatives vanish. The linear terms in $B$ on the second line can be arranged noticing that $g^{\mu\nu}\partial_{\mu}B_{\nu}-B_{\rho}\Gamma^{\rho}_{\mu\nu}g^{\mu\nu}=\partial_{\mu}B^{\mu}+B^{\sigma}\Gamma_{\mu\sigma}^{\sigma}=\partial_{\mu}B^{\mu}+B^{\sigma}(\tilde \Gamma_{\mu\sigma}^{\sigma}-6\partial_{\sigma}r)$, where $\tilde \Gamma_{\mu\nu}^{\rho}$ are the Christoffel symbols computed from the metric $\tilde g_{\mu\nu}$. Using $\tilde \Gamma_{\mu\sigma}^{\mu}=\frac{1}{\sqrt{-\tilde g}}\partial_{\sigma}\sqrt{-\tilde g}$, integrating by part and again discarding a total derivative $\partial_{\sigma}$, we get the final result:
\begin{eqnarray}\label{eq:final_result_1}
\frac{1}{2}\hat N\sqrt{|\hat N^8\hat g|}\left(K^2-K_{\mu\nu}K^{\mu\nu}\right)&=&e^{-\sqrt{3}\alpha|y|}e^{-3r}\sqrt{-\tilde g}\left\{2\alpha^2+2\sqrt{3}\alpha\sgn yr^{'}+\frac{3}{2}(r^{'})^2\right.\nonumber\\
&+&\left.B^{\mu}\partial_{\mu}\left[-\frac{3}{2}r^{'}+\sqrt{3}\alpha\sgn y r+\frac{3}{2}r^{'}r\right]\right.\\
&-&\left.\frac{1}{4}g^{\mu\nu}g^{\rho\sigma}F_{\mu\rho}F_{\nu\sigma}-\frac{\sqrt{3}\alpha}{2}\sgn y(B^2)^{'}+2\alpha^2B^2\right\}+\text{h.o.t.}\nonumber
\end{eqnarray}
Finally, the $d$-dimensional Ricci scalar \eqref{eq:d_dim_ricci_scalar},
\begin{eqnarray}
R^{(d)}[\hat N^2\hat g_{\mu\nu}]&=&\frac{1}{2}\partial_{\mu}\left(\sqrt{-\det(\hat N^2\hat g_{\mu\nu})}\hat N^{-2}\hat g^{\mu\nu}\hat N\right)\partial_{\nu}\ln \left(-\det(\hat N^2\hat g_{\mu\nu})\right)\nonumber\\
&&\quad+\partial_{\mu}\hat N\partial_{\nu}\left(\sqrt{-\det(\hat N^2\hat g_{\mu\nu})}\hat N^{-2}\hat g^{\mu\nu}\right)\nonumber\\
&&\qquad-\frac{1}{2}\hat N\sqrt{-\det(\hat N^2\hat g_{\mu\nu})}\hat \Gamma_{\mu\nu}^{\rho}\partial_{\rho}\left(\hat N^{-2}\hat g^{\mu\nu}\right),
\end{eqnarray}
expressed in terms of the variables $N$, $g_{\mu\nu}$ and $B_{\rho}$, reads:
\begin{multline}\label{eq:full_Ricci_scalar}
R^{(d)}[N^{-2}(g_{\mu\nu}+B_{\mu}B_{\nu})]=\frac{1}{2}\partial_{\nu}\ln \left(-\det(N^{-2}(g_{\mu\nu}+B_{\mu}B_{\nu}))\right)\\
\times\partial_{\mu}\left(\sqrt{-\det(N^{-2}(g_{\mu\nu}+B_{\mu}B_{\nu}))}N^2(g^{\mu\nu}-xB^{\mu}B^{\nu})N^{-1}(1+B^2)^{-1/2}\right)\\
+\partial_{\mu}\left(N^{-1}(1+B^2)^{-1/2}\right)\partial_{\nu}\left(\sqrt{-\det(N^{-2}(g_{\mu\nu}+B_{\mu}B_{\nu}))}(N^2(g^{\mu\nu}-xB^{\mu}B^{\nu}))\right)\\
-\frac{1}{2}N^{-1}(1+B^2)^{-1/2}\sqrt{-\det(N^{-2}(g_{\mu\nu}+B_{\mu}B_{\nu}))}\Gamma_{\mu\nu}^{\rho}\partial_{\rho}\left(N^2(g^{\mu\nu}-xB^{\mu}B^{\nu})\right),
\end{multline}
with $\Gamma_{\mu\nu}^{\rho}$ the Christoffel symbols computed from the metric $N^{-2}(g_{\mu\nu}+B_{\mu}B_{\nu})$. Although rather involved, it is easy to see that this expression does not bring additional $B_{\mu}$ terms up to quadratic order in $B$: in the first two terms of \eqref{eq:full_Ricci_scalar}, quadratic terms in $B$ always appear in the form $\sim B^2\partial_{\mu}N\sim B^2\partial_{\mu}r$, which is an interaction term, while in the third term, quadratic terms in $B$ are multiplied by Christoffel symbols, and are thus again of the form $B^2\times\text{fluctuations}$. One can hence set $B_{\mu}$ to zero in the previous expression, and the computation then gives:
\begin{equation}\label{eq:final_result_2}
\frac{1}{2}R^{(d)}[N^{-2}g_{\mu\nu}]=-\frac{3}{4}e^{-\sqrt{3}\alpha |y|}\sqrt{-\tilde g}\tilde g^{\mu\nu}\partial_{\mu}r\partial_{\nu}r+\text{h.o.t.}
\end{equation}
Combining the results \eqref{eq:final_result_1} and \eqref{eq:final_result_2}, we deduce the total gravitational action, up to quadratic order in the KK vector $B_{\mu}$:
\begin{eqnarray}\label{eq:final_gravitational_action}
S_G&=&\int d^5xe^{-\sqrt{3}\alpha|y|}\sqrt{-\tilde g}\left\{-\frac{3}{4}(\partial_{\mu}r)^2+e^{-3r}\left[2\alpha^2+2\sqrt{3}\alpha\sgn yr^{'}+\frac{3}{2}(r^{'})^2\right]\right.\nonumber\\
&&\qquad\left.+\tilde B^{\mu}\partial_{\mu}\left(-\frac{3}{2}r^{'}+\sqrt{3}\alpha\sgn y r+\frac{3}{2}r^{'}r\right)\right.\nonumber\\
&&\qquad\left.-\frac{1}{4}e^{3r}\tilde g^{\mu\nu}\tilde g^{\rho\sigma}F_{\mu\rho}F_{\nu\sigma}-\frac{\sqrt{3}\alpha}{2}\sgn y(\tilde B^2)^{'}+2\alpha^2\tilde B^2\right\}+\text{h.o.t.},
\end{eqnarray}
where we have replaced $B^{\mu}=g^{\mu\nu}B_{\nu}=e^{3r}\tilde g^{\mu\nu}B_{\nu}\equiv e^{3r}\tilde B^{\mu}$, $B^2=g^{\mu\nu}B_{\mu}B_{\nu}=e^{3r}\tilde g^{\mu\nu}B_{\mu}B_{\nu}\equiv e^{3r}\tilde B^2$. 

\section{$\mathcal{N}=2$, $D=5$ supergravity in a nutshell}
\label{sect:append_5D_sugra}
\subsection{The ungauged theory}
\label{sect:append_ungauged_sugra}
Pure $\mathcal{N}=2$, $D=5$ supergravity is built out of the supergravity multiplet
\begin{equation}\label{eq:sugra_multiplet}
\left(e_{M}^m~,~\psi_M^i~,~A_M^0\right),
\end{equation}
which contains, in addition of the graviton $e_{M}^m$, two symplectic-Majorana gravitini $\psi_M^i$, with $i=1,2$ the $SU(2)_R$ index, and a vector field $A_M^0$, called the graviphoton. The most general $\mathcal{N}=2$, $D=5$ matter-coupled supergravity can then be obtained by coupling the supergravity multiplet with a given numbers of vector, tensor and hypermultiplets. In this paper, we will not consider the case of tensor and hypermultiplets and refer to \cite{General_matter_coupled_sugra_1,General_matter_coupled_sugra_2} for a general description of $\mathcal{N}=2$, $D=5$ supergravity coupled to these matter multiplets. Instead, we focus on $\mathcal{N}=2$ Maxwell-Einstein supergravity, obtained by coupling a given number $n_V$ of vector multiplets to the supergravity multiplet \eqref{eq:sugra_multiplet} \cite{Ungauged_5D_sugra}. A $\mathcal{N}=2$ vector multiplet, 
\begin{equation}
\left(A_M~,~\lambda^i~,~\varphi\right),
\end{equation}
contains a vector field $A_M$, a $SU(2)_R$ doublet of symplectic-Majorana spin-${1}/{2}$ fermions $\lambda^i$, called dilatini, and a real scalar $\varphi$. The total field content of the theory is thus
\begin{equation}
e_M^m~,~\psi_M^i~,~A_M^I~,~\lambda_i^{\tilde a}~,~\varphi^x,
\end{equation}
with $I=0,1,...,n_V$. The real scalars $\varphi^x$ describe a real $n_V$-dimensional manifold $\mathcal{M}$, whose structure has been called very special real geometry, and whose coordinate and local frame indices are respectively written $x=1,...,n_V$ and $\tilde a=1,...,n_V$. $\mathcal{M}$ is equipped with a metric $g_{xy}$ and a $n_V$-bein $f_x^{\tilde a}$, related through:
\begin{equation}\label{eq:nv_bein_def}
g_{xy}=f_x^{\tilde a}\delta_{\tilde a\tilde b}f_y^{\tilde b}.
\end{equation}
It turns out that $\mathcal{M}$ is better described as a submanifold of a $(n_V+1)$-dimensional Riemannian space, with coordinates $h^I(\varphi^x)$, with an embedding defined through the constraint
\begin{equation}\label{eq:embedding_constraint}
\mathcal{F}\equiv C_{IJK}h^Ih^Jh^K=1.
\end{equation}
$C_{IJK}$ are completely symmetric real constants, which will turn out to uniquely determine the whole theory. From $C_{IJK}$ and $h^I(\varphi^x)$, we define another set of variables $h_{I}(\varphi^x)$ through
\begin{equation}\label{eq:h_I}
h_I\equiv\frac{1}{3}\frac{\partial}{\partial h^I}C_{JKL}h^Jh^Kh^L=C_{IJK}h^Jh^K,
\end{equation}
so that $h_Ih^I=1$, as well as a symmetric tensor $G_{IJ}(\varphi^x)$ which can be seen as the metric tensor of the embedding $(n_V+1)$-dimensional space, raising and lowering the indices $I,J...$ according to
\begin{equation}
h_I\equiv G_{IJ}h^J,\qquad h^I\equiv G^{IJ}h_J.
\end{equation}
$G_{IJ}(\varphi^x)$ will appear to be the kinetic matrix of the vector fields. From the additional condition that $G_{IK}G^{KJ}=\delta_I^J$, it is easy to check that it can be written as
\begin{equation}\label{eq:G_IJ}
G_{IJ}=-2C_{IJK}h^K+3h_Ih_J.
\end{equation}
Introducing the quantities
\begin{equation}
h^I_x\equiv-\sqrt{\frac{3}{2}}\partial_xh^I,\qquad h_{Ix}\equiv G_{IJ}h^J_x=\sqrt{\frac{3}{2}}\partial_xh_I,
\end{equation}
the metric $g_{xy}(\varphi^x)$ of $\mathcal{M}$ is then defined as being the pullback of $G_{IJ}(\varphi^x)$ to $\mathcal{M}$:
\begin{equation}\label{eq:g_xy}
g_{xy}\equiv G_{IJ}h^I_xh^J_y=-2C_{IJK}h^I_xh^J_kh^K.
\end{equation}
Finally, we define the symmetric $T_{xyz}(\varphi^x)$ tensor by
\begin{equation}\label{eq:T_xyz}
T_{xyz}\equiv\sqrt{\frac{3}{2}}h_{Ix;y}h^I_z=-\sqrt{\frac{3}{2}}h_{Ix}h^{I}_{y;z}=C_{IJK}h^I_xh^J_yh^K_z,
\end{equation}
where a semicolon ";" denotes the covariant derivative associated with the Levi-Civita connection on $\mathcal{M}$, such that $g_{xy;z}=0$, as well as the quantity $\Phi_{Ixy}(\varphi^x)$ symmetric in its last two indices:
\begin{equation}\label{eq:Phi_Ixy}
\Phi_{Ixy}\equiv\sqrt{\frac{2}{3}}\left(\frac{1}{4}g_{xy}h_I+T_{xyz}h^z_I\right).
\end{equation}

With the formalism and notations introduced above, the Lagrangian of $\mathcal{N}=2$, $D=5$ supergravity coupled to $n_V$ vector multiplets can then be written as \cite{Ungauged_5D_sugra}: 
\begin{eqnarray}\label{eq:ungauged_lagrangian}
e^{-1}\mathcal{L}^{(0)}&=&\frac{1}{2}\mathcal{R}^{(5)}-\frac{1}{2}g_{xy}\partial_M\phi^x\partial^M\phi^y-\frac{1}{4}G_{IJ}F^I_{MN}F^{MNJ}+\frac{e^{-1}}{6\sqrt{6}}C_{IJK}\epsilon^{MNPQR}F_{MN}^IF_{PQ}^JA_R^K\nonumber\\
&-&\frac{1}{2}\bar{\psi}^i_M\gamma^{MNP}D_N\psi_{Pi}-\frac{1}{2}\bar{\lambda}^{i\tilde a}\gamma^M\left(D_M\delta^{\tilde a\tilde b}+\Omega_x^{\tilde a \tilde b}\partial_M\phi^x \right)\lambda_i^{\tilde b}-\frac{i}{2}\bar{\lambda}^{i\tilde a}\gamma^M\gamma^N\psi_{Mi}f^a_x\partial_N\phi^x\nonumber\\
&+&\frac{1}{4}h_I^{\tilde a}\bar{\lambda}^{i\tilde a}\gamma^M\gamma^{NP}\psi_{Mi}F_{NP}^I+\frac{i}{4}\Phi_{I\tilde a\tilde b}\bar{\lambda}^{i\tilde a}\gamma^{MN}\lambda_{i}^{\tilde b}F_{MN}^I\\
&-&\frac{3i}{8\sqrt{6}}h_I\left(\bar{\psi}^i_M\gamma^{MNPQ}\psi_{Ni}F^I_{PQ}+2\bar{\psi}^{Mi}\psi_i^NF^I_{MN}\right)+\mathcal{L}_{\text{4-fermions}}.\nonumber
\end{eqnarray}
The action $S=\int d^5x \mathcal{L}^{(0)}$ is invariant under the following $\mathcal{N}=2$ supersymmetry transformations:
\begin{eqnarray}
\label{eq:original_graviton_susy_transfo}
\delta e_M^m&=&\frac{1}{2}\bar{\epsilon}^i\gamma^m\psi_{Mi},\\
\label{eq:original_gravitini_susy_transfo}
\delta\psi_{Mi}&=&D_{M}(\hat\omega)\epsilon_i+\frac{i}{4\sqrt{6}}h_Ie_M^m\left(\gamma_{mnl}\epsilon_i-4\eta_{mn}\gamma_l\epsilon_i\right)\hat{F}^{nlI}-\frac{1}{12}e_M^m\gamma_{mn}\epsilon^j\bar{\lambda}_i^{\tilde a}\gamma^n\lambda_j^{\tilde a}\nonumber\\
&+&\frac{1}{48}e_M^m\gamma_{mnl}\epsilon^j\bar{\lambda}_i^{\tilde a}\gamma^{nl}\lambda_j^{\tilde a}+\frac{1}{6}e_M^m\epsilon^j\bar{\lambda}_i^{\tilde a}\gamma_m\lambda_j^{\tilde a}-\frac{1}{12}e^m_M\gamma^n\epsilon^j\bar{\lambda}_i^{\tilde a}\gamma_{mn}\lambda_j^{\tilde a},\\
\label{eq:original_dilaton_susy_transfo}
\delta\varphi^x&=&\frac{i}{2}\bar{\epsilon}^i\lambda_i^{\tilde a}f_{\tilde a}^x,\\
\label{eq:original_vector_susy_transfo}
\delta A_M^I&=&-\frac{1}{2}e_M^m\bar{\epsilon}^i\gamma_m\lambda_i^{\tilde a}h_{\tilde a}^I+\frac{i\sqrt{6}}{4}\bar{\psi}^i_M\epsilon_ih^I,\\
\label{eq:original_dilatini_susy_transfo}
\delta \lambda_i^{\tilde a}&=&-\frac{i}{2}f^{\tilde a}_x\gamma^M(\hat{\partial}_M\varphi)^x\epsilon_i-\delta\varphi^x\Omega_x^{\tilde a \tilde b}\lambda_i^{\tilde b}+\frac{1}{4}h_I^{\tilde a}\gamma^{mn}\epsilon_i\hat{F}_{mn}^I\nonumber\\
&-&\frac{i}{4\sqrt{6}}T^{\tilde a}_{\tilde b\tilde c}\left(-3\epsilon^j\bar{\lambda}_i^{\tilde b}\lambda_j^{\tilde c}+\gamma_m\epsilon^j\bar{\lambda}_i^{\tilde b}\gamma^m\lambda_j^{\tilde c}+\frac{1}{2}\gamma_{mn}\epsilon^j\bar{\lambda}_i^{\tilde b}\gamma^{mn}\lambda_j^{\tilde c}\right).
\end{eqnarray}
The hatted quantities $\hat X$ are the supercovariantization of the unhatted ones $X$, namely:
\begin{eqnarray}
\hat\omega_{Mmn}(e)&\equiv&\omega_{Mmn}(e)-\frac{1}{4}\left(\bar{\psi}_n^i\gamma_M\psi_{mi}+2\bar{\psi}_M^i\gamma_{[n}\psi_{m]i}\right),\\
D_M(\hat\omega)\epsilon_i&\equiv&\partial_M\epsilon_i+\frac{1}{4}\hat\omega_M^{mn}(e)\gamma_{mn}\epsilon_i,\\
\hat F_{MN}^I&\equiv&F_{MN}^I+\frac{i\sqrt{6}}{4}h^I\bar{\psi}^i_{[M}\psi_{N]i}+h_{\tilde a}^I\bar{\psi}^j_{[M}\gamma_{N]}\lambda^{\tilde a}_j,\\
(\hat{\partial}_M\varphi)^x&\equiv&\partial_M\varphi^x-\frac{i}{2}f_{\tilde a}^x\bar{\psi}^j_M\lambda^{\tilde a}_j.
\end{eqnarray}
Knowing the symmetric constants $C_{IJK}$, one can find the functions $h^I(\varphi^x)$ by solving the constraint \eqref{eq:embedding_constraint}, then deduce the functions $h_I(\varphi^x)$, $G_{IJ}(\varphi^x)$, $g_{xy}(\varphi^x)$, $T_{xyz}(\varphi^x)$ and $\Phi_{Ixy}(\varphi^x)$ using Eqs. \eqref{eq:h_I}, \eqref{eq:G_IJ}, \eqref{eq:g_xy}, \eqref{eq:T_xyz} and \eqref{eq:Phi_Ixy} respectively, and thus completely determine the above Lagrangian and susy transformations. Therefore, even for a fixed number $n_V$ of vector multiplets, several matter-coupled $\mathcal{N}=2$, $D=5$ supergravity theories are possible, depending on the geometry of the scalar manifold $\mathcal{M}$, in turn determined by the constants $C_{IJK}$. In Section \ref{sec:runaway_scal_pot}, we list the different $\mathcal{N}=2$, $D=5$ supergravity theories coupled to $n_V=1$ vector multiplet whose $U(1)_R$ gauging produces the runaway scalar potential of the non-critical string, and we now recall the main ideas of the $U(1)_R$ gauging of $\mathcal{N}=2$, $D=5$ supergravity.

\subsection{$U(1)_R$ gauging of $\mathcal{N}=2$, $D=5$ supergravity}
\label{sect:append_gauged_sugra}
The global symmetry group $G$ of the Lagrangian \eqref{eq:ungauged_lagrangian} can be written as $G=H\times SU(2)_R$, where $H$ is the group of linear transformations acting on $h^I$ and leaving $C_{IJK}$ invariant\footnote{Note that in the most general case, $H$ is only a subgroup of the isometry group of $\mathcal{M}$.}, and $SU(2)_R$ the $R$-symmetry group acting on the fermions $\psi_M^i$ and $\lambda_i^{\tilde a}$. One can then arbitrarily choose to gauge the $U(1)_R$ subgroup of $SU(2)_R$\footnote{Since the vector fields are invariant under $SU(2)_R$, they cannot be used as non-Abelian gauge fields for $SU(2)_R$, and the full $SU(2)_R$ group cannot be gauged.}, a subgroup of $H$ or a combination of both. These general gaugings have been described in \cite{General_gaugings}, and we will only consider in the following the simplest case of the $U(1)_R$ gauging alone, following \cite{GST, Gauged_5D_sugra}.

The gauging along the $U(1)_R$ subgroup of the $SU(2)_R$ $R$-symmetry group is achieved by defining the $U(1)_R$ gauge field as a linear combination of the $n_V+1$ vector fields $A_M^I$,
\begin{equation}\label{eq:direction_gauging}
A_M\equiv v_IA_M^I,
\end{equation} 
with $v_I$ a set of $n_V+1$ real constants. In the same time, we promote the Lorentz covariant derivatives of the fermionic fields to Lorentz-$U(1)_R$ covariant derivatives\footnote{The scalars $\varphi^x$ are kept uncharged under the $U(1)_R$, and their partial derivatives are thus not replaced by covariant derivatives in the gauging procedure.}
\begin{subequations}\label{eq:fermion_cov_deriv}
\begin{align}
D_M\lambda^{\tilde ai}&\rightarrow(\mathcal{D}_M\lambda^{\tilde a})^i\equiv D_M\lambda^{\tilde ai}+gA_M\delta^{ij}\lambda_j^{\tilde a},\\
D_M\psi_N^i&\rightarrow(\mathcal{D}_M\psi_N)^i\equiv D_M\psi_N^i+gA_M\delta^{ij}\psi_{Nj},
\end{align}
\end{subequations}
where $g$ is the $U(1)_R$ coupling constant. These replacements in the original Lagrangian $\mathcal{L}^{(0)}$ \eqref{eq:ungauged_lagrangian} and in the susy transformations \eqref{eq:original_graviton_susy_transfo}-\eqref{eq:original_dilatini_susy_transfo} will break the supersymmetry. It can be recovered via the addition to $\mathcal{L}^{(0)}$ of a scalar potential $P$ and fermion mass terms, given by
\begin{equation}\label{eq:additional_lagrangian}
e^{-1}\mathcal{L}^{'}=-g^2P-\frac{i\sqrt{6}}{8}g\bar{\psi}^i_M\gamma^{MN}\psi_N^j\delta_{ij}P_0-\frac{g}{\sqrt{2}}\bar{\lambda}^{i\tilde{a}}\gamma^M\psi_M^j\delta_{ij}P_{\tilde a}+\frac{ig}{2\sqrt{6}}\bar{\lambda}^{i\tilde{a}}\lambda^{j\tilde{b}}\delta_{ij}P_{\tilde a \tilde b},
\end{equation}
as well as adding new $g$-dependent parts to the original susy transformations \eqref{eq:original_gravitini_susy_transfo} and \eqref{eq:original_dilatini_susy_transfo} of the gravitini and dilatini, of the form:
\begin{eqnarray}
\delta{'}\psi_{Mi}&=&-\frac{ig}{2\sqrt{6}}P_0\gamma_M\epsilon_{ij}\delta^{jk}\epsilon_k,\\
\delta{'}\lambda_i^{\tilde a}&=&-\frac{g}{\sqrt{2}}P^{\tilde a}\epsilon_{ij}\delta^{jk}\epsilon_k.
\end{eqnarray}
Supersymmetry then requires the new functions $P(\varphi^x)$, $P_0(\varphi^x)$, $P^{\tilde a}(\varphi^x)$ and $P_{\tilde a \tilde b}$ to satisfy
\begin{eqnarray}
P&=&-P_0^2+P_{\tilde a}P^{\tilde a},\\
P_0&=&2h^Iv_I,\\
P^{\tilde a}&=&\sqrt{2}h^{I\tilde a}v_I,\\
P_{\tilde a\tilde b}&=&\frac{1}{2}\delta_{\tilde a\tilde b}P_0+2\sqrt{2}T_{\tilde a\tilde b\tilde c}P^{\tilde c}.
\end{eqnarray}

\section{Effective theory of the heterotic string}
\label{sect:append_heterotic_string}
In this appendix, we check that the supergravity theory introduced in Section \ref{sec:minimal_susy_extension} is an effective theory of an heterotic string theory, in agreement with \cite{heterotic_string_dual_5D}. We recall the Lagrangian \eqref{eq:total_bosonic_action} of its bosonic sector, in the Einstein frame:
\begin{eqnarray}\label{eq:append_total_bosonic_action}
{e_E}^{-1}\mathcal{L}^{\text{bos}}&=&\frac{1}{2}\mathcal R^{(5)}[G_{MN}^E]-\frac{1}{2}\partial_M\phi\partial^M\phi-e^{\frac{2}{\sqrt{3}}\phi}\Lambda\nonumber\\
&&\qquad-\frac{1}{8}e^{\frac{4}{\sqrt{3}}\phi}F_{MN}^0F^{MN0}-\frac{1}{4}e^{-\frac{2}{\sqrt{3}}\phi}F_{MN}^1F^{MN1}\\
&&\qquad\qquad+\frac{{e_E}^{-1}}{6\sqrt{6}}C_{011}\epsilon^{MNPQR}\left(A_M^0F_{NP}^1F_{QR}^1+2A_M^1F_{NP}^1F_{QR}^0\right),\nonumber
\end{eqnarray}
where $e_E$ stands for the $5D$ Einstein frame f\"{u}nfbein, while $e_S$ will denote later its string frame counterpart. In order to show the heterotic nature of this action, one must dualize the graviphoton $A_M^0$ into the Kalb-Ramond (KR) two-form $B_{MN}$, whose completely antisymmetric three-form field strength will be written $H_{MNP}=\partial_{[M}B_{NP]}$. To this purpose, we consider the action as being a functionnal of $F_{MN}^0$ rather than $A_M^0$, and add the Lagrange multiplier term
\begin{equation}
\mathcal{L}_{LM}=\frac{1}{4}\epsilon^{MNPQR}F_{MN}^0H_{PQR},
\end{equation}
so that the equation of motion of $B_{MN}$ enforces $F^0$ to be closed. On the other hand, the equation of motion for $F_{MN}^0$, 
\begin{equation}
F^{MN0}={e_E}^{-1}e^{-\frac{4}{\sqrt{3}}\phi}\epsilon^{MNPQR}\left(H_{PQR}+\frac{2C_{011}}{\sqrt 6}A_P^1F_{QR}^1\right),
\end{equation}
can be used in order to completely eliminate it in $\mathcal{L}^{\text{bos}}+\mathcal{L}_{LM}$, leading to:
\begin{eqnarray}\label{eq:einstein_frame_lagrangian}
{e_E}^{-1}\mathcal{L}^{\text{bos}}+\mathcal{L}_{LM}&=&\frac{1}{2}\mathcal R^{(5)}[G_{MN}^E]-\frac{1}{2}\partial_M\phi\partial^M\phi-\frac{3}{2}e^{-\frac{4}{\sqrt{3}}\phi}\left(H_{PQR}+\frac{2C_{011}}{\sqrt 6}A_{[P}^1F_{QR]}^1\right)^2\nonumber\\
&&\qquad-\frac{1}{4}e^{-\frac{2}{\sqrt{3}}\phi}F_{MN}^1 F^{MN1}-e^{\frac{2}{\sqrt{3}}\phi}\Lambda.
\end{eqnarray}
In order to move from the Einstein to the string frame, we perform the Weyl transformation
\begin{equation}
G_{MN}^E=e^{-2\sigma}G_{MN}^S,
\end{equation}
which leads, in $D$ space-time dimensions, to the well-known relations
\begin{eqnarray}
e_E&=&e^{-D\sigma}e_S,\\
\mathcal R^{(D)}[G_{MN}^E]&=&e^{2\sigma}\left[\mathcal R^{(D)}[G_{MN}^S]+2(D-1)\Box\sigma-(D-1)(D-2)G_S^{MN}\partial_M\sigma\partial_N\sigma\right]\!\!.
\end{eqnarray}
For $\sigma=\frac{\phi}{\sqrt{3}}$, $D=5$ and after discarding a total derivative, we get the string frame Lagrangian
\begin{eqnarray}
\mathcal{L}^{\text{bos}}_S&=&e_Se^{-\sqrt{3}\phi}\left\{\frac{1}{2}\mathcal{R}^{(5)}[G_{MN}^S]+\frac{3}{2}\partial_M\phi\partial^M\phi-\frac{3}{2}\left(H_{PQR}+\frac{2C_{011}}{\sqrt 6}A_{[P}^1F_{QR]}^1\right)^2\right.\nonumber\\
&&\qquad\left.-\frac{1}{4}F_{MN}^1F^{MN1}-\Lambda\right\},
\end{eqnarray}
which is indeed the effective Lagrangian density of an heterotic superstring theory. The topological term present in \eqref{eq:append_total_bosonic_action} in the vector formulation is translated in the KR formulation into a gauge Chern-Simons term $\omega_3=A^1\wedge dA^1$, which combines with $H_{PQR}$ to form the generalized field strength $\mathcal{H}_3\sim H_3+\omega_3$ satisfying the modified Bianchi identity $d\mathcal{H}_3\sim dA^1\wedge dA^1$.

This analysis therefore shows that the $4D$ vector $A_{\mu}^0$, which becomes massive after compactification on a LD background as shown in Section \ref{sect:compactified_theory}, is the dual of the vector $B_{\mu 5}$ coming from the dimensional reduction of the $5D$ KR two-form $B_{MN}$.

\acknowledgments
Work partially performed by I.A. as International Professor of the Francqui Foundation, Belgium.



\begin{thebibliography}{99}

\bibitem{Berkooz:1997cq}
  M.~Berkooz, M.~Rozali and N.~Seiberg,
  ``Matrix description of M theory on T**4 and T**5,''
  Phys.\ Lett.\ B {\bf 408} (1997) 105
  [hep-th/9704089];\\
  N.~Seiberg,
  ``New theories in six-dimensions and matrix description of M theory on T**5 and T**5 / Z(2),''
  Phys.\ Lett.\ B {\bf 408} (1997) 98
  [hep-th/9705221].

\bibitem{Review_LST}
For a review see: O.~Aharony, ``A brief review of `little string theories' '', Class.\ Quantum Grav.\  {\bf 17} (2000) 929
[arXiv:hep-th/9911147];\\
D.~Kutasov, ``Introduction to Little String Theory'', Superstrings and related matters. Proceedings, Spring School, Trieste, Italy, April 2-10, 2001 {\bf 7} (2002) 165-209

\bibitem{Duality_LST_LD}
O.~Aharony, M.~Berkooz, D.~Kutasov and N.~Seiberg, ``Linear Dilatons, NS5-branes and Holography'', JHEP {\bf 9810} (1998) 004
[arXiv:hep-th/9808149v2]

\bibitem{Antoniadis:2001sw}
  I.~Antoniadis, S.~Dimopoulos and A.~Giveon,
  ``Little string theory at a TeV,''
  JHEP {\bf 0105} (2001) 055
  [hep-th/0103033].

\bibitem{Pheno_LST}
I.~Antoniadis, A.~Arvanitaki, S.~Dimopoulos and A.~Giveon, ``Phenomenology of TeV Little String Theory from Holography'', Phys.~Rev.~Lett. {\bf 108} (2012) 081602
[arXiv:1102.4043]

\bibitem{Baryakhtar:2012wj}
  M.~Baryakhtar,
  ``Graviton Phenomenology of Linear Dilaton Geometries,''
  Phys.\ Rev.\ D {\bf 85} (2012) 125019
  [arXiv:1202.6674 [hep-ph]].

\bibitem{Radion_pheno}
P.~Cox and T.~Gherghetta, ``Radion Dynamics and Phenomenology in the Linear Dilaton Model'', JHEP {\bf 1205} (2012) 149
[arXiv:1203.5870v2 [hep-ph]]

\bibitem{Choi:2015fiu}
  K.~Choi and S.~H.~Im,
  ``Realizing the relaxion from multiple axions and its UV completion with high scale supersymmetry,''
  JHEP {\bf 1601} (2016) 149
  [arXiv:1511.00132 [hep-ph]];\\
  D.~E.~Kaplan and R.~Rattazzi,
  ``Large field excursions and approximate discrete symmetries from a clockwork axion,''
  Phys.\ Rev.\ D {\bf 93} (2016) no.8,  085007
  [arXiv:1511.01827 [hep-ph]].

\bibitem{Giudice:2016yja}
  G.~F.~Giudice and M.~McCullough,
  ``A Clockwork Theory,''
  JHEP {\bf 1702} (2017) 036
  [arXiv:1610.07962 [hep-ph]];\\
  G.~F.~Giudice, Y.~Kats, M.~McCullough, R.~Torre and A.~Urbano,
  ``Clockwork/linear dilaton: structure and phenomenology,''
  JHEP {\bf 1806} (2018) 009
  [arXiv:1711.08437 [hep-ph]].
  
  \bibitem{Craig:2017cda}
  N.~Craig, I.~Garcia Garcia and D.~Sutherland,
  ``Disassembling the Clockwork Mechanism,''
  JHEP {\bf 1710} (2017) 018
  [arXiv:1704.07831 [hep-ph]].


\bibitem{ADM}
R.~L.~Arnowitt, S.~Deser and C.~W.~Misner, ``The dynamics of general relativity'', Gen.~Rel.~Grav. {\bf 40} (2008) 1997-2027
[arXiv:gr-qc/0405109]

\bibitem{Kehagias:2017grx}
  A.~Kehagias and A.~Riotto,
  ``The Clockwork Supergravity,''
  JHEP {\bf 1802} (2018) 160
  [arXiv:1710.04175 [hep-th]].

\bibitem{IA_CM}
I.~Antoniadis, A.~Delgado, C.~Markou and S.~Pokorski, ``The effective supergravity of Little String Theory'', Eur.\ Phys.\ J.\ C {\bf 78} (2018) 146
[arXiv:1710.05568 [hep-th]]

\bibitem{Ungauged_5D_sugra}
M.~G\"unaydin, G.~Sierra and P.~K.~Townsend, ``The geometry of $\mathcal{N}=2$ Maxwell-Einstein supergravity and Jordan algebras'', Nucl.\ Phys.\ {\bf B242} (1984) 89

\bibitem{GST}
M.~G\"unaydin, G.~Sierra and P.~K.~Townsend, ``Vanishing potentials in gauged $\mathcal{N}=2$ supergravity: an application of Jordan algebras'', Phys.\ Lett.\ B {\bf 144} (1984) 41-45

\bibitem{Gauged_5D_sugra}
M.~G\"unaydin, G.~Sierra and P.~K.~Townsend, ``Gauging the $D=5$ Maxwell-Einstein supergravity theories: more on Jordan algebras'', Nucl.\ Phys.\ {\bf B253} (1985) 573

\bibitem{Randall:1999ee}
  L.~Randall and R.~Sundrum,
  ``A Large mass hierarchy from a small extra dimension,''
  Phys.\ Rev.\ Lett.\  {\bf 83} (1999) 3370
  [hep-ph/9905221].

\bibitem{Farakos}
K.~Farakos, A.~Kehagias, G.~Koutsoumbas, ``Gauge Field Localization in the Linear Dilaton Background'', Phys.\ Lett.\ B {\bf 807} (2020) 135549
[arXiv:2004.14856 [hep-ph]].

\bibitem{Clockwork_SM}
Y.-J.~Kang, S.~Kim and H.~M.~Lee, ``The Clockwork Standard Model'', JHEP {\bf 09} (2020) 005
[arXiv:2006.03043v3 [hep-ph]].

\bibitem{Constraints}
C.~Csaki, M.~L.~Graesser and G.~D.~Kribs, ``Radion Dynamics and Electroweak Physics'', Phys.~Rev.~D {\bf 63} (2001) 065002
[arXiv:hep-th/0008151v2]

\bibitem{Physical_scalar}
L.~Kofman, J.~Martin and M.~Peloso, ``Exact identification of the radion and its coupling to the observable sector'', Phys.~Rev.~D {\bf 70} (2004) 085015
[arXiv:hep-ph/0401189]

\bibitem{heterotic_string_dual_5D}
I.~Antoniadis, S.~Ferrara and T.~R.~Taylor, ``$\mathcal{N}=2$ heterotic superstring and its dual theory in five-dimensions'', Nucl.\ Phys.\  {\bf B460} (1996) 489-505
[arXiv:hep-th/9511108]


\bibitem{Ferrara:1983gn}
S.~Ferrara and P.~van Nieuwenhuizen,
``Noether Coupling of Massive Gravitinos to $N=1$ Supergravity,''
Phys. Lett. B \textbf{127} (1983), 70-74



\bibitem{Ferrara:1995gu}
S.~Ferrara, L.~Girardello and M.~Porrati,
``Minimal Higgs branch for the breaking of half of the supersymmetries in N=2 supergravity,''
Phys. Lett. B \textbf{366} (1996), 155-159
[arXiv:hep-th/9510074 [hep-th]].



\bibitem{Ferrara:1995xi}
S.~Ferrara, L.~Girardello and M.~Porrati,
``Spontaneous breaking of N=2 to N=1 in rigid and local supersymmetric theories,''
Phys. Lett. B \textbf{376} (1996), 275-281
[arXiv:hep-th/9512180 [hep-th]].



\bibitem{FVP}
D.~Z.~Freedman and A.~Van Proeyen, ``Supergravity,''
Cambridge University Press, Cambridge, 2012.



\bibitem{General_matter_coupled_sugra_1}
A.~Ceresole and G.~Dall'Agata, ``General matter-coupled $\mathcal{N}=2$, $D=5$ gauged supergravity'', Nucl.\ Phys.\  {\bf B585} (2000) 143-170
[arXiv:hep-th/0004111v2]

\bibitem{General_matter_coupled_sugra_2}
E.~Bergshoeff, S.~Cucu, T.~de Wit, J.~Gheerardyn, S.~Vandoren and A.~Van Proeyen, ``$\mathcal{N}=2$ supergravity in five dimensions revisited'', Class.\ Quantum Grav.\  {\bf 21} (2004) 3015
[arXiv:hep-th/0403045]

\bibitem{General_gaugings}
M.~G\"unaydin and M.~Zagermann, ``The gauging of five-dimensional, $\mathcal{N}=2$ Maxwell-Einstein supergravity theories coupled to tensor multiplets'', Nucl.\ Phys.\  {\bf B572} (2000) 131-150
[arXiv:hep-th/9912027]



\end{thebibliography}


\end{document}